\def\year{2022}\relax
\documentclass[letterpaper]{article} 
\usepackage{aaai22}  
\usepackage{times}  
\usepackage{helvet}  
\usepackage{courier}  
\usepackage[hyphens]{url}  
\usepackage{graphicx} 
\usepackage{natbib}  
\usepackage{caption} 

\usepackage{tabularx}

\DeclareCaptionStyle{ruled}{labelfont=normalfont,labelsep=colon,strut=off} 
\usepackage{algorithm}
\usepackage{algorithmic}
\usepackage{subcaption}

\usepackage{xcolor}
\newcommand{\answerYes}[1]{\textcolor{blue}{#1}} 
\newcommand{\answerNo}[1]{\textcolor{teal}{#1}} 
\newcommand{\answerNA}[1]{\textcolor{gray}{#1}} 
 
\usepackage{newfloat}
\usepackage{listings}
\floatstyle{ruled}
\newfloat{listing}{tb}{lst}{}
\floatname{listing}{Listing}
\title{Examining (Political) Content Consumption on Facebook Through Data Donation}

\author{
    Joao Couto\textsuperscript{\rm 1},
    Kiran Garimella\textsuperscript{\rm 1}
}
\affiliations{
    \textsuperscript{\rm 1}Rutgers University \\

    jm3175@rutgers.edu, 
    kiran.garimella@rutgers.edu
}

\begin{document}

\maketitle
\thispagestyle{plain}
\pagestyle{plain}

\begin{abstract}
This paper investigates the usage patterns of Facebook among different demographics in the United States, focusing on the consumption of political information and its variability across age, gender, and ethnicity. Employing a novel data donation model, we developed a tool that allows users to voluntarily share their interactions with public Facebook groups and pages, which we subsequently enrich using CrowdTangle. This approach enabled the collection and analysis of a dataset comprising over 1,200 American users. Our findings indicate that political content consumption on Facebook is relatively low, averaging around 17\%, and exhibits significant demographic variations. Additionally, we provide insights into the temporal trends of these interactions. The main contributions of this research include a methodological framework for studying social media usage in a privacy-preserving manner, a comprehensive dataset reflective of current engagement patterns, and descriptive insights that highlight demographic disparities and trends over time. This study enhances our understanding of social media’s role in information dissemination and its implications for political engagement, offering a valuable resource for researchers and policymakers in a landscape where direct data access is diminishing.

\end{abstract}

\section{Introduction}

Understanding social media usage patterns is pivotal in the digital age, where platforms such as Facebook significantly influence public opinion, cultural norms, and political landscapes. Facebook, utilized by approximately 70\% of Americans, many of whom consume news and other information through this medium, serves as a central hub for digital interactions~\cite{pewresearchAmericansSocial}. Despite its widespread use, comprehensive, empirical studies that unravel how different demographics engage with content on Facebook are scarce. This research aims to bridge this gap by exploring critical questions about individual usage patterns, including the consumption of political information and how these patterns vary across different age groups, genders, and ethnicities, as well as their evolution over time.

The necessity to understand these patterns stems not only from Facebook's substantial user base but also from its role in disseminating information. However, a significant challenge in studying Facebook usage is the restricted access to user data, which is predominantly controlled by the platform itself. This limitation poses a significant barrier for independent researchers who seek to analyze the platform’s impact without access to internal data, as traditional methods often rely on public data that is either incomplete or not representative of broader user activities.

Historically, studies exploring social media usage were conducted in the earlier phases of these networks' development and did not capture data at the granularity required today. Moreover, with growing concerns about misinformation and data privacy, there is an increased need for updated methodologies that respect user privacy while providing valuable insights.

To overcome these challenges, this paper introduces a novel approach utilizing a data donation model, where users voluntarily share their interactions with public Facebook groups and pages. This data is then enriched using CrowdTangle, a tool that allows for an extended analysis of public posts and interactions. Our methodology has enabled us to compile a comprehensive dataset from over 1,200 American users, encompassing a wide range of age groups, genders, and ethnic backgrounds. 
Our analysis reveals that political content consumption on Facebook is relatively low, averaging around 17\% overall, which underscores the modest role that political engagement plays within the broader spectrum of social media activity. This level of engagement exhibits notable demographic variations: men are found to consume more political content, while Hispanics and younger individuals aged 18-24 consume the least, with the latter group's consumption nearly half that of those in the 65+ age bracket. Besides political content, our findings indicate that users predominantly engage with lifestyle, entertainment, and religious content, which collectively form a significant part of their social media diet. Importantly, our data collection methodology provides a nuanced view of Facebook usage, capturing intricate patterns of content consumption and enabling detailed analyses over time. These insights are critical for understanding the broader implications of social media on public discourse and misinformation.



While our study provides extensive insights, it is important to note that our results are descriptive and rely on data that represents only a segment of Facebook's total activity. Despite this, the implications of our research are profound, offering a foundation for further studies in a post-API era where direct access to platform data is increasingly restricted. Additionally, in line with our commitment to transparency and furthering research in this field, we will make our data and code available to the public, adhering to the policies set by Facebook/CrowdTangle. This research not only advances our understanding of social media dynamics but also equips researchers, policymakers, and the public with the knowledge needed to navigate and potentially regulate the digital landscape more effectively.

\section{Background and Related Work}

\noindent\textbf{Data donation}.
Data donation, as explored by Ohme et al.~\cite{ohme2023digital}, encompasses a variety of advanced data collection techniques that are crucial for capturing and analyzing social media content effectively. These techniques are designed to address the challenges of gathering raw social media data and transforming it into meaningful metrics that reflect the nuanced impacts of social media usage across different demographic and psychographic levels. This approach allows for a detailed examination of how users interact with content and how these interactions influence broader social and behavioral outcomes.

Keusch et al.~\cite{keusch2023you} further investigate the specific nuances of data donation on Facebook, highlighting the factors that influence the success or failure of these initiatives. Their findings emphasize the importance of trust in the research team and distrust in the platform (Facebook) as significant predictors of user participation in data donation schemes. Interestingly, the manner in which data donation requests are framed appears to have minimal impact on participation rates, suggesting that other factors such as user engagement with the platform and their perceptions of privacy and data security are more critical.

Due to the changes in access by various social media platforms, which researchers term the dawn of the post-API age~\cite{freelon2018computational}, the implementation of data donation practices on a larger scale, such as in projects by the National Internet Observatory, are increasingly becoming common~\cite{meyer2023enhancing}. Despite its increasing popularity, the quantitative exploration of data donation, especially on platforms like Facebook, remains limited. This gap highlights the novelty of the current study, which aims to operationalize data donation in innovative ways that could set a precedent for future research in a landscape constrained by limited access to traditional API-based data collection methods.

While data donation offers a promising avenue for obtaining high-quality, consent-based data with a well-defined sampling frame, there are inherent challenges and limitations to this approach. Conducting data donation at a large scale presents logistical challenges, including ensuring sufficient user participation and dealing with potential non-cooperation from social media platforms, which may hinder data collection efforts. Additionally, the significant upfront effort required by researchers may not always result in successful data acquisition, illustrating the non-trivial nature of this method compared to more traditional data collection techniques.

Nevertheless, the benefits of data donation are substantial. This method ensures the collection of high-quality data that is both ethical, given its opt-in nature and clear consent processes, and practical, allowing for the integration of additional research tools such as surveys~\cite{de2017linking}. The ability to capture public data without casting an excessively wide net is particularly valuable for studies aimed at estimating prevalence and understanding behaviors at a population scale.
By building on the foundations laid by previous studies and addressing the unique challenges of the post-API era, this paper contributes to the ongoing dialogue on how best to harness the power of user-generated content for scientific research.

\noindent\textbf{Audit studies}.
In the evolving landscape of social media research, audit studies represent a crucial methodology for understanding platform mechanisms and their broader societal impacts. Here were explore the contributions and methodologies of recent audit studies on various social platforms, distinguishing how these approaches complement and diverge from our data donation-based research.

A clear extension and application of data donations in auditing is the concept of `end-user audits', proposed by Lam et al.~\cite{lam2022end}. These audits involve typical platform users in the auditing process to perform certain tasks which collectively can help in auditing a socio technical system.
The field of algorithmic auditing, although not the focus of our research, provides valuable insights into the internal workings of social media platforms. Metaxa et al.~\cite{metaxa2021auditing} highlight the significance of understanding the algorithms that underpin user interactions and content delivery on these platforms. While our study does not perform a direct algorithm audit, the methodologies developed for such audits are informative for our approach.

Audit studies on various platforms have been extensively documented. For example, Robertson et al.\cite{robertson2018auditing} audit Google search algorithms to understand their operational biases and outcomes. Similarly, audits on YouTube by Hosseinmardi et al.\cite{hosseinmardi2024causally} explore biases in video recommendations, while Huszar et al.\cite{huszar2022algorithmic} focus on Twitter to analyze content dissemination practices. Beyond social networks, platforms like Amazon and Uber have also been subjects of audit studies, with works by Juneja et al.\cite{juneja2021auditing} and Chen et al.~\cite{chen2015peeking} examining marketplace biases and service dynamics, respectively. These studies typically employ methodologies involving artificial accounts (such as bots or sock puppets) or collaborate directly with the platform to gather data.

Our current research, centered on data donation, does not directly engage in such audits. However, the model and data collected through our methodology could be readily adapted for auditing specific platform features, such as group recommendations. The involvement of real users in our study provides a unique opportunity to extend the research to include user surveys, which can offer deeper insights into user perceptions and experiences, similar to those gathered in end-user audits. By leveraging both donated data and direct user feedback, our study could potentially bridge the gap between traditional audit studies and innovative data donation approaches, offering a comprehensive view of platform dynamics and user interactions.


\noindent\textbf{Use of CrowdTangle}.
In examining the landscape of social media research, particularly the consumption of news and political content on platforms like Facebook, it becomes imperative to explore the methodologies and findings from existing studies. This subsection delves into how Facebook data has been utilized in research, the limitations of current data collection methods, and the implications of these studies for understanding social media dynamics.

Since CrowdTangle, a tool that allows researchers to track interactions and trends on Facebook~\cite{crowdtangleAboutCrowdTangle}, became available to journalists, access to Facebook data has significantly expanded. A search for ``CrowdTangle" on Google Scholar returns a substantial number of papers (2,740), underscoring its widespread use in academic research. However, a major limitation of CrowdTangle -- and indeed, much of the data from Facebook -- is the absence of demographic details related to content consumption. This gap poses a significant challenge for researchers aiming to understand how different groups engage with content on the platform.

\noindent\textbf{Studies on other platforms}.
The reliance on survey-based methods for estimating social media usage is prevalent, with instruments such as those deployed by Pew Research providing insights into these patterns~\cite{pewresearchAmericansSocial}. However, such surveys often suffer from biases related to self-reporting. Furthermore, existing studies provide a fragmented view of social media engagement. For instance, Bestvater et al.~\cite{bestvater2022politics} report that one-third of tweets from U.S. adults on Twitter are political, with those aged 50 and older contributing to 78\% of all political tweets. This highlights the uneven distribution of content production across different age groups.

Similarly, Allen et al.~\cite{allen2020evaluating} explore news consumption across various media, finding that only a maximum of 14\% of Americans' media diets consist of news content. Moreover, television remains the dominant source of news, overshadowing online sources by a substantial margin. This points to the significant role traditional media still plays in news consumption, despite the growing influence of digital platforms.

TODO: \cite{wojcieszak2009online} show that around 17\% of the people who were online had some political discussion on discussion boards in 2009.

These findings are typically derived from large-scale nationally representative surveys or data obtained from professional panel companies. Both sources are prohibitively expensive for most researchers, which limits the scope and frequency of such studies. This financial barrier underscores the need for more accessible, cost-effective methods for gathering and analyzing social media data, particularly as it relates to demographic segmentation and content type.





\section{Data Donation}

\begin{figure*}
\centering 
    \includegraphics[width=\textwidth,trim=10 180 10 190, clip]{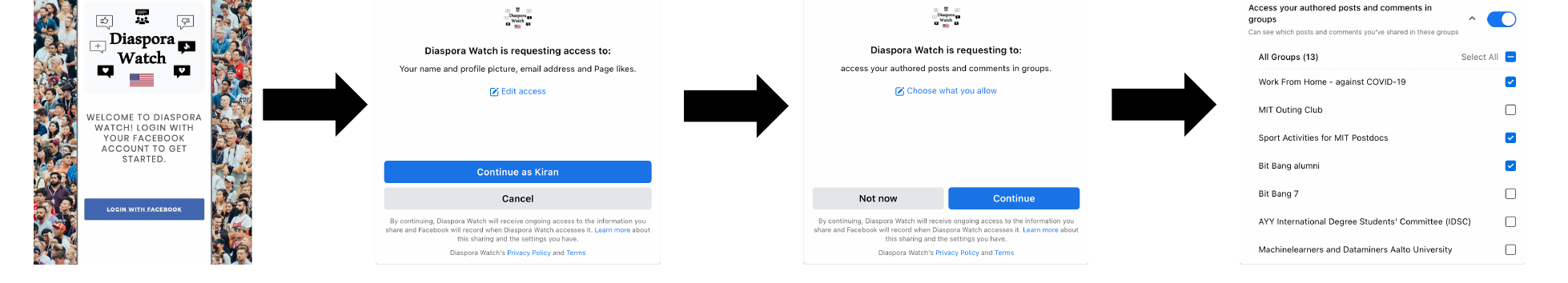}
    \caption{Facebook data donation flowchart. Panel 1 -- Welcome page where the user sees a `Login with Facebook' button. Panels 2--4 allow the users to select what public Pages and Groups they wish to donate. Identifying details have been anonymized.}
    \label{fig:fb_data_donation}
\end{figure*}

To collect the data, we developed a Facebook application similar to popular platforms like Farmville~\cite{burroughs2014facebook}, designed specifically for the ethical collection of social media data through user consent. Users log in via our website using their Facebook credentials, selectively granting access to their interactions with groups and pages. This process leverages the Facebook Graph API, particularly focusing on endpoints such as user\_likes and groups\_access\_member\_info. Given the tightened security protocols post the Cambridge Analytica scandal~\cite{heawood2018pseudo}, our application underwent a rigorous manual approval process to justify each permission request, ensuring compliance with privacy standards.
The design and flow of data collection through our tool are illustrated in Figure~\ref{fig:fb_data_donation}.
The technical backbone of the application is built using Django, a high-level Python web framework that efficiently handles user data interactions and API requests. 
%

For each user, we get a list of Facebook groups and pages along with their basic demographics such as age, gender and ethnicity. 
We used CrowdTangle to obtain the posts shared in these pages/groups.
Apart from the demographics and the list of groups/pages for each user, we do not collect or store any other personally identifying information about the users donating the data. Our data collection received approval from the Institutional Review Board (IRB) at our university.

%
%
Even though there is a vast amount of public information on Facebook, including public pages and groups, accessing it presents significant challenges due to the exclusive control platforms have over data concerning the content consumed by users. Our data donation model allows us to bypass these platform restrictions and gain deeper insights into user behavior by allowing us access to get information on pages/groups from which users get their information.
This data donation approach also helps overcome limitations seen in existing tools like CrowdTangle. While CrowdTangle provides some access to Facebook data, it often fails in offering comprehensive coverage and timely detection of critical content. This shortfall was notably evident during the events leading up to significant incidents, such as the coordination by ``Stop the Steal" groups prior to the January 6th Capitol riot, where relevant data was not promptly accessible.\footnote{\url{https://www.npr.org/2021/10/22/1048543513/facebook-groups-jan-6-insurrection}} By focusing on public data and implementing an opt-in model, our tool not only reduces the volume of data collected but also significantly enhances our capacity to analyze public sentiment and detect emerging trends. This methodology provides a crucial advantage in understanding and responding to dynamics on social platforms in real-time.


%

Our method primarily captures potential content consumption based on user interactions with public pages and groups, which, while not a direct measure of news feed content, serves as a strong indicator of user interests and preferences. This approach aligns with strategies used by researchers to infer social media behavior in the absence of direct data access from platforms, as discussed in studies such as those by Eady et al.~\cite{eady2023exposure}. Despite its limitations in capturing private group data and potential biases, this methodology offers valuable insights into the public dimensions of platform engagement, contributing to our understanding of digital communication landscapes.
With the implementation of GDPR and the forthcoming Digital Services Act (DSA), there are already regulations in place that empower users by mandating platforms to offer mechanisms for data access, such as the ability for users to download their own data held by these platforms. This provision enables researchers to utilize such data for analytical purposes.

%

%

We recruited users through an online survey panel company PureSpectrum. Purespectrum provides high quality panels and have been used in other academic studies~\cite{lazer2020covid}. The users donated their groups and took a 5 question survey asking them about demographics.
The entire process takes less than 3-5 minutes for the user. The entire data collection cost USD \$2,500.

We obtained data from 1,261 users, which included a total of 251,220 pages and groups -- approximately 210,000 pages and 42,000 groups. 
Our objective was to extract posts from these entities using CrowdTangle. However, due to CrowdTangle's limitation of 25,000 total entities per dashboard, we randomly sampled 10\% of both groups and pages, uploading them into separate lists on separate dashboards.\footnote{Note: we use the term Pages for simplicity and not Pages/Groups every time.}
A note on coverage by CrowdTangle: According to the FAQ page~\cite{crowdtangleWhatData}, any page that is not private is accessible via the CrowdTangle API. Thus, our dataset covers all public pages irrespective of their following. We miss a small fraction of pages (around 10\%) which are private.
Navigating CrowdTangle's official API to extract posts from our sample, limited to six calls per minute per dashboard, was challenging. Each call to the `/posts/' endpoint returns a JSON containing 100 posts, their metadata, and a pagination URL for the next batch of 100 posts. However, these pagination chains are limited to 50,000 posts. Upon reaching this limit, our extraction script executes a relocation query: it identifies the date of the latest post received and initiates a new `/posts/' query starting one second earlier. From the resulting 100 posts, it locates the initial `latest' post to determine its offset within the result set. The new chain then begins with the same relocation query, now including the offset as a parameter, ensuring no posts are lost or duplicated.

The final dataset, comprises 32,026,862 posts published between October 2022 and September 2023.\footnote{The data collection is ongoing and will be extended beyond this period, potentially until it CrowdTangle will be shut down in August 2024. Due to the strict rate limits on CrowdTangle, we only present analysis for this one year period in the current draft of the paper.} Of those, 16,633,666 are tagged as English according to CrowdTangle's `language\_code' field. It is noteworthy that the second largest language code in the dataset is `und' (undefined). Within the `und' posts, there are many written in English; however, this study only considers English posts.



\begin{figure}[ht]
    \centering
    \includegraphics[width=\linewidth]{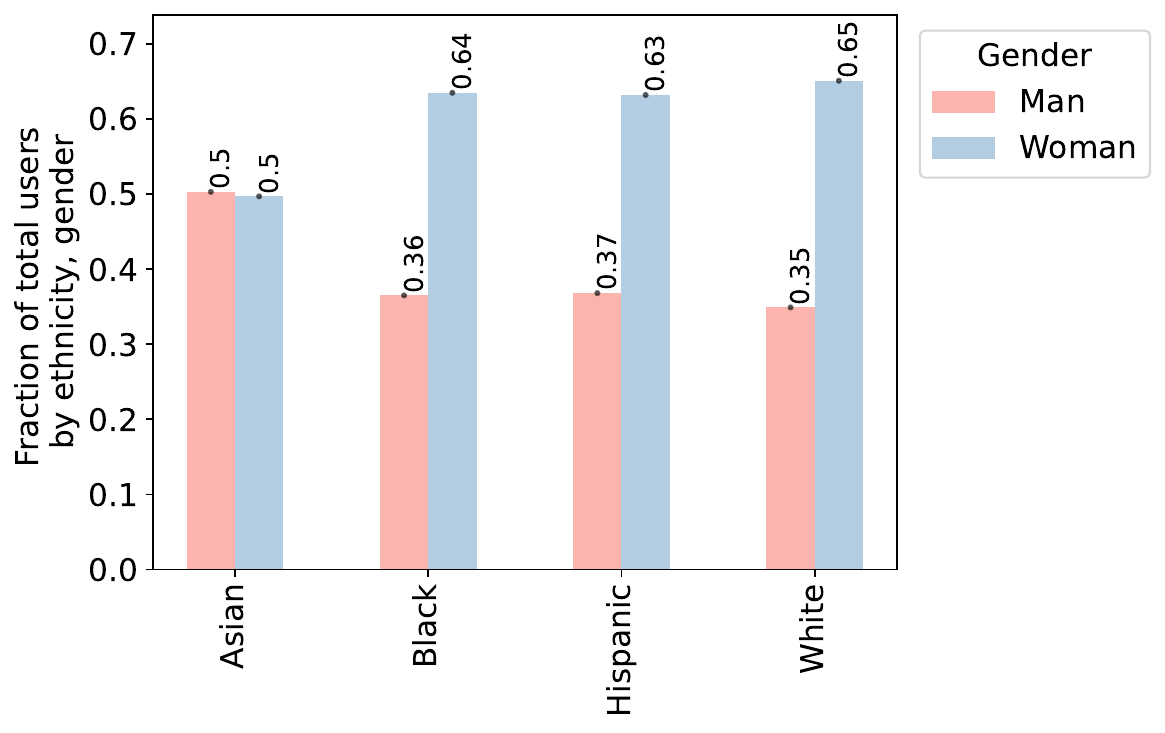}
    \caption{Fraction of users by ethnicity and gender.}
    \label{fig:frac_users_ethnicity_gender}
\end{figure}

\begin{figure}[ht]
    \centering
    \includegraphics[width=\linewidth]{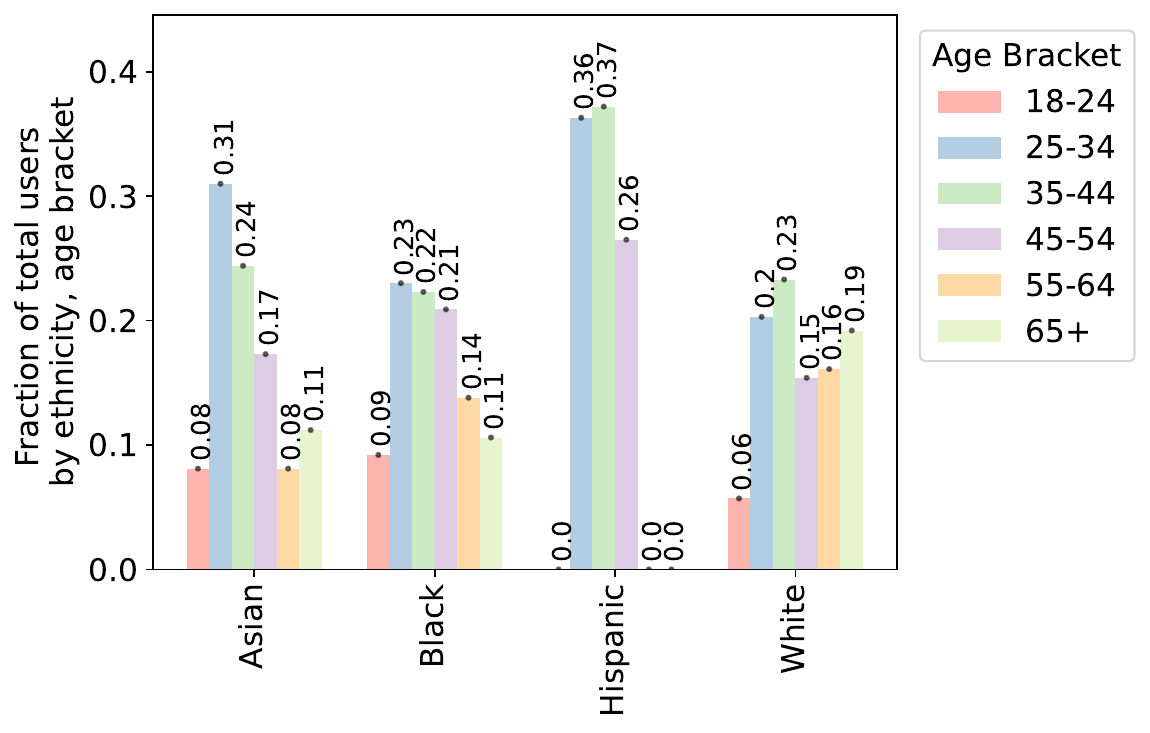}
    \caption{Fraction of users by ethnicity and age.}
    \label{fig:frac_users_ethnicity_age}
\end{figure}

The ethnic, gender and age distribution of our data is shown in Figures~\ref{fig:frac_users_ethnicity_gender},~\ref{fig:frac_users_ethnicity_age}.
We can see that we are oversampling women in most ethnicities and under sampling users in the 18-24 age category.
This is due to the convenient nature of our sampling. To compensate for the lack of representative data, we use reweighting techniques in our estimates in Section~\ref{sec:results}.

\section{Data Processing}

In this section, we describe the various pre-processing steps taken before our analysis. We first annotated a subset of the groups for high precision political content.

\subsection{Labeling of Political and Non-Political Content}
To systematically investigate prevalence of political content in various groups and pages, we started with extracting the top 1,000 unigrams and bigrams present in the titles of the pages. Each of these n-grams was carefully evaluated and classified as either politically relevant or not. Political n-grams encompassed references to specific politicians (e.g., `Trump,' `Obama') and broader political concepts (e.g., `Senate,' `Congress', `Republican'). Conversely, non-political n-grams were culled from a manually curated list of the 1,000 most common unigrams and bigrams, including terms such as `buy,' `sell,' `fan club,' `real estate,' `Taylor Swift,' etc. While not exhaustive, this list provided a comprehensive overview of non-political discourse.

Following this classification, pages that featured any identified political (non-political) n-gram in their names were designated as explicitly political (non-political). This thorough review process yielded a dataset comprising 1,578 political and 3,678 non-political pages. Utilizing CrowdTangle, we then collected all posts from these designated pages over the period from January 2021 through December 2023.

Subsequently, this `clean' dataset served as the foundation for training a classifier aimed at detecting political content.
All posts obtained from explicitly political pages were labeled as political, while those from non-explicitly political pages were labeled as non-political.\footnote{Though this is a strong assumption, we verified this manually by randomly sampling 250 posts from each set and found a 97\% of the posts from political pages were political and 98\% of the non political page posts were non political.} The textual content of these labeled posts was defined as the concatenation of the title, description, and message fields obtained from the CrowdTangle API. Using this labeled dataset, we aimed to create a classifier capable of distinguishing political from non-political content in our larger unlabeled dataset.

\subsection{Detection of Political Content}
\label{political_model}
We built our political or not classifier by fine tuning an existing classifier from~\cite{POLUSA} on this clean political posts dataset from Facebook. The classifier is trained on articles labeled in the ``politics" category in various news sources such as the BBC and Huffington Post. Even though the base classifier already performs with a high F1 (94.4), we wanted the classifier to be particularly good at detecting content from our Facebook sample. We began by loading the model structure and weights provided by~\cite{POLUSA} and fine-tuned the model over 25 epochs, employing an early stopping validation loss callback with a patience of three epochs to prevent over fitting. The resulting fine-tuned model achieved an AUC-ROC of 0.951 on a previously unseen test set comprising 20\% of all labeled posts. Finally, the model was used to predict the probability each data point in our unlabeled dataset contains political content.

Despite the fact that text posts constitute only a minor fraction of the overall content (as illustrated in Figure~\ref{fig:content_types_age_groups}), nearly all types of posts on Facebook are accompanied by some form of text. Consequently, the coverage of our political content classifier remains highly effective, covering approximately 85\% of the posts across various formats. This includes posts where the primary content may be an image or video; the accompanying text captions are utilized for classification purposes, ensuring broad applicability of our analysis tools regardless of the post type.



\subsection{Topics}
\label{subsec:bertopic}
To analyze the content in our unlabeled dataset of 16 million posts, we utilized BERTopic for topic modeling~\cite{grootendorst2022bertopic}. Given the large size of our dataset, critical decisions regarding the model's configuration were necessary to ensure feasibility. The training process of a BERTopic model involves four primary customizable submodules: embeddings, dimensionality reduction, clustering, and tokenization.
For embeddings, we used the sentence-transformer's `all-MiniLM-L6-v2' model, as recommended by BERTopic defaults. For dimensionality reduction, we employed UMAP with a `low\_memory' flag, which trades high memory usage for a slightly more computationally intensive k-nearest neighbors step. We used a count vectorizer limited to a n-gram range of two for tokenization, instead of the typically recommended range of three, as well as a min\_df of two to prevent large sparse matrices for single-occurrence words. Crucially, for the HDBSCAN clustering module, we performed a comprehensive grid search to determine the optimal minimum cluster size, which directly influenced the model's computational feasibility and topic accuracy.

The BERTopic default parameters and FAQ suggest a minimum cluster size of 10 to 250 for large datasets. However, in our use case, this range was insufficiently low and resulted in extremely high computational loads due to the excessive number of topics generated. To address this, we sampled 20\% of our dataset and trained several dozen models using a grid search pattern with varying minimum cluster sizes as percentages of our dataset. Initial experiments with 1\%, 0.1\%, and 0.01\% helped us narrow down the optimal range for grid search to between 0.01\% and 0.05\%. 

To quantify and compare the quality of these models we employed GenSim's coherence model~\cite{vrehuuvrek2011gensim}, which compares each topic's representation words with size one and two n-grams obtained from their documents to give a final `coherence' score. The higher this score the more closely related the topics are to the documents labeled under them and the better the model is. In our dataset the highest coherence score was obtained with a minimum cluster size of 0.025\% of the dataset sample used for topic modelling. The resulting model identified 333 high-quality cohesive topic clusters.
After obtaining the topics, we manually annotated them and identified the prevalence of various topics across various demographics.

\subsection{Re-weighting to obtain population level estimates}

In our study, we employed the `balance' library from Facebook~\cite{sarig2023balance} to address the challenge of biased sampling in our dataset. 
This process involves generating propensity scores based on available demographic information such as age, gender, and ethnicity. These scores are then used to adjust the sample distribution to more closely resemble the target population distribution.
We overcome the convenience sampling biases by generating weights for each data unit using various methods like inverse propensity scoring, and evaluating the bias and variance post-adjustment. This approach is crucial for our analysis as it allows us to derive population-level statistics from data that may otherwise be skewed due to non-random sampling processes.
The findings in the subsequent sections are all weighted and represent population level consumption estimates.

Note on reproducibility: The code for the data donation tool, the dataset containing the demographics, names/IDs of pages/groups, keywords used for identifying political/non-political pages, along with fine-tuned models for detecting political content will be released upon paper acceptance.
Unfortunately CrowdTangle's terms of service prohibit the release of the full raw dataset we collected,\footnote{\url{https://help.crowdtangle.com/en/articles/3192685-citing-crowdtangle-data}} but we hope the IDs and other artefacts might be valuable and can be used on Facebook's new open research and transparency tools for academics: FORT.\footnote{\url{https://fort.fb.com/}}

\section{Political content}
\label{sec:results}

We divide our analysis into political content and overall content. First, in this section, we show the prevalence of political content across various demographic groups, look for trends over time and dig deeper into the subcategories of political topics.
In the next section, we get a look at the overall content consumption by Facebook users.



\subsection{Political content prevalence}

This section examines the distribution and prevalence of political content consumption among various demographic groups. Utilizing advanced statistical techniques, we weighted all figures to provide population-level estimates, enhancing the generalizability of our findings.
In each plot, the dotted gray line indicates the overall mean. The error bars indicate 95\% confidence intervals.

Overall, around 17\% of the content being consumed is political content.
These findings align with broader media consumption trends identified in other studies. A representative survey by Pew Research found that about a third of tweets on Twitter were political~\cite{bestvater2022politics}, while a large-scale study by Allen et al. reported an overall `news' content prevalence of about 14\% across various platforms, suggesting that most political content consumption still occurs through television~\cite{allen2020evaluating}. This comparative perspective highlights that while political engagement on Facebook is less than on Twitter, it is comparable to traditional media like television, suggesting that different platforms serve distinct roles in political communication ecosystems.

\begin{figure*}[ht]
\centering
\begin{subfigure}[b]{0.24\textwidth}
    \includegraphics[width=\textwidth]{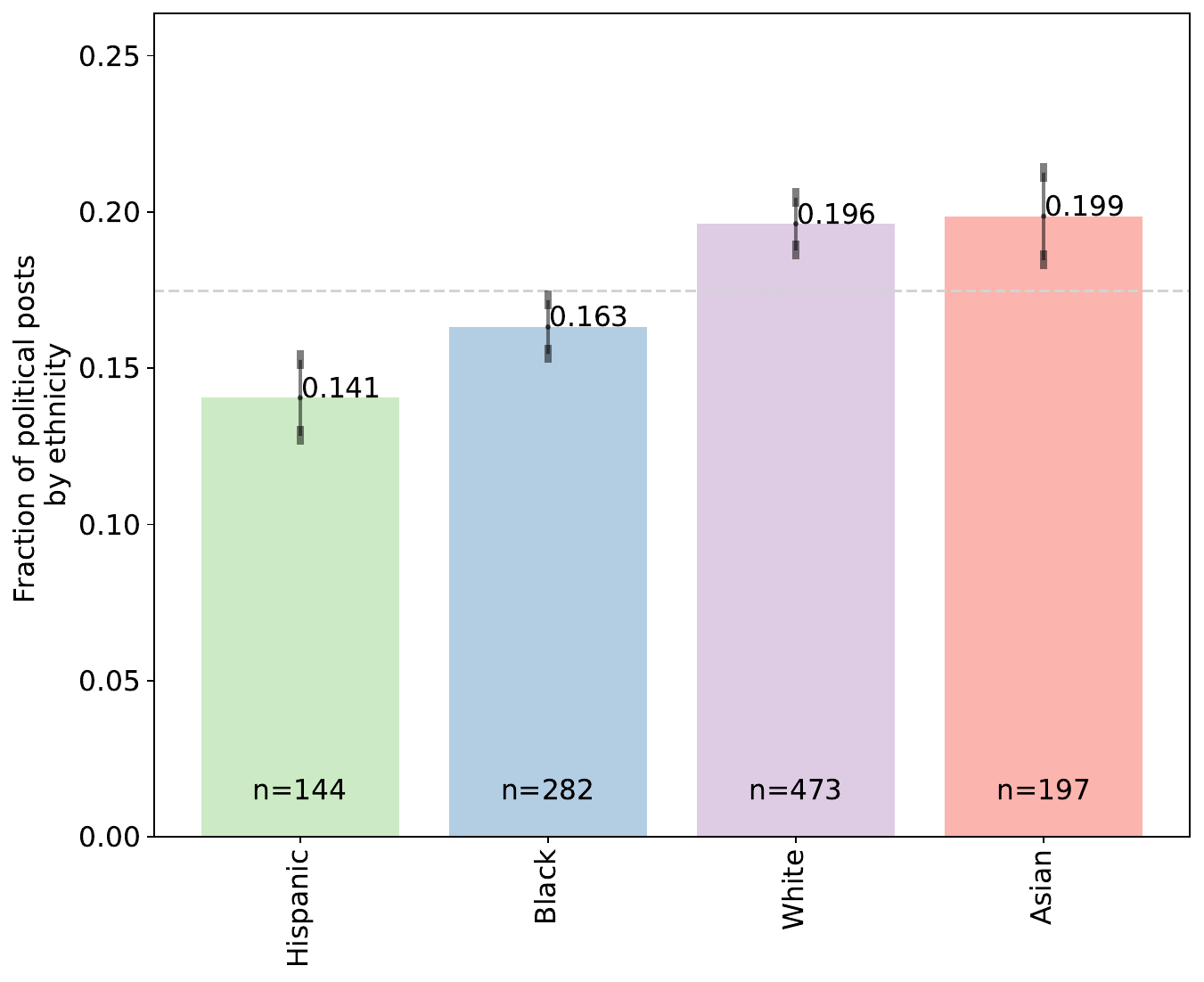}
    \caption{Ethnicity}
    \label{fig:political_ethnicity}
\end{subfigure}
\hfill
\begin{subfigure}[b]{0.24\textwidth}
    \includegraphics[width=\textwidth]{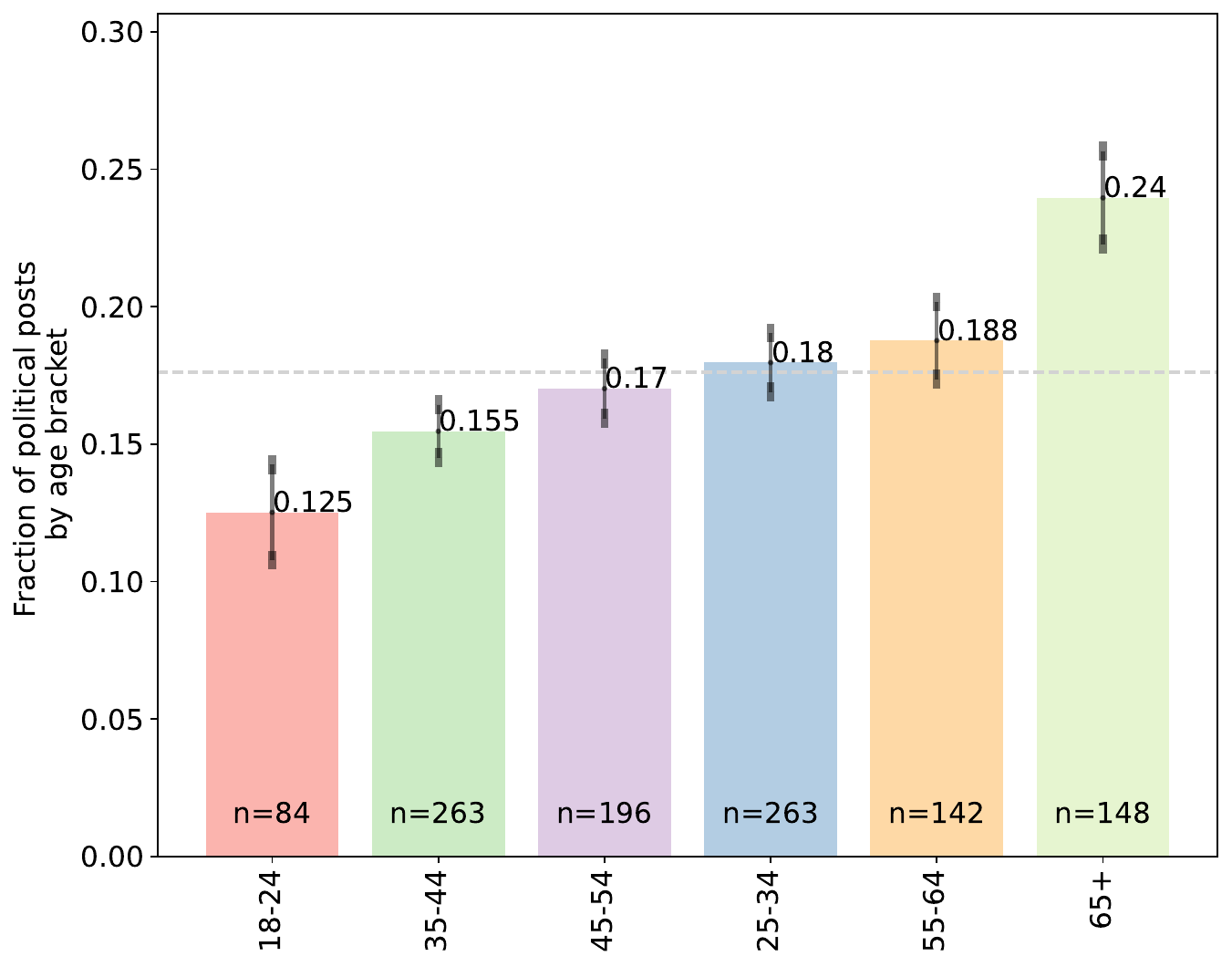}
    \caption{Age group}
    \label{fig:political_age}
\end{subfigure}
\hfill
\begin{subfigure}[b]{0.24\textwidth}
    \includegraphics[width=\textwidth]{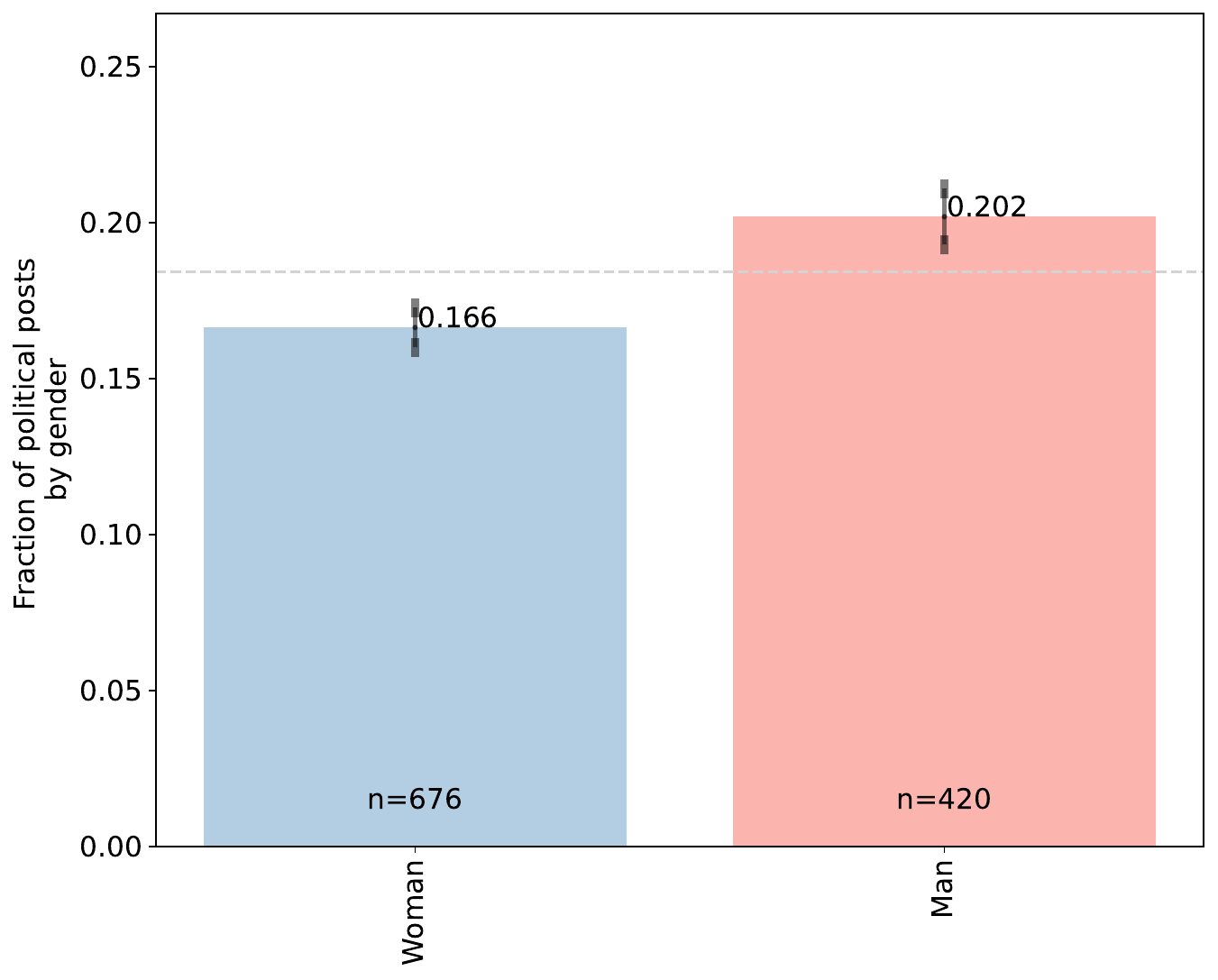}
    \caption{Gender}
    \label{fig:political_gender}
\end{subfigure}
\hfill
\begin{subfigure}[b]{0.24\textwidth}
    \includegraphics[width=\textwidth]{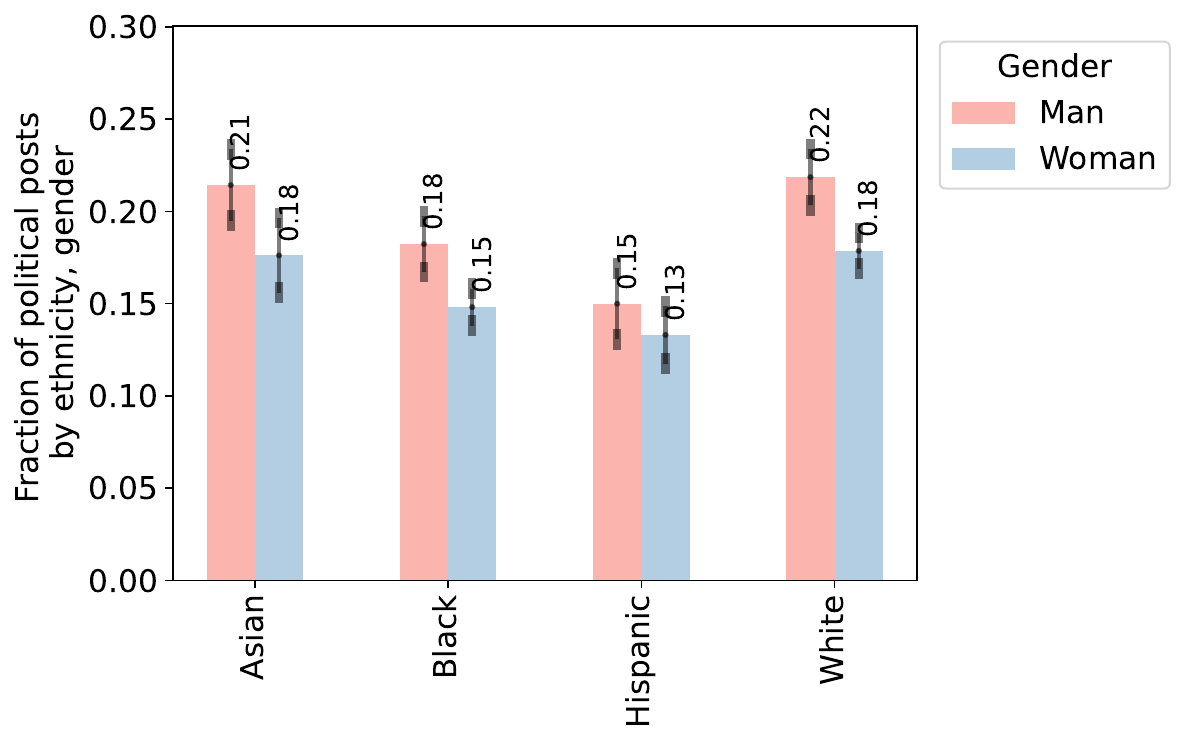}
    \caption{Ethnicity and gender}
    \label{fig:political_gender_ethnicity}
\end{subfigure}
\caption{Political content consumption by: (a) ethnicity, (b) age group, (c) gender, (d) gender and ethnicity.}
\label{fig:political_content}
\vspace{-\baselineskip}
\end{figure*}

Figure~\ref{fig:political_ethnicity} reveals a differentiated pattern in the consumption of political content among ethnic groups. Hispanics engage with political content significantly less, at about 14\%, compared to Asians and Whites, each around 20\%. 
This disparity could be partly explained by the exclusion of Spanish language content in our analysis, though not significantly, given that only 5\% of the content was in Spanish (while Hispanics made up 12\% of our sample).
As depicted in Figure~\ref{fig:political_age}, age correlates strongly with political content consumption. Younger individuals (18-24 years) show a markedly lower engagement rate at 12\%, in contrast to older users (65+ years), who engage at almost a double that rate, of 24\%. This finding is in line with previous research which showed that older users are significantly more active in consuming misinformation on Facebook~\cite{guess2019less}. The same pattern across age groups also holds for a `News' content -- as defined by Facebook's page/group categorization. See Figure~\ref{fig:news_consumption_age} in the Appendix.
Figure~\ref{fig:political_gender} indicates that men consume more political content than women.
This study provides the first detailed estimates of political content consumption on Facebook, revealing that a majority of the content consumed by users is not political. 
The granularity and quality of the data also allow us to slice content further, say, among gender and ethnicity. For instance, Figure~\ref{fig:political_gender_ethnicity} shows that men are more likely to consume political content across ethnicity.

These findings are crucial, particularly when considering the broader context of political polarization and misinformation. They suggest that users, especially young people, are not as inundated with political content as might be feared. Instead, as detailed in Section~\ref{sec:overall_consumption}, the predominant categories of content consumed are related to religion, entertainment, and lifestyle. This distribution indicates that while political content has a significant presence, it does not dominate the social media landscape to the extent often perceived in public discourse. Understanding these dynamics is essential for addressing issues of political engagement, polarization and the spread of misinformation effectively.

\subsection{Political content consumption over time}

\begin{figure*}[ht]
\centering
\begin{subfigure}[b]{0.32\textwidth}
    \includegraphics[width=\textwidth]{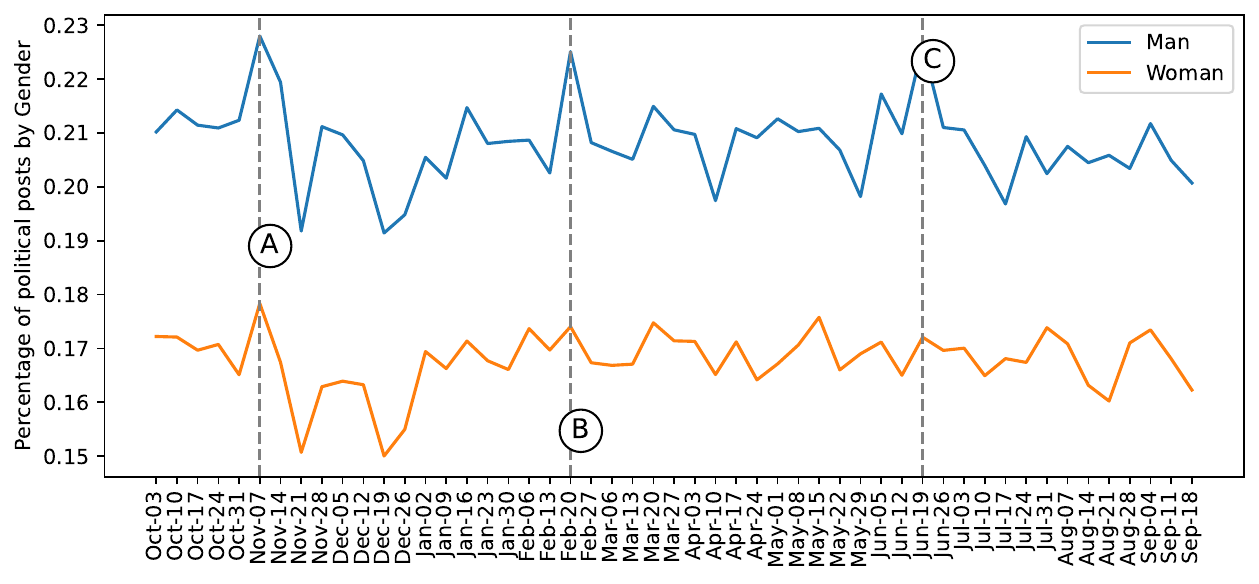}
    \caption{Gender}
    \label{timeseries_political_gender}
\end{subfigure}
\hfill
\begin{subfigure}[b]{0.32\textwidth}
    \includegraphics[width=\textwidth]{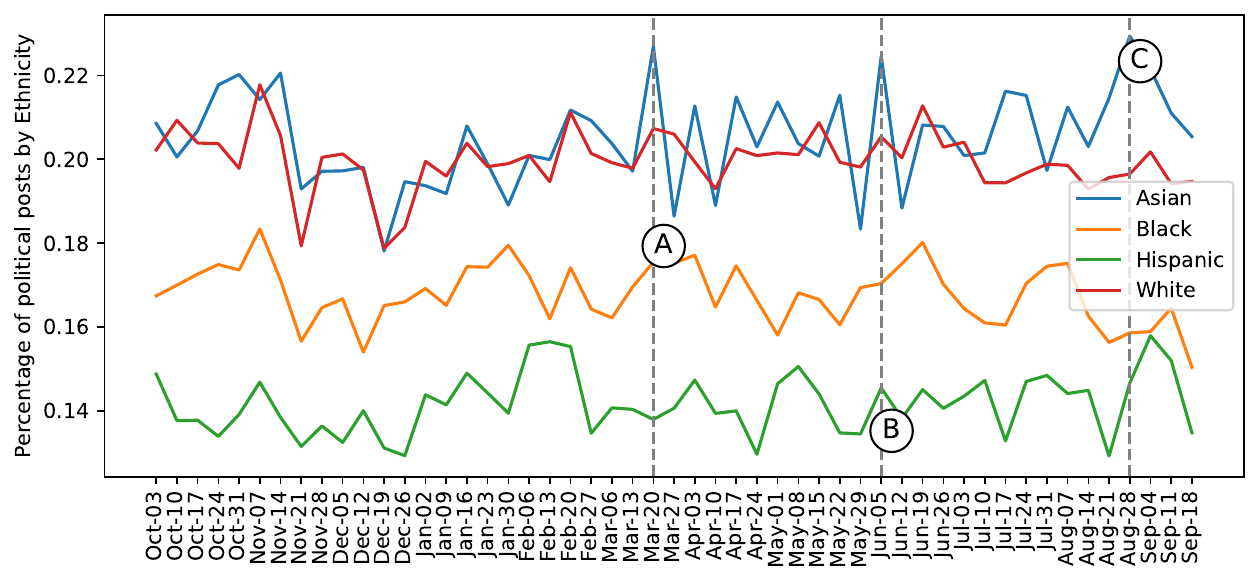}
    \caption{Ethnicity}
    \label{fig:2}
\end{subfigure}
\hfill
\begin{subfigure}[b]{0.32\textwidth}
    \includegraphics[width=\textwidth]{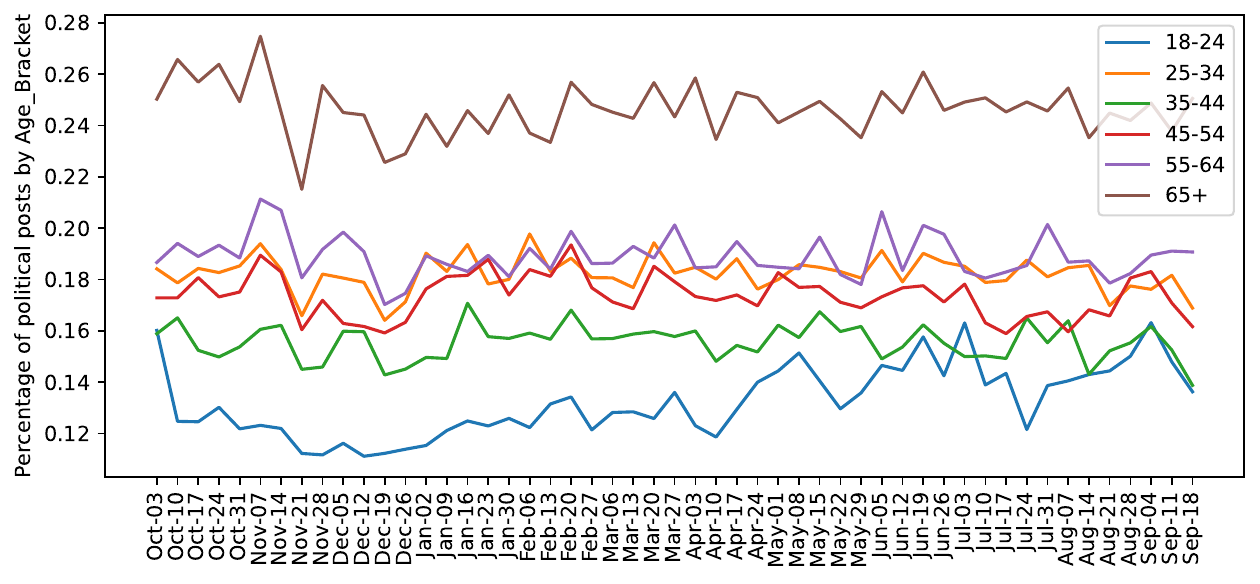}
    \caption{Age group}
    \label{fig:3}
\end{subfigure}
\caption{Time series. The data spans from October 2022 to October 2023.}
\label{fig:timeseries}
\vspace{-\baselineskip}
\end{figure*}

Next, we go into further details of political consumption, looking at trends in political consumption over time.
Figure~\ref{fig:timeseries} shows the result. As we see, there are significant periods of ebbs and flows of political content corresponding to external events.
Firstly, we see that certain patterns stay consistent even over time -- for instance, Hispanic and women and younger users consistently have a significantly lower political consumption fraction.


To analyze the nature of posts contributing to peaks in political content, we developed a separate BERTopic model. This model was trained exclusively on posts with over 50\% probability of containing political content, as assessed by the fine-tuned neural network described in Section \ref{political_model}. The new model identified 211 topic clusters, representing significantly more specific political themes. For instance, while the BERTopic model in Section~\ref{subsec:bertopic} grouped all US politics-related documents under a single topic, this new model created distinct clusters for specific events, such as Donald Trump's indictment and the Joe Biden classified documents incident.

To elucidate the topical explanations for the peaks observed in Figure~\ref{fig:timeseries}, we identified the top five topics with the highest positive delta in their prevalence each week, relative to the total number of posts in each demographic segment. We then isolated the topics that appeared consistently in the top five rankings within the same demographic category (e.g., male, female within the gender category). Our analysis reveals distinct peaks in political engagement that correlate with significant societal and political events.

Across all plots in Figure~\ref{fig:timeseries}, we clearly see a drop off in political content after the November 8th 2022 mid terms. Our data's capacity to monitor content consumption patterns at a high level of granularity can be illustrated via some examples of peaks in the gender-bound timeseries in Figure~\ref{fig:timeseries} (a). We observe notable peaks in political content during the weeks starting on November 7th, February 20th, and June 19th. These peaks were driven by topics consistently ranked among the top five rising topics for both men and women. The peak on February 20th was entirely explained by a surge in discussions about the Russia-Ukrainian war, corresponding with the one-year anniversary of the Russian invasion of Ukraine.
 
In Figure~\ref{fig:timeseries} (b), three primary peaks of political content are observed on March 20th, June 5th, and August 28th. During the week starting March 20th, three of the top five rising political topics were common across various ethnicity demographic segments: TikTok Ban (White +0.69\%, Hispanic +0.42\%, Black +0.43\%, Asian +1.24\%), Donald Trump Indictment (White +1.05\%, Black +0.47\%, Asian +0.64\%), and LGBT Affirmative Action (White +0.5\%, Black +0.5\%). Similarly, the June 5th peak saw four shared topics: Canadian Wildfires (White +1.6\%, Hispanic +0.8\%, Black +1.0\%, Asian +1.3\%), TikTok Ban (White +0.47\%, Hispanic +0.78\%, Black +0.2\%, Asian +0.83\%), Donald Trump Indictment (White +1.16\%, Hispanic +0.84\%, Black +0.9\%), and LGBT Affirmative Action (Hispanic +1.22\%, Black +1.41\%). The peak on August 28th was driven by increased prevalence of the following topics: Hurricane Ian (White +0.62\%, Hispanic +0.60\%, Black +0.42\%, Asian +0.30\%) and Joe Biden Campaign (White +0.27\%, Black +0.42\%).

\subsection{Political leanings of various demographics}
Analyzing the political orientation of users through the bias of URLs shared in posts is a viable approach, despite link content making up only  27\% of the overall content. While this sample may not be entirely representative, it allows us to estimate user biases, following methodologies employed in previous studies such as those by~\citet{allcott2017social} and~\citet{grinberg2019fake}. However, this approach does not account for biases potentially conveyed through images, videos, and other media formats. Initially, our goal was to extend this analysis to such content types to classify political support more comprehensively, but this proved to be more challenging than anticipated so we instead opted to use the domain level leaning from Robertson et al.~\cite{robertson2018auditing}. 

To visualize the distribution of these bias scores across the dataset, we plotted the histogram of the URL bias distribution and estimated the density using kernel density estimation (KDE) techniques. KDE helps in producing smoother probability density curves, offering a clearer view of the data distribution compared to traditional histograms.

\begin{figure*}[ht]
\centering
\begin{subfigure}[b]{0.24\textwidth}
    \includegraphics[width=\textwidth]{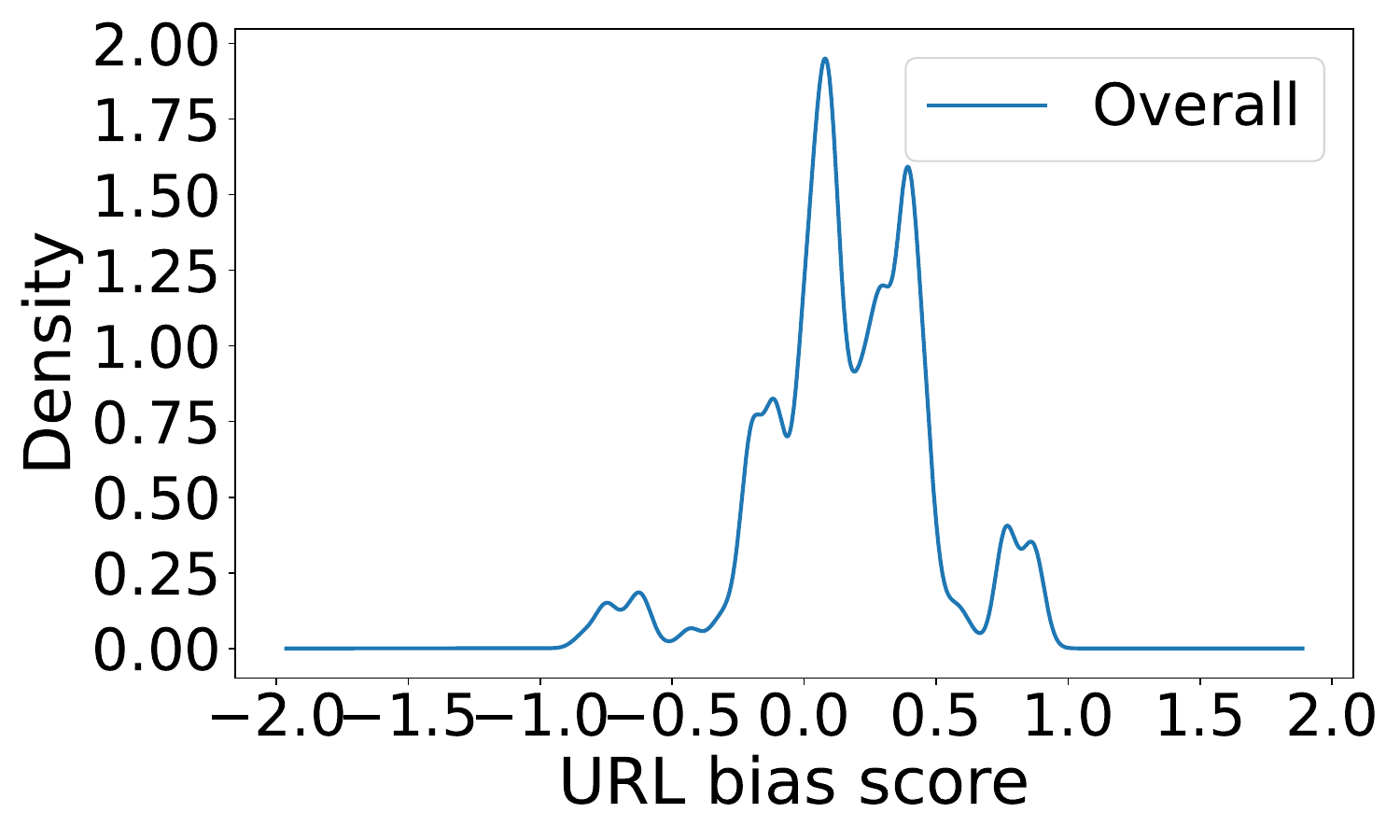}
    \caption{Overall}
    \label{fig:1}
\end{subfigure}
\hfill
\begin{subfigure}[b]{0.24\textwidth}
    \includegraphics[width=\textwidth]{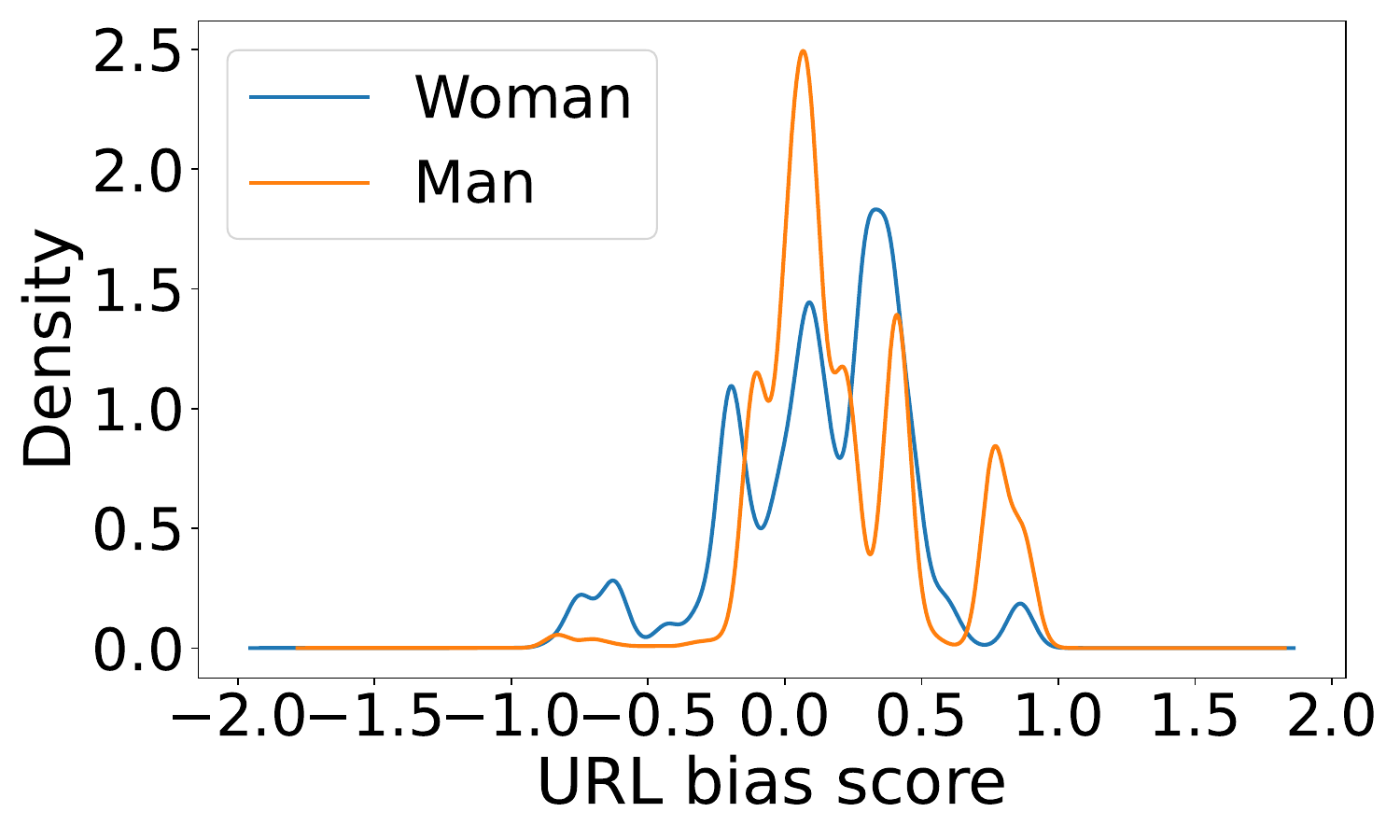}
    \caption{Gender}
    \label{fig:2}
\end{subfigure}
\hfill
\begin{subfigure}[b]{0.24\textwidth}
    \includegraphics[width=\textwidth]{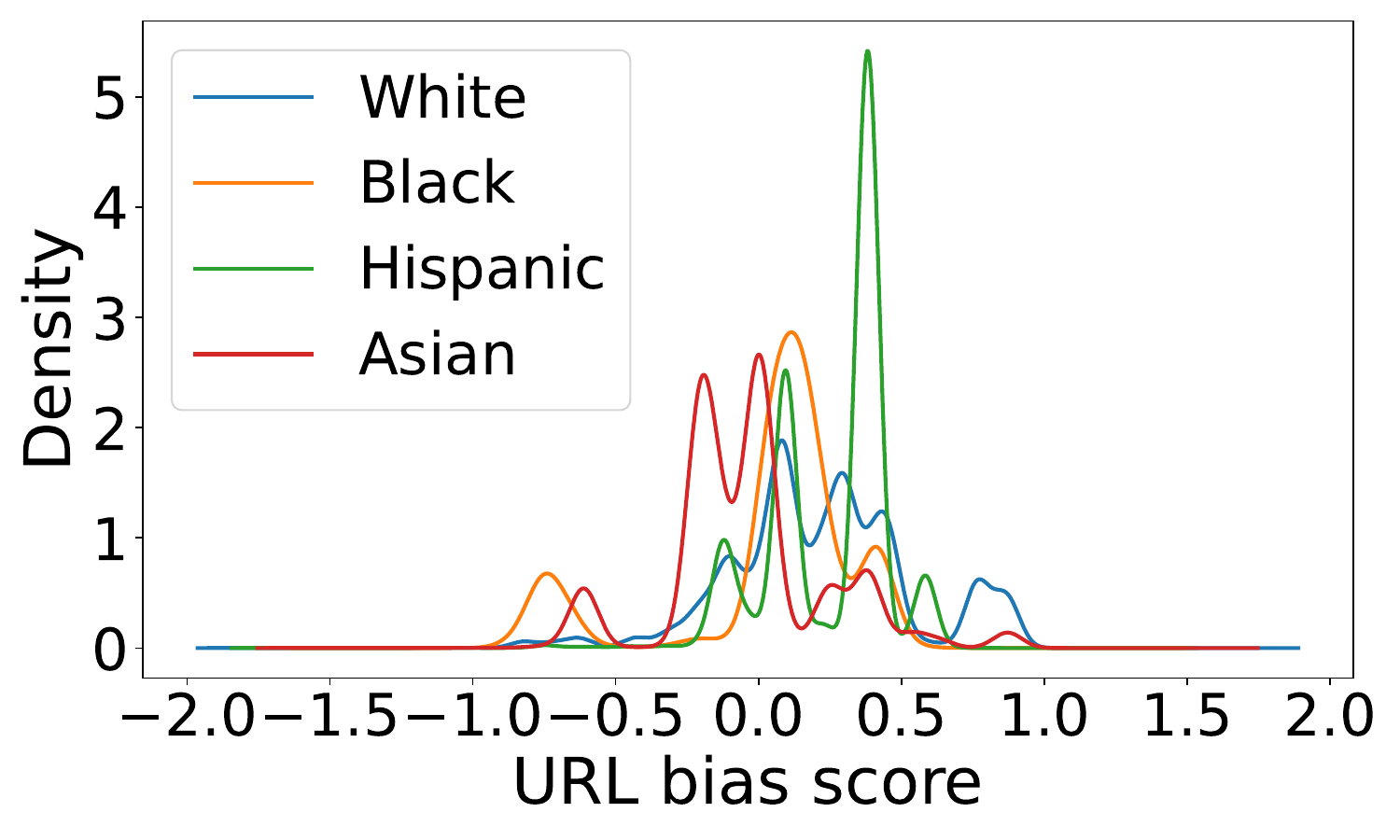}
    \caption{Ethnicity}
    \label{fig:3}
\end{subfigure}
\hfill
\begin{subfigure}[b]{0.25\textwidth}
    \includegraphics[width=\textwidth]{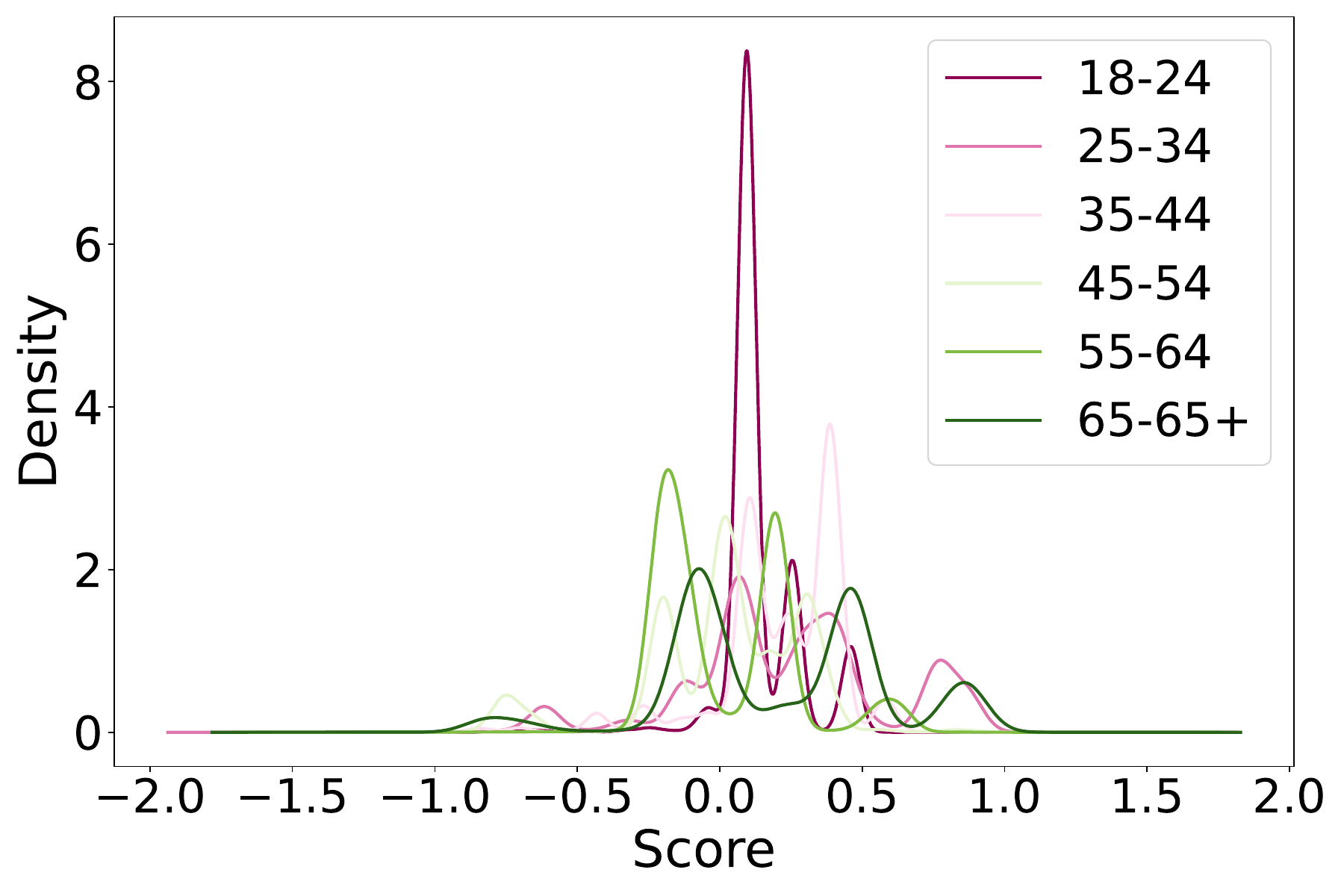}
    \caption{Age groups}
    \label{fig:4}
\end{subfigure}
\caption{Political bias in URL consumption. Negative numbers indicate a liberal bias.}
\label{fig:urls_bias}
\vspace{-\baselineskip}
\end{figure*}

Figure~\ref{fig:urls_bias} illustrates the variation in URL bias across different demographics, where a score of -1 indicates liberal domains and +1 indicates conservative domains. Our findings reveal a predominantly centrist distribution with a slight rightward skew (Figure~\ref{fig:urls_bias} (a)). Women tend to lean slightly right, whereas men generally appear centrist (Figure~\ref{fig:urls_bias} (b)). Ethnic disparities were also notable; Asian demographics mostly leaned left, while Hispanic groups leaned slightly rightward (Figure~\ref{fig:urls_bias} (c)). Age also played a critical role in bias distribution: younger individuals tended to cluster around the center, whereas older demographics displayed a bimodal distribution, with peaks on both ends of the political spectrum.

Additionally, we investigated the prevalence of low-quality or `fake news' URLs using sources listed by Media Bias Fact Check~\cite{weld2021political}, finding them to constitute less than 0.01\% of the total urls shared. This minimal presence suggests effective moderation by Facebook of domains known for low-quality content, aligning with the platform's policies against misinformation.

\section{Overall content consumption}
\label{sec:overall_consumption}

The analyses in the previous section indicated that no more than 20\% of the content that people engage with is political in nature. This prompts an investigation into the nature of the remaining 80\% of content that captures the majority of audience interest. To this end, we conducted a detailed examination of content preferences across various demographic groups, focusing specifically on the top 10 topics by volume for each demographic.

We manually coding the content topics for each demographic segment to ascertain the prevalence and distribution of interest areas. The findings reveal a compelling trend: the majority of demographic groups typically feature only one or two political topics within their top 10, with many groups showing none at all. However, there are notable exceptions. Individuals aged 18-24 showed a distinct interest in the Russia-Ukraine conflict, while demographics identified as White, Black, men, and those aged 65 and older exhibited a higher engagement with US politics. Additionally, the topic of crime and shootings was predominantly favored among Black audiences.

\begin{figure}
    \centering
    \includegraphics[width=\linewidth]{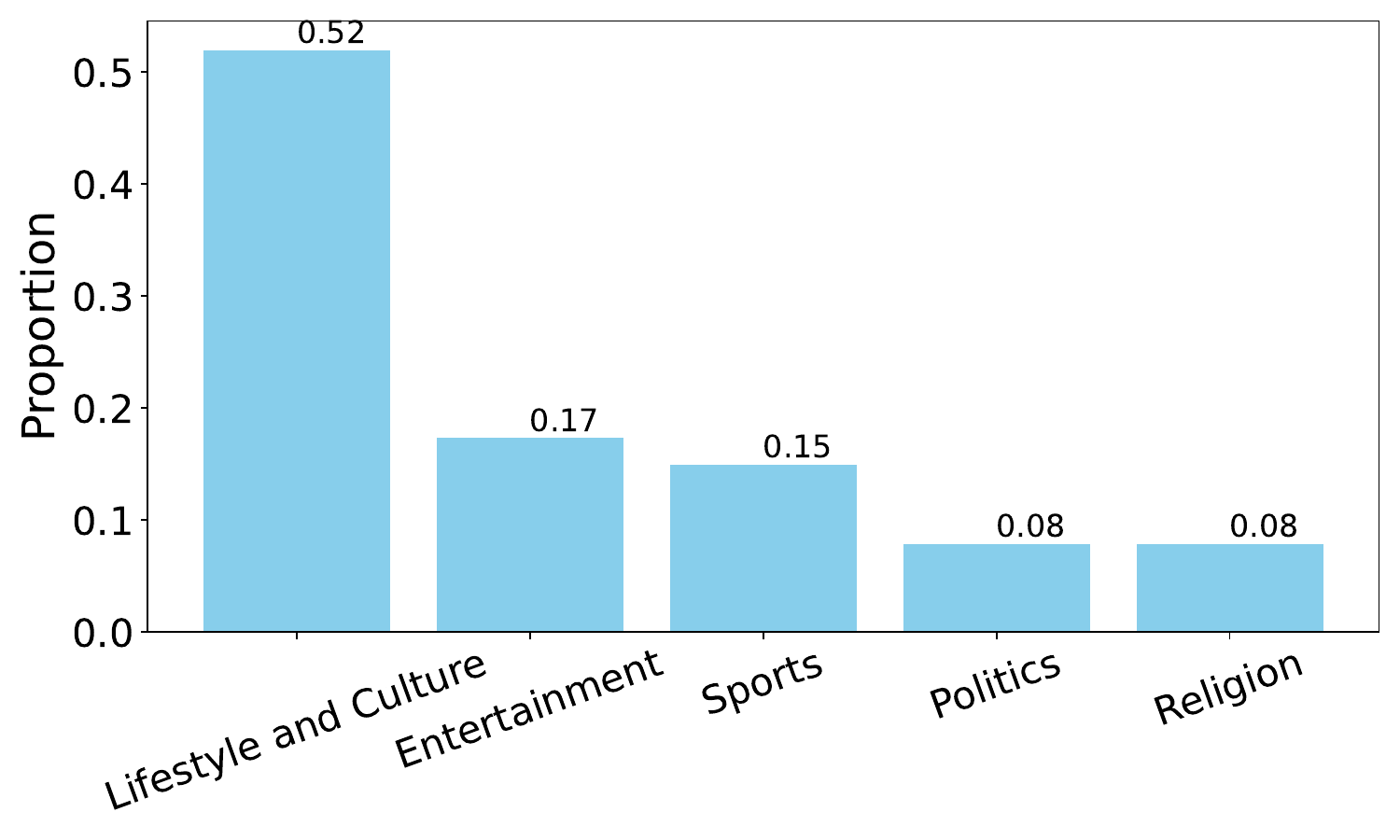}
    \caption{Top level aggregated categories. It is clear that in the top 10 categories across demographics, politics is rare.}
    \label{fig:top_level_categories}
    \vspace{-\baselineskip}
\end{figure}

The vast majority of content that dominates people's consumption falls into broader categories such as sports, lifestyle and culture, religion, and entertainment. These categories represent a diverse range of interests that reflect more everyday activities and personal preferences rather than political engagement. Figure~\ref{fig:top_level_categories} provides a distribution of these top-level categories across the top 10 topics for all demographics analyzed. This visualization underscores the varied nature of content that resonates with different audience segments, highlighting the substantial influence of non-political topics in shaping public discourse and media consumption.




\subsection{Topics disproportionately consumed by specific demographics}

In this section, we identify topics that are consumed disproportionately by specific demographic groups, which showed statistically significant differences compared to others. These differences illuminate the unique preferences and informational needs that characterize diverse demographic segments.
Table~\ref{tab:interest_topics} provides a summary of these differences, only including on the youngest (18-24) and oldest (65+) for age groups. 
For instance, Figure~\ref{fig:topic_plots} panels (a–c) (Appendix) exemplify the variations in topic interest across demographics, revealing a statistically significant higher fraction of engagement among male, Asian, and 65+ users for certain topics.
%
%
The findings provide interesting and several surprising insights, while also confirming various stereotypes. For instance, white users were significantly more interested in activities such as art, bird watching, beer brewing and entertainment.
Black users were significantly more interested in sports, family and civil rights. 

It is important to note that our reporting only includes topics where the interest was statistically significant. We abstain from discussing topics where the differences were not substantial, to keep the discussion clear. The comprehensive data, including raw plots for all demographic categories, is detailed in the Appendix for those interested in a deeper dive into the full range of content consumption patterns observed.

This analysis not only helps in understanding the diverse content preferences across demographic groups but also aids stakeholders in tailoring communication strategies effectively. By understanding these preferences, content creators and policymakers can better address the unique needs of different demographic groups, enhancing engagement and information dissemination.


\begin{table}[h]
\centering
\begin{tabularx}{\columnwidth}{|l|X|}
\hline
\textbf{Demographic} & \textbf{Topics of Interest} \\ \hline
White & Art, Bird watching, Beer brewing, British royalty, Film entertainment, Parks, Social issues, US politics, Weather \\ \hline
Black & Basketball, Wrestling, Film entertainment, Kardashians, Civil rights, College sports, Family \\ \hline
Asians & Bollywood, Cricket \\ \hline
Hispanics & Horoscope, British royalty (least) \\ \hline
\hline
Men & Cars, Basketball, Crypto currencies, Gadgets \\ \hline
Women & Bags and accessories, Jewelry, Skin care, Baking, Cooking, Home decor, Family, Horoscope, Kardashians, Animals \\ \hline
\hline
18-24 & Cute babies, Horoscope, Friendship, Climate change \\ \hline
65+ & Migrants on the border, Budlight boycott, Christianity, Lottery, Animals \\ \hline
\end{tabularx}
\caption{Interest Topics by Demographic.}
\label{tab:interest_topics}
\vspace{-\baselineskip}
\end{table}

\subsection{Heterogeneity in content consumption across groups}

This section delves into the varied modalities through which different demographic groups consume information, highlighting significant differences in content preferences that are not only relevant but also consequential for studies related to information dissemination and misinformation. 

\begin{figure}[ht]
    \centering
    \includegraphics[width=\linewidth]{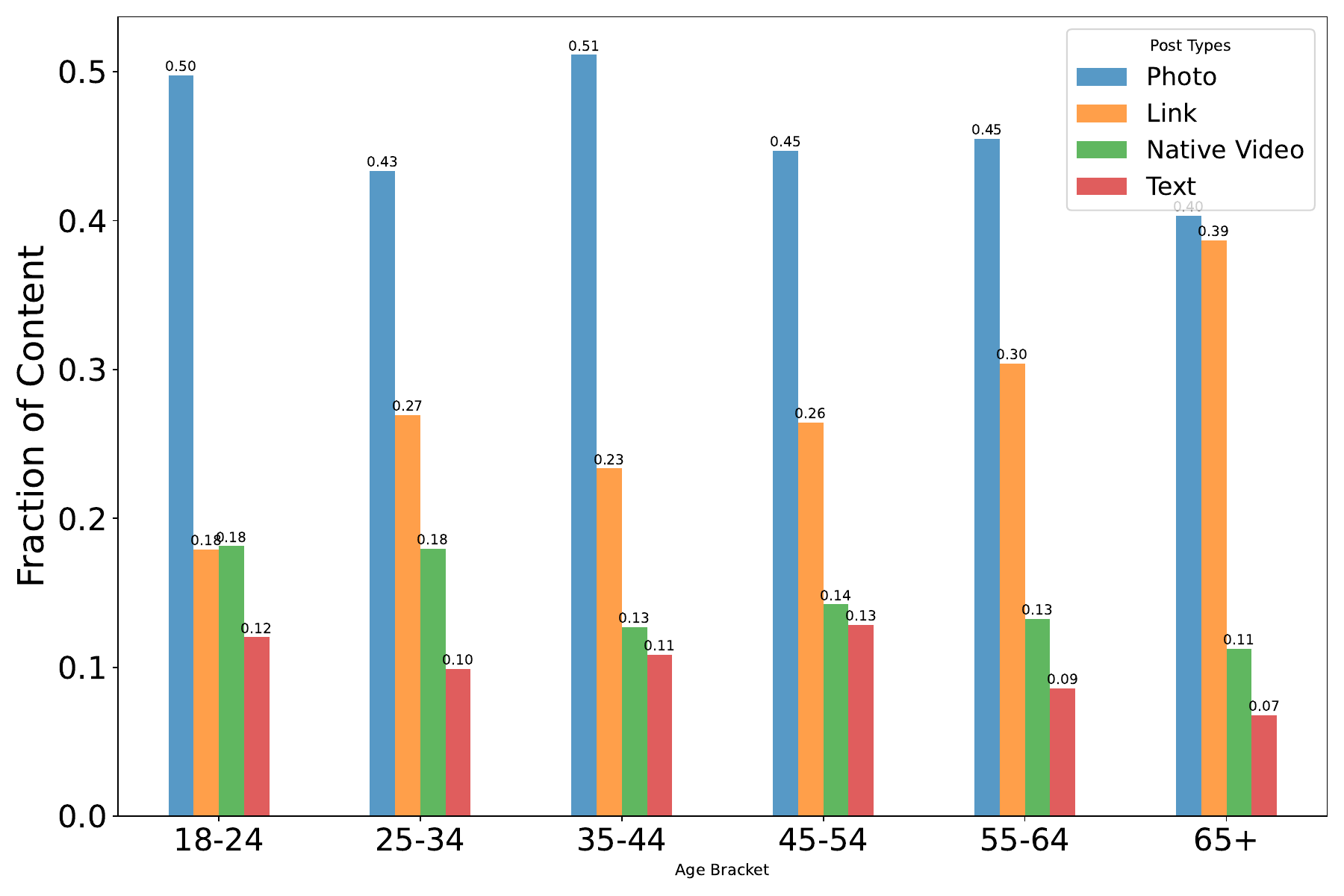}
    \caption{Content types by age group.}
    \label{fig:content_types_age_groups}
    \vspace{-\baselineskip}
\end{figure}

We observed distinct patterns in the way content is consumed by different age groups. As illustrated in Figure~\ref{fig:content_types_age_groups}, older users, particularly those aged 65 and above, exhibit a pronounced preference for link-based content, consuming nearly double the amount of such content compared to the 18-24 age group. In contrast, younger users show a substantial inclination towards video content, reflecting a dynamic shift in engagement as technology and media consumption habits evolve.

\begin{figure}[ht]
    \centering
    \includegraphics[width=\linewidth]{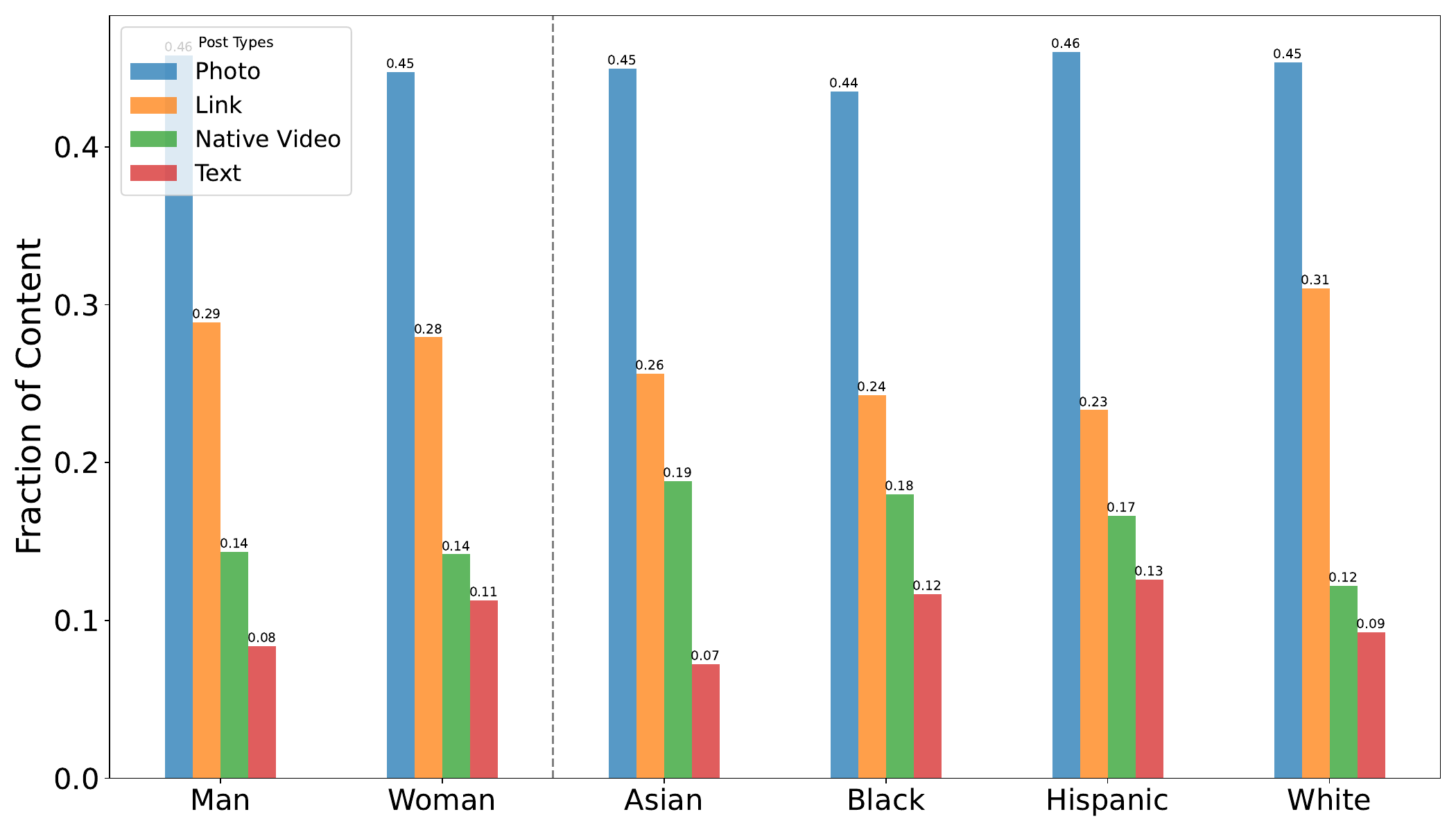}
    \caption{Content types by gender and ethnicity.}
    \label{fig:content_types_gender_ethnicity}
    \vspace{-\baselineskip}
\end{figure}

The heterogeneity extends beyond age and into ethnic differences in content consumption. Figure~\ref{fig:content_types_gender_ethnicity} reveals that White users are less likely to consume video content, accounting for only 12\% of their consumption, compared to 18-19\% for other ethnic groups. Conversely, White users engage more frequently with link-based content, at a rate 5-8\% higher than that observed in other demographics. Intriguingly, our analysis indicates no significant disparities in content consumption patterns across genders.

These observed disparities are critical when considering the broader implications for misinformation and its moderation. Previous research on misinformation primarily focuses on the annotation of low-quality or `fake news' domains~\cite{allcott2017social,grinberg2019fake}. However, the effectiveness of such interventions may vary significantly across different demographic groups due to their divergent content consumption habits. For instance, strategies that are effective in studying misinformation among older adults may not be as successful with younger audiences who prefer different content modalities.


%


\section{Discussion}

This paper introduces a pioneering data donation methodology aimed at collecting and analyzing public Facebook preferences across a variety of demographics. By employing reweighting techniques, we offer population-level estimates, marking this as one of the inaugural studies estimating both political and non-political content consumption on public Facebook pages. Our findings highlight interesting insights, notably the prevalence of political content and the demographic differences therein. Furthermore, our results align with previous research that minimizes the scale and significance of political content consumption online.

The significance of data donation as a platform-independent method is evident through this work. However, a major challenge when working directly with platforms is that researchers often find themselves at the mercy of these platforms. Notably, we anticipate that the application used for this research might be removed following the publication of this paper.
While some of our findings may appear `trivial,' they underscore the utility of descriptive analyses in unveiling nuanced consumption patterns. With this methodology, researchers are empowered to uncover and understand diverse behavioral insights that would otherwise remain obscured.

The generalizability of this data collection model is considerable. It can be applied to various widely used platforms such as YouTube and Telegram, which, despite having robust APIs, often offer data of substandard quality unless the specific content relevant to the research is precisely identified. For example, during the outbreak of the Russia-Ukraine conflict, it was commonly known that Telegram was a significant source of information, yet researchers had no means to determine which Telegram channels to monitor. A data donation approach modeled on the one presented in this paper, where users could donate just the list of public groups and channels they consume information from could help pinpoint these critical sources of information.

As the utility of platforms like CrowdTangle wanes --a tool previously central to social media research-- questions about the continuing relevance of this work emerge. Although this research is replicable with our current data, its future extensions might be constrained, as discussed in recent studies~\cite{substackCrowdTangleDead}. Despite these challenges, data donation methodologies offer a robust framework for future research endeavors. Moreover, even as tools like CrowdTangle become obsolete, alternatives such as the Facebook Open Research and Transparency (FORT) tools provide viable means for data acquisition, accessible to academics.

\noindent\textbf{Future work}. The potential applications of our data donation methodology extend far beyond the scope of this initial study. One promising direction for future research involves leveraging this data to track how user preferences evolve over time. With our methodology, once a user consents to data collection, we can continuously monitor the pages and groups they follow until they revoke this consent. This continuous data stream opens up the possibility of studying how specific events, such as elections, influence users to engage with certain types of content. Understanding these dynamics could provide invaluable insights into the shifting landscape of public opinion and media consumption.

Moreover, the data collected through this methodology could be used to develop and train what we refer to as `artificial silicon samples'~\cite{argyle2023out} or bots. These bots, based on real user behaviors, could simulate future user interactions and preferences, providing a powerful tool for predictive analytics and behavioral modeling in social media ecosystems~\cite{hosseinmardi2024causally}.

Another avenue for future work involves designing targeted interventions. By gaining a nuanced understanding of which demographics consume specific types of content, we can craft personalized strategies aimed at enhancing information literacy and countering misinformation. For example, if we identify that a particular demographic predominantly consumes content that is prone to misinformation, targeted educational content could be designed to enhance critical thinking and fact-checking skills among that group.

While our current research has focused on textual content, the dataset we have compiled is rich with other forms of media, including images and videos, which we have yet to explore. Future studies could extend our analysis to these mediums, potentially uncovering new patterns of engagement and preference that are not visible through text-based analysis alone.
Lastly, the topic of misinformation, while touched upon briefly in this paper, warrants a deeper investigation. Our methodology provides a unique opportunity to examine the spread and impact of misinformation across different demographic groups. Future research could utilize this data to develop more effective ways of identifying and combating false information, thus contributing to healthier public discourse.






\noindent\textbf{Limitations}. One of the significant limitations of the study is our focus on English-language content. The development of a general-purpose political classifier for content in languages other than English, particularly Spanish, presented substantial challenges that we could not surmount within the scope of this project. As a result, our findings may exhibit a certain degree of bias, reflecting the content preferences and engagement patterns primarily of English-speaking users. Efforts are underway to extend our methodology to encompass additional languages, which will help mitigate this limitation in future research.

Additionally, our analysis is constrained to public content on Facebook, which does not necessarily encompass the entirety of content users encounter on their feeds. Facebook has increasingly prioritized content from social contacts over public pages and groups, and recent changes towards a TikTok-like algorithmic feed further complicate this issue. This new model promotes content from pages that users have not explicitly followed, potentially skewing the visibility and engagement of public content. Consequently, the data extracted from public pages and groups may not accurately represent the full spectrum of information that users are exposed to in their daily interactions on the platform.

Moreover, as Facebook continues to evolve its content delivery algorithms, the gap between the content that users actually see and the content available for analysis through public pages and groups may widen. This discrepancy underscores the limitations inherent in our approach, as accessing comprehensive user `diets' from an external perspective remains a challenge. Despite these constraints, our study provides a critical step toward understanding content consumption dynamics on social media platforms, offering a foundation for more nuanced analyses in the future.

Finally, while the data donation approach offers a robust methodology for studying user preferences and behaviors on social media, it also presents potential risks if misused. The voluntary nature of data donation could lead to privacy breaches if the donated data is not handled  securely. For instance, without stringent protocols on what data is collected and anonymized, malicious entities could potentially access or infer sensitive information about individuals’ political inclinations, personal interests, or affiliations from the collected data. Furthermore, there is a risk that the data, although donated for research purposes, could be repurposed for commercial or political campaigns without the explicit consent of the users (like what happened with Cambridge Analytica). This not only violates ethical standards but also undermines public trust in data donation initiatives. To mitigate these risks, it is imperative to implement rigorous data management protocols that ensure data anonymity and restrict access to authorized personnel only, thereby preserving the integrity and ethical foundation of the research.



\bibliographystyle{aaai22}
\bibliography{biblio}

\clearpage
\appendix

\section*{Appendix}

\begin{figure}[ht]
    \centering
    \includegraphics[width=\linewidth]{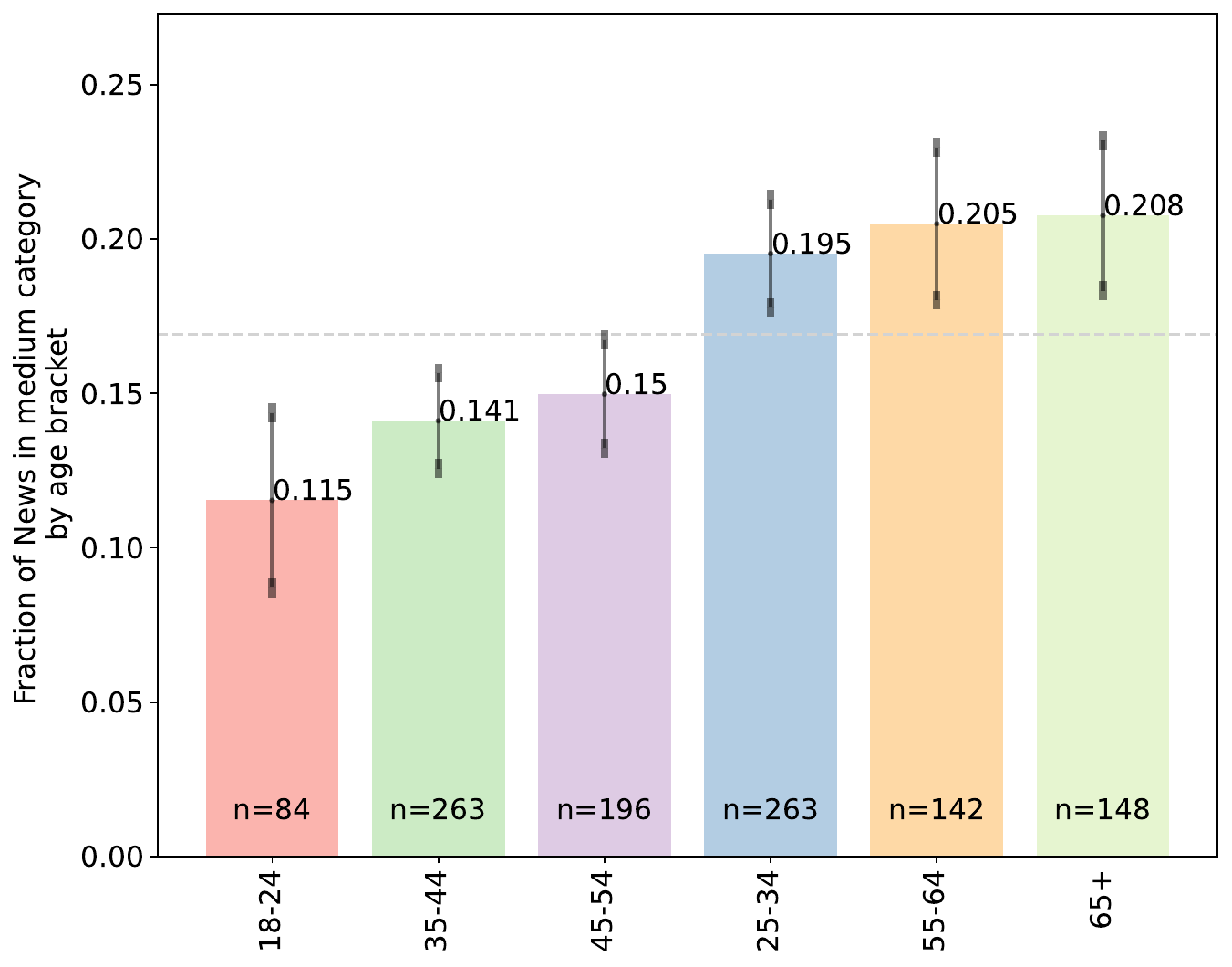}
    \caption{News consumption by age group.}
    \label{fig:news_consumption_age}
\end{figure}

Young people consume less ``News". these are according to categories for pages/groups assigned by Facebook. Figure~\ref{fig:news_consumption_age} shows the numbers. The numbers closely reflect the estimates we have in Figure~\ref{fig:political_age}.

\begin{figure}[ht]
    \centering
    \includegraphics[width=\linewidth]{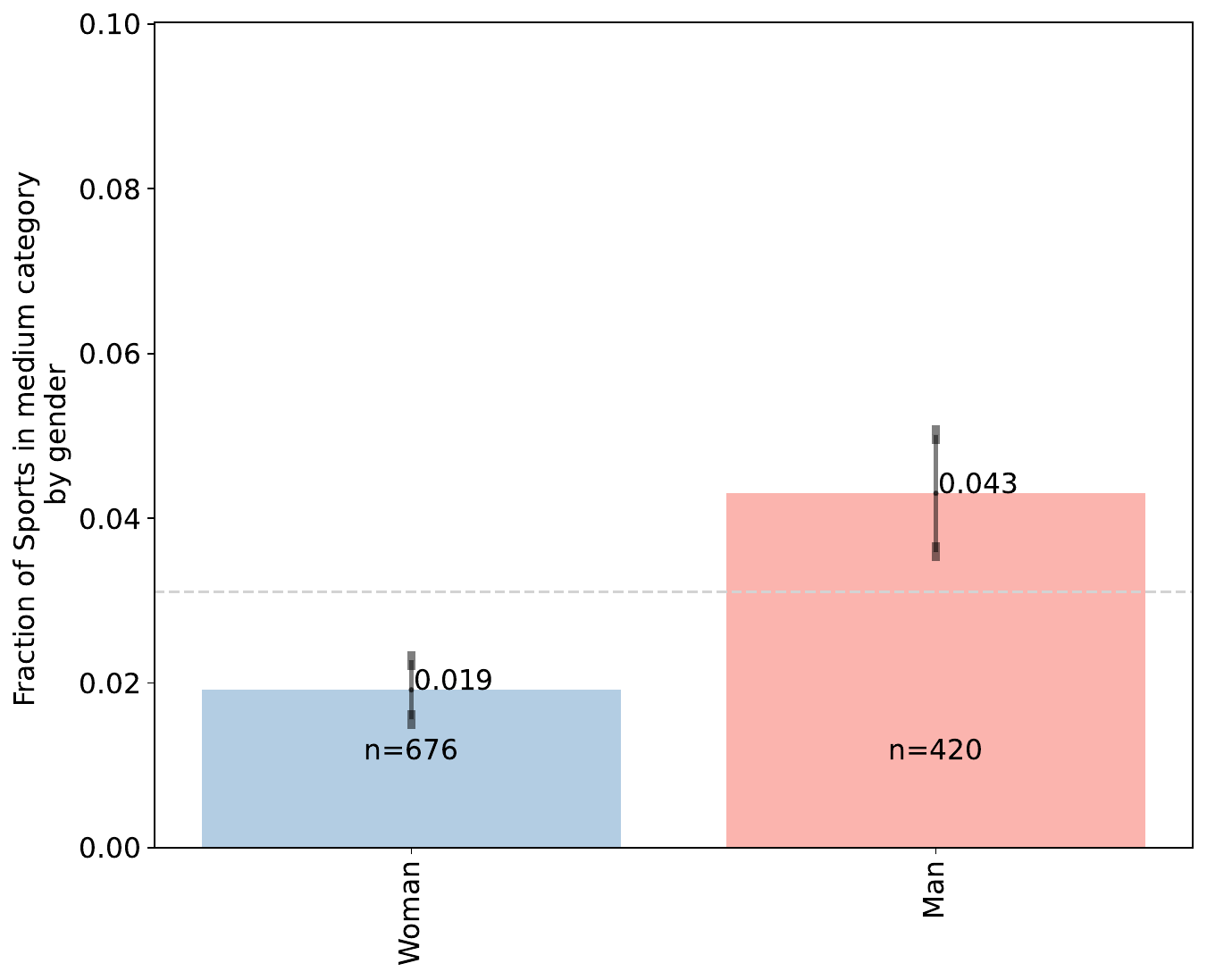}
    \caption{Sports consumption by gender.}
    \label{fig:sports_consumption_gender}
\end{figure}

Men consume so much more sports: Figure~\ref{fig:sports_consumption_gender}

\begin{figure}[ht]
    \centering
    \includegraphics[width=\linewidth]{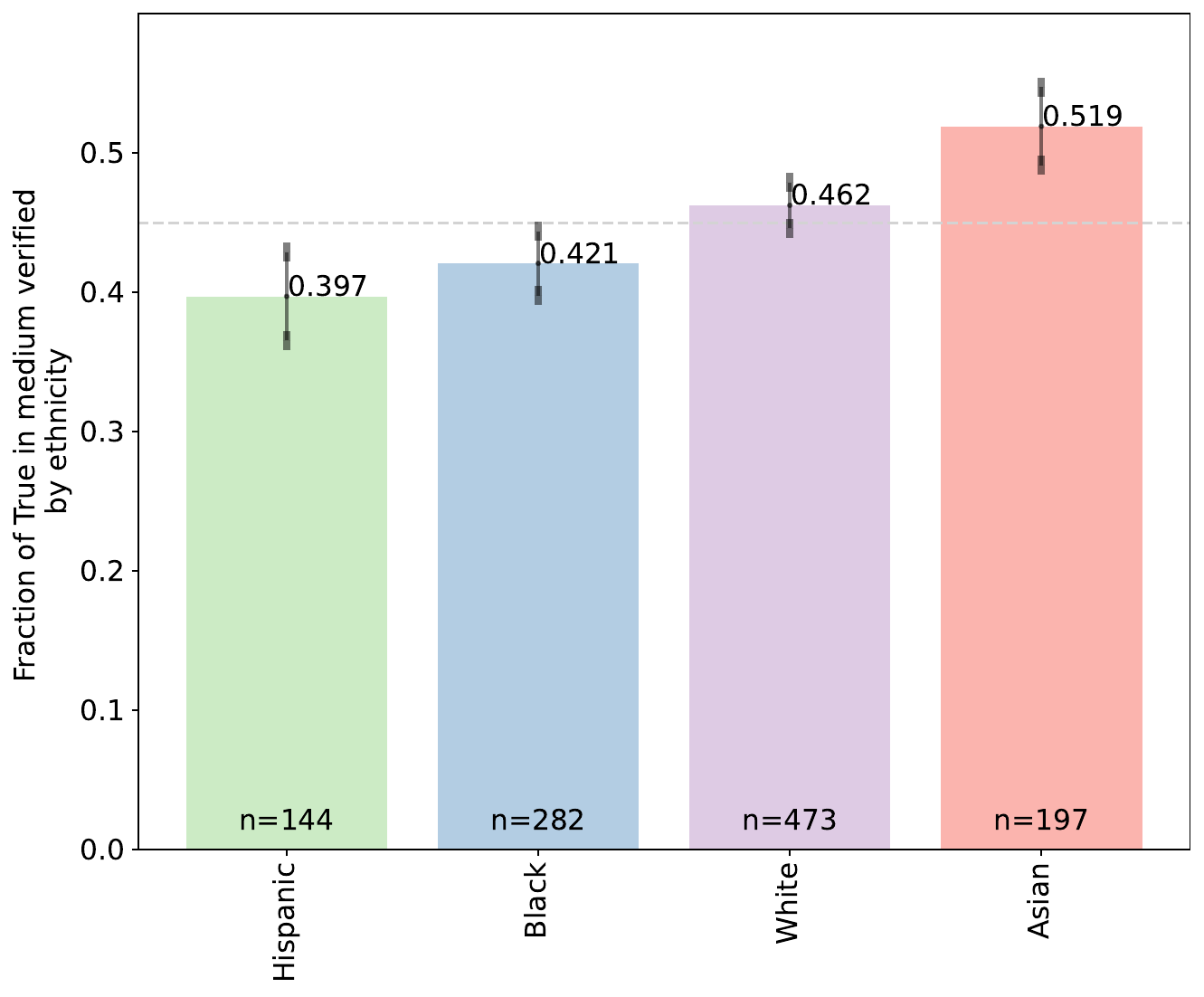}
    \caption{Verified content consumption by ethnicity.}
    \label{fig:verified_consumption_ethnicity}
\end{figure}

Asians consume significantly higher content from verified sources (compared to Hispanics). Figure~\ref{fig:verified_consumption_ethnicity}
Over 50\% of the content Asians consume is from verified pages, where as its close to 40\% for Hispanics.

\begin{figure*}
\centering
\begin{subfigure}[b]{0.30\textwidth}
    \includegraphics[width=\textwidth]{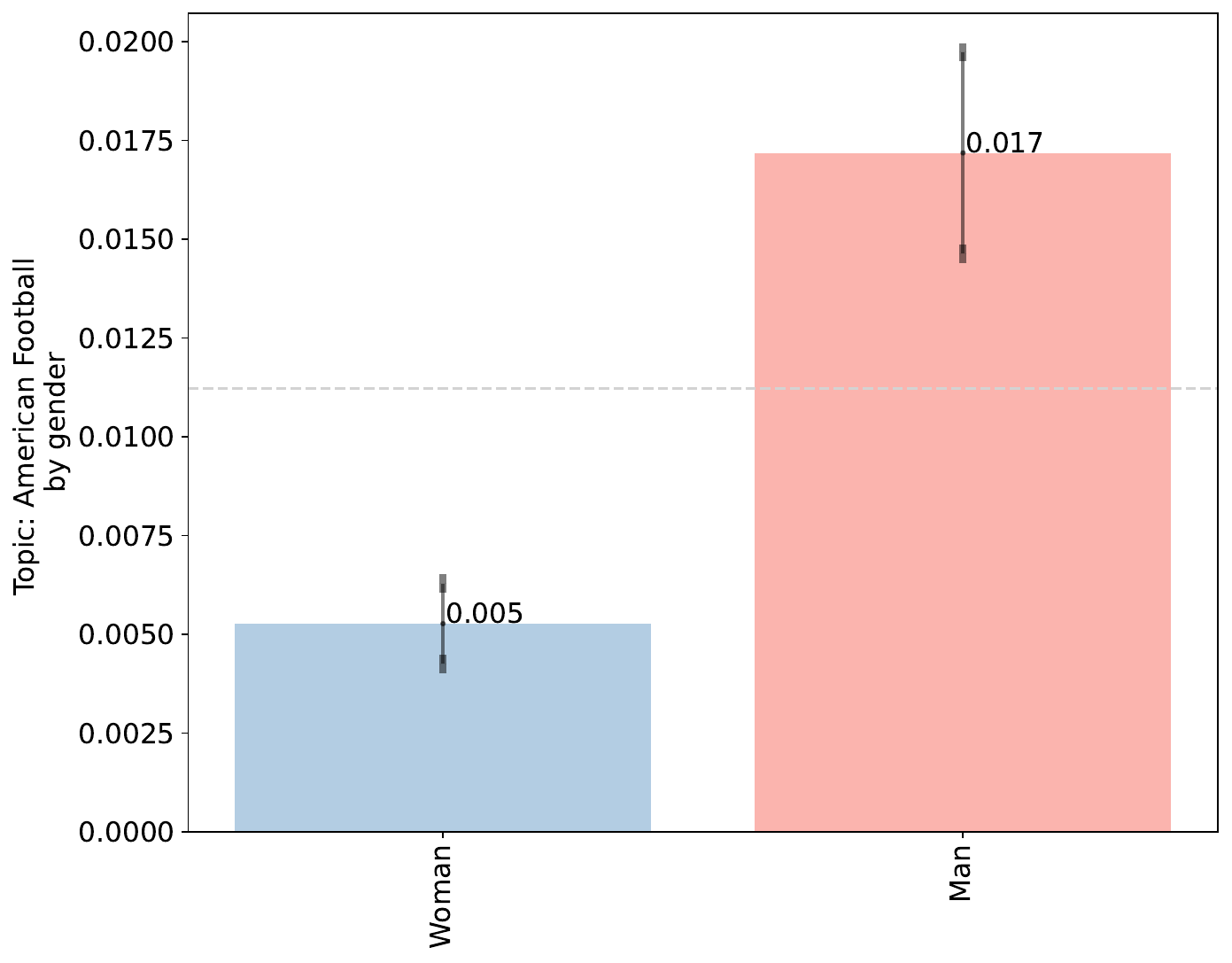}
    \caption{}
    \label{fig:1}
\end{subfigure}
\hfill
\begin{subfigure}[b]{0.30\textwidth}
    \includegraphics[width=\textwidth]{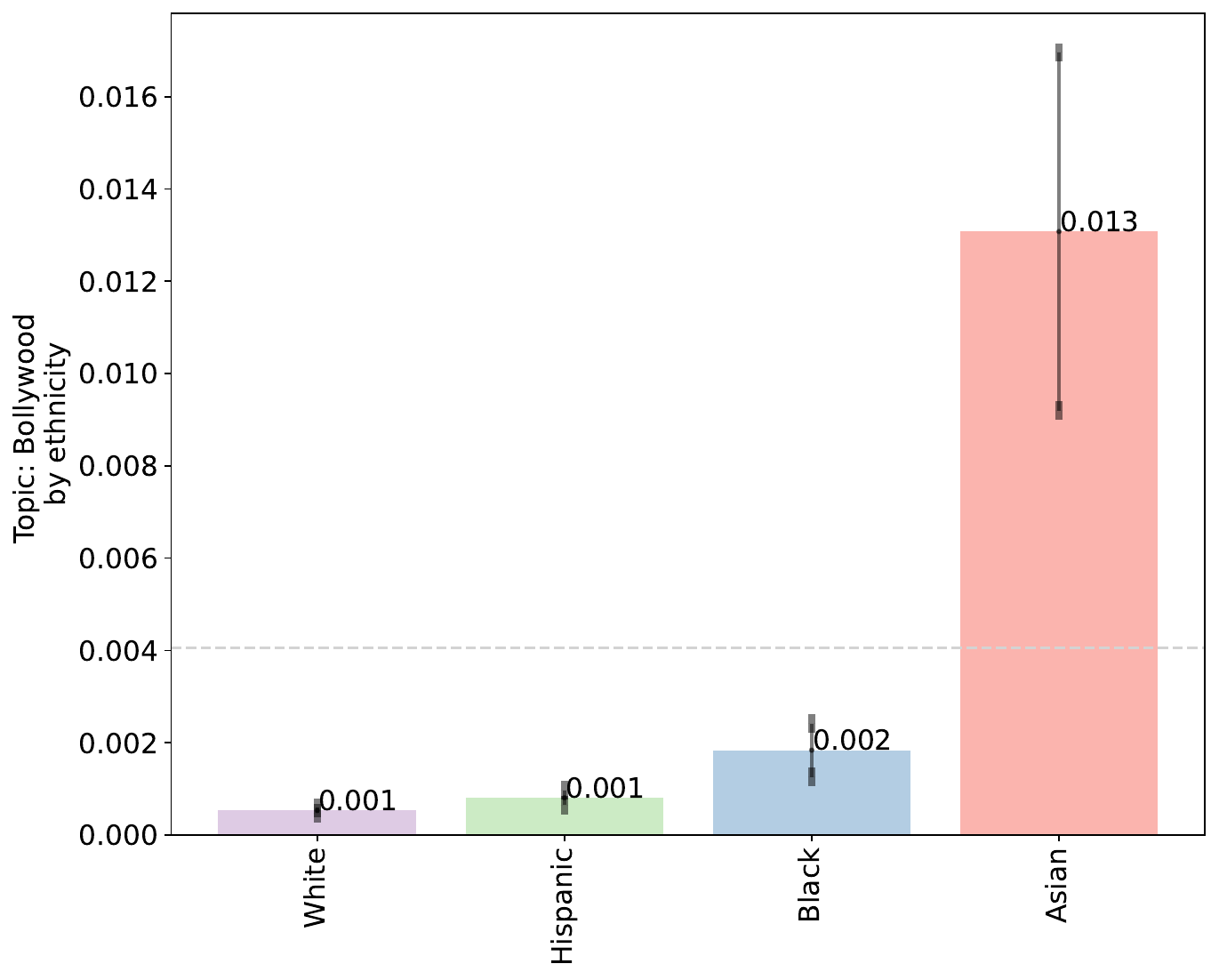}
    \caption{}
    \label{fig:2}
\end{subfigure}
\hfill
\begin{subfigure}[b]{0.30\textwidth}
    \includegraphics[width=\textwidth]{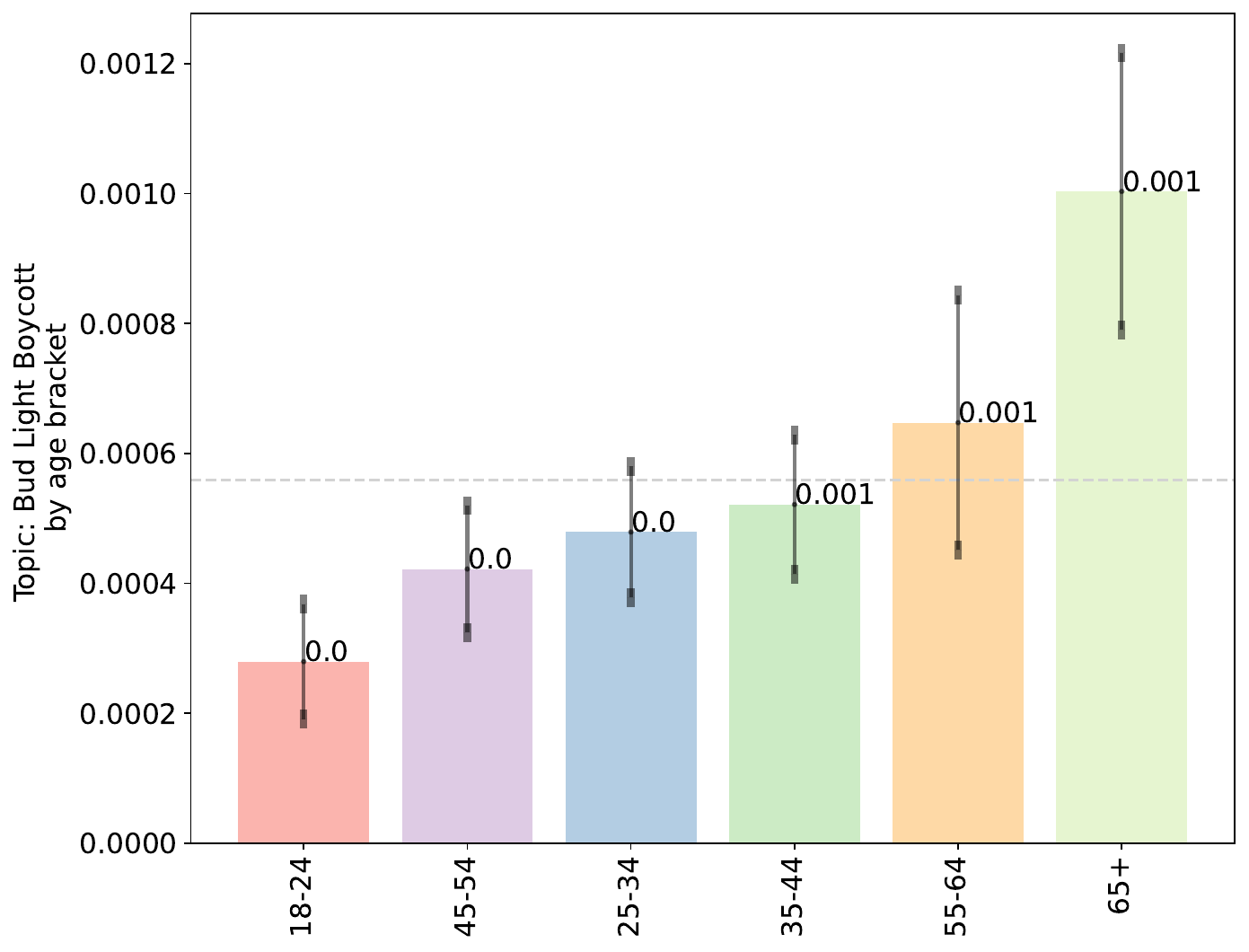}
    \caption{}
    \label{fig:3}
\end{subfigure}
\caption{(a) Topic: American Football proportion by gender, (b) Topic: Bollywood proportion by ethnicity, (c) Topic: Budlight boycott proportion by age group}
\label{fig:topic_plots}
\vspace{-\baselineskip}
\end{figure*}

\begin{figure}[ht]
    \centering
    \includegraphics[width=\linewidth]{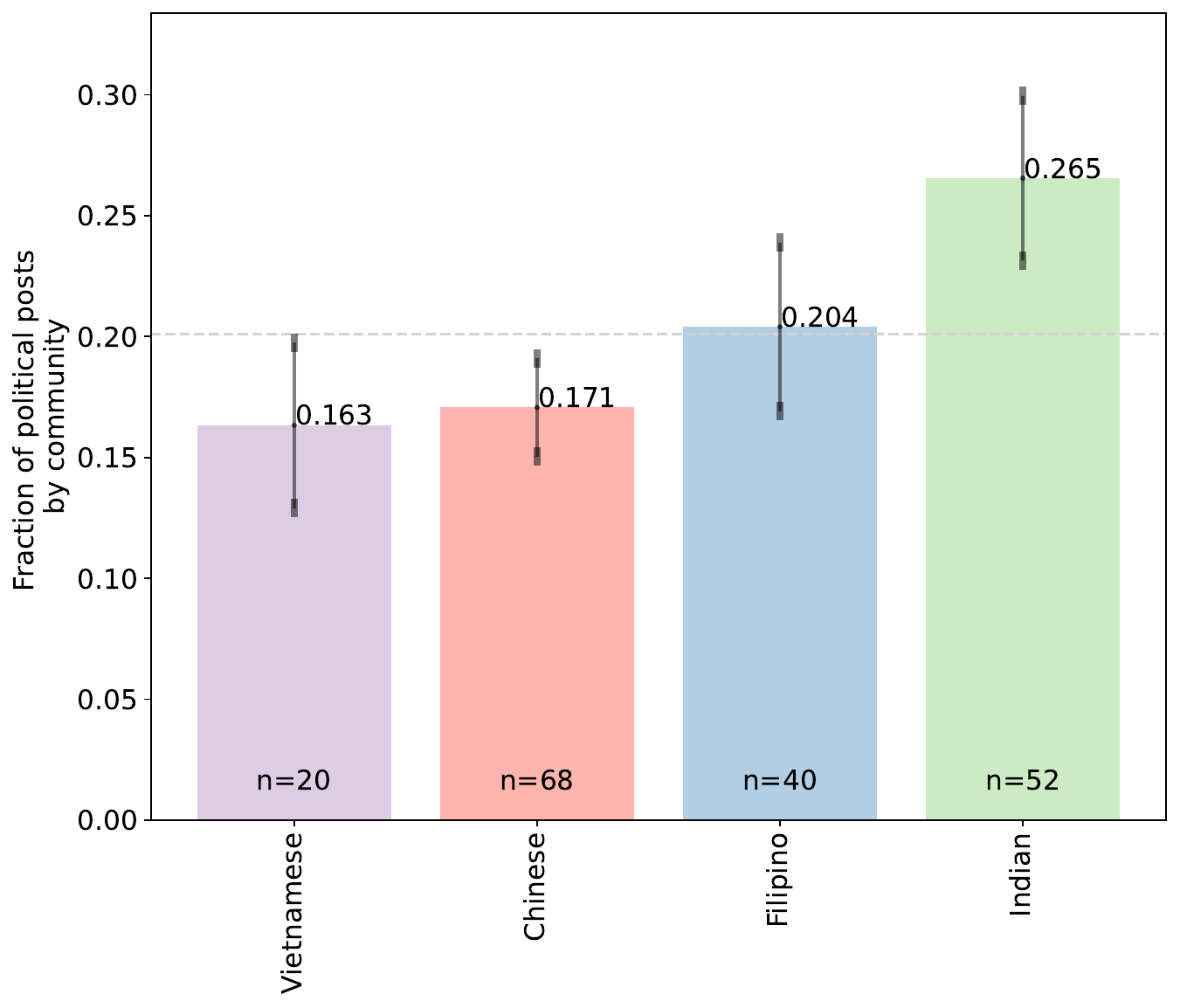}
    \caption{Political content consumption by community within Asian American.}
    \label{fig:political_community}
\end{figure}

Political content consumption by community within Asian American (unweighted). This granular level of detail shows the value of our data donation model. Figure~\ref{fig:political_community}

\begin{figure}[ht]
    \centering
    \includegraphics[width=\linewidth]{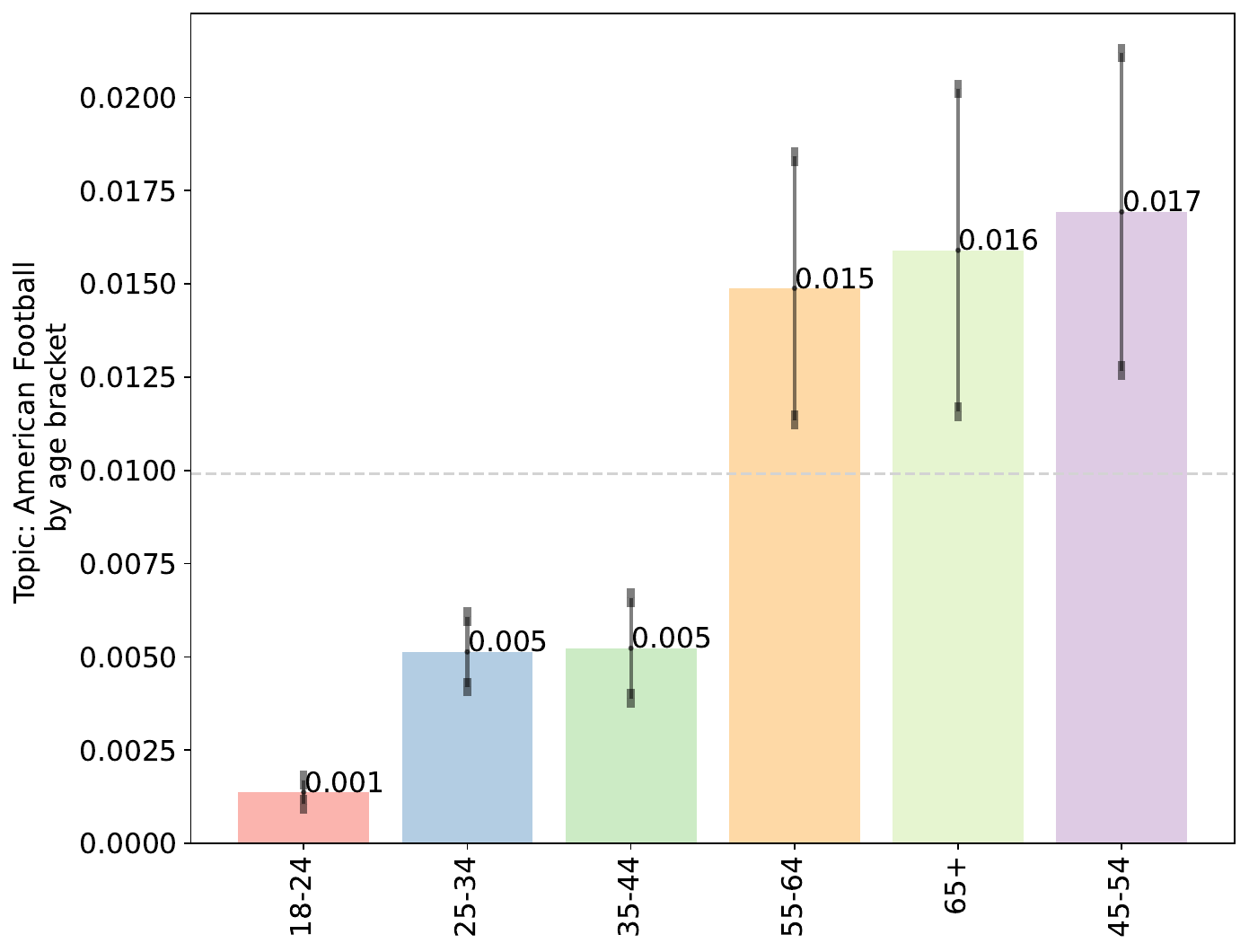}
    \caption{American football by age group}
    \label{fig:topic_american_football_age}
\end{figure}

American football is mostly consumed by older people. younger people do not consume it. Figure~\ref{fig:topic_american_football_age}.
Same trend for Baseball, college sports,

\begin{figure}[ht]
    \centering
    \includegraphics[width=\linewidth]{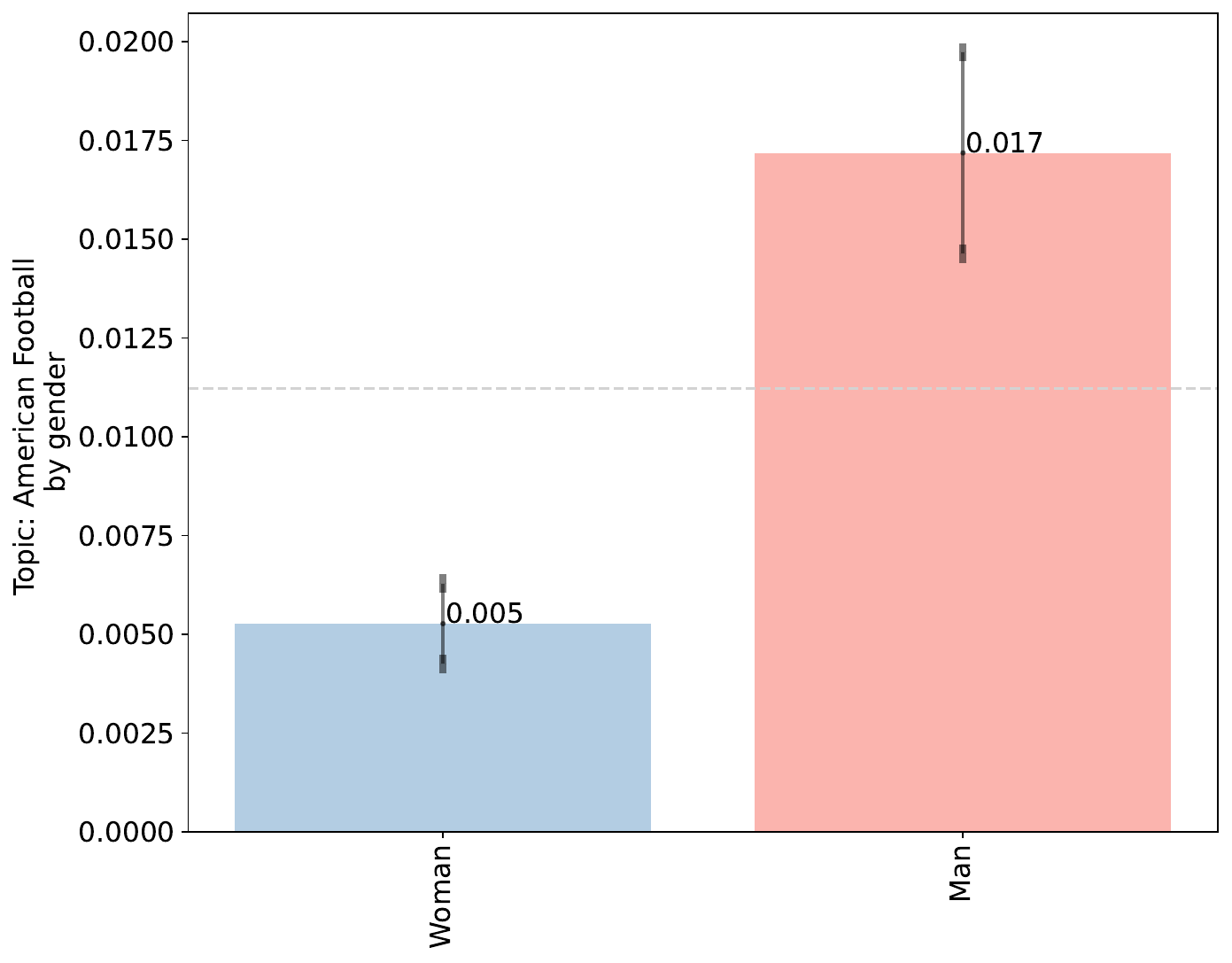}
    \caption{American football by gender}
    \label{fig:topic_american_football_gender}
\end{figure}

American football (and basketball) consumption (similar to sports) is much higher in men. Figure~\ref{fig:topic_american_football_gender}.

\begin{figure*}[ht]
\centering
\begin{subfigure}[b]{0.24\textwidth}
    \includegraphics[width=\textwidth]{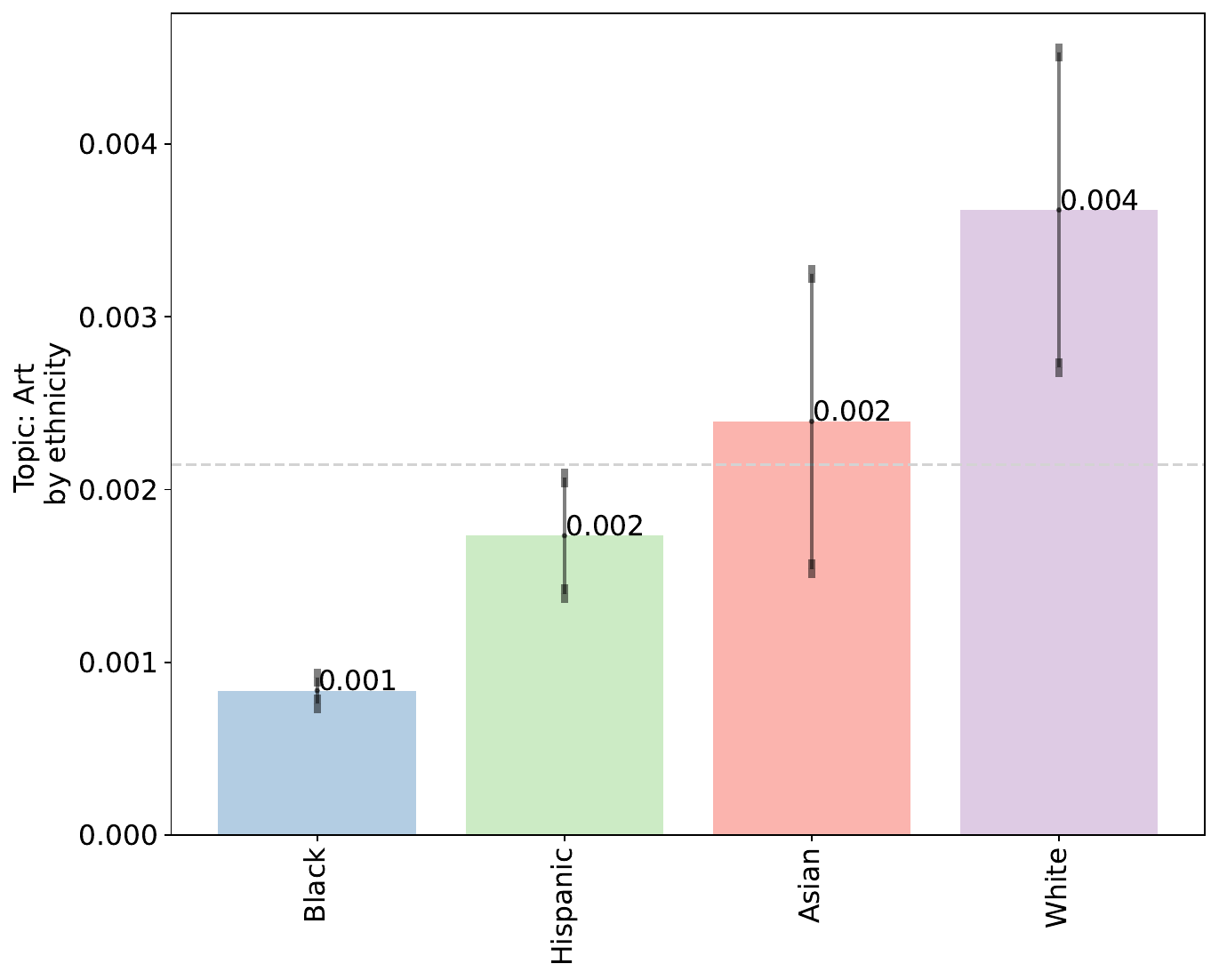}
    \caption{Art}
    \label{fig:1}
\end{subfigure}
\hfill
\begin{subfigure}[b]{0.24\textwidth}
    \includegraphics[width=\textwidth]{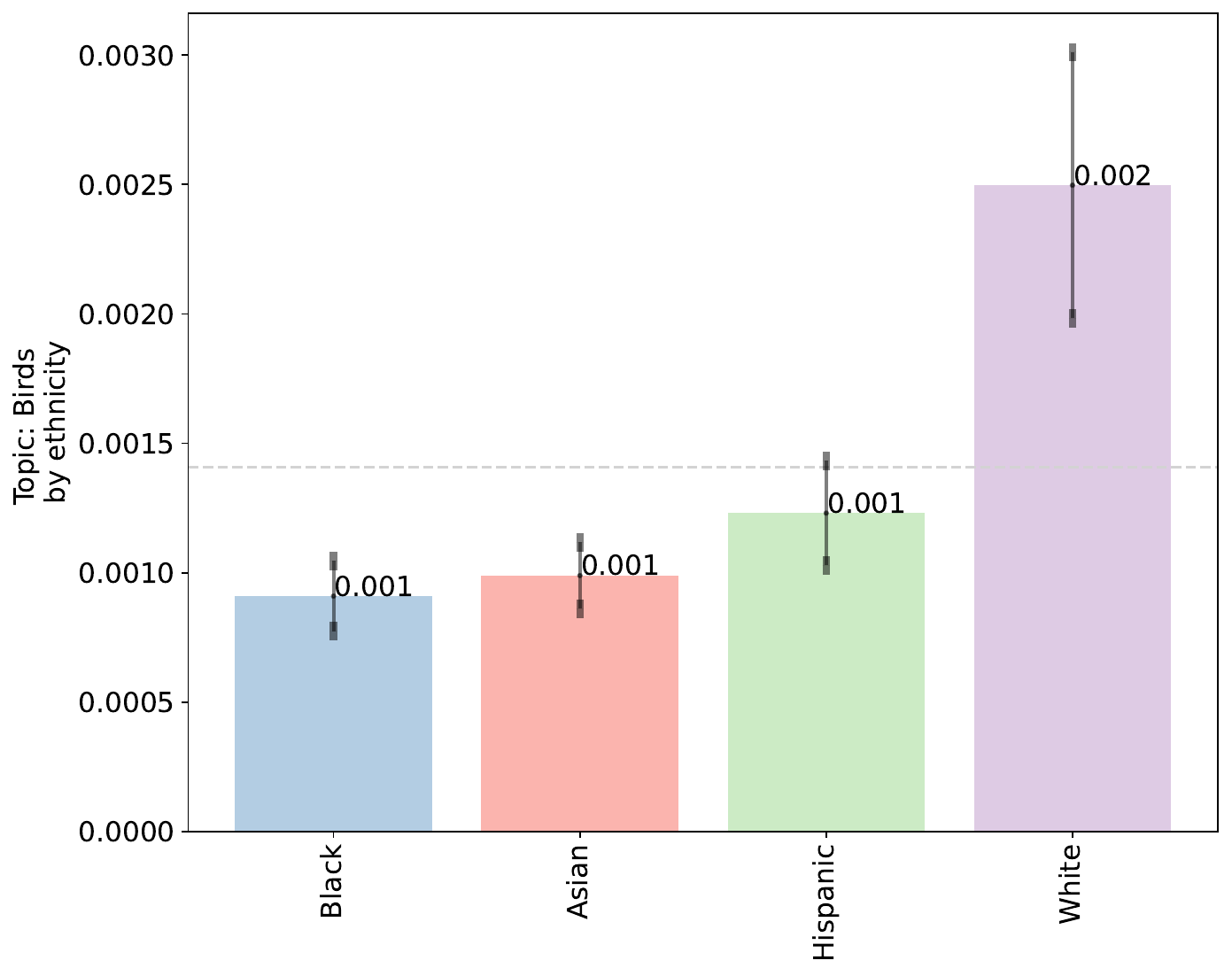}
    \caption{Bird watching}
    \label{fig:2}
\end{subfigure}
\hfill
\begin{subfigure}[b]{0.24\textwidth}
    \includegraphics[width=\textwidth]{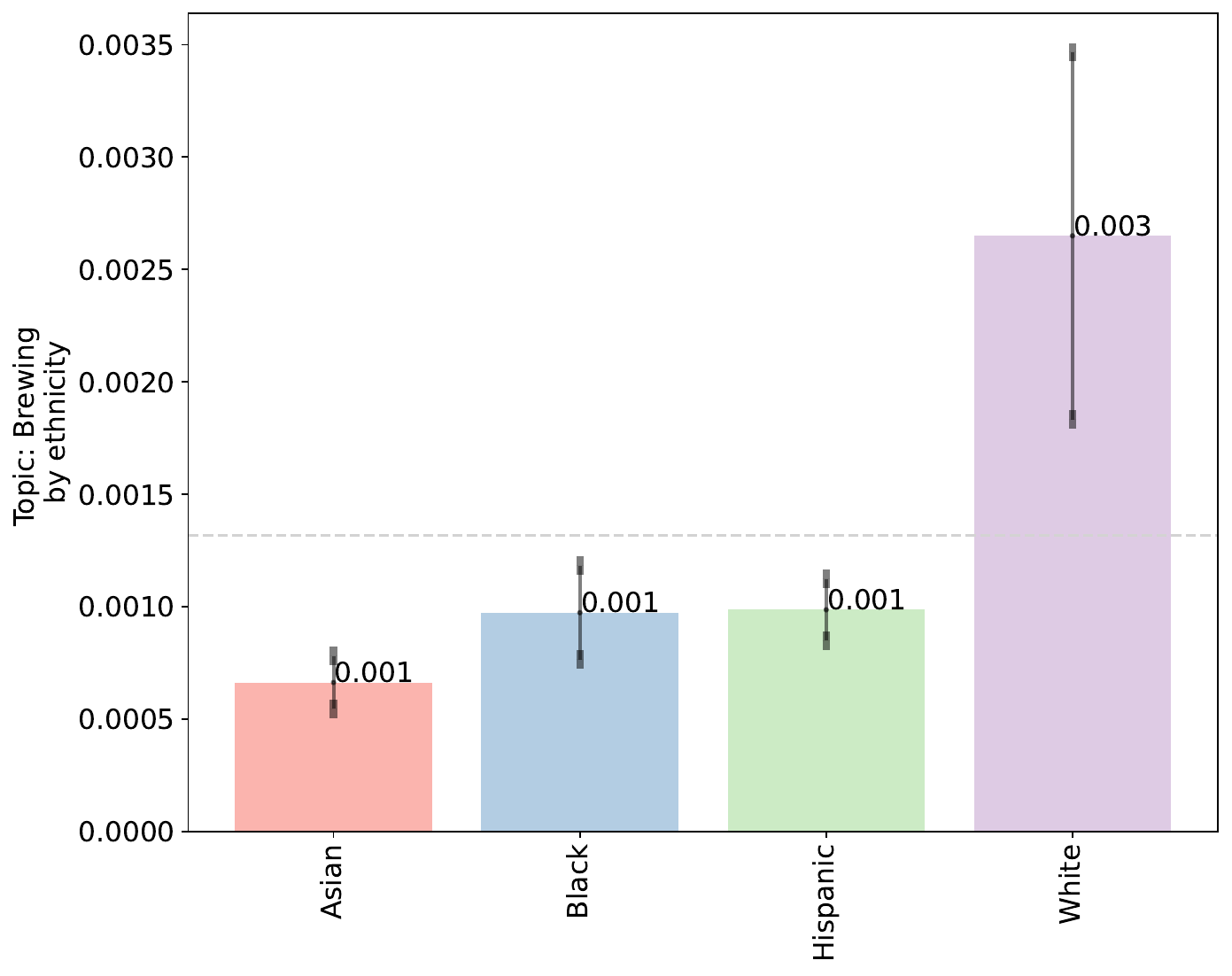}
    \caption{Beer brewing}
    \label{fig:3}
\end{subfigure}
\hfill
\begin{subfigure}[b]{0.25\textwidth}
    \includegraphics[width=\textwidth]{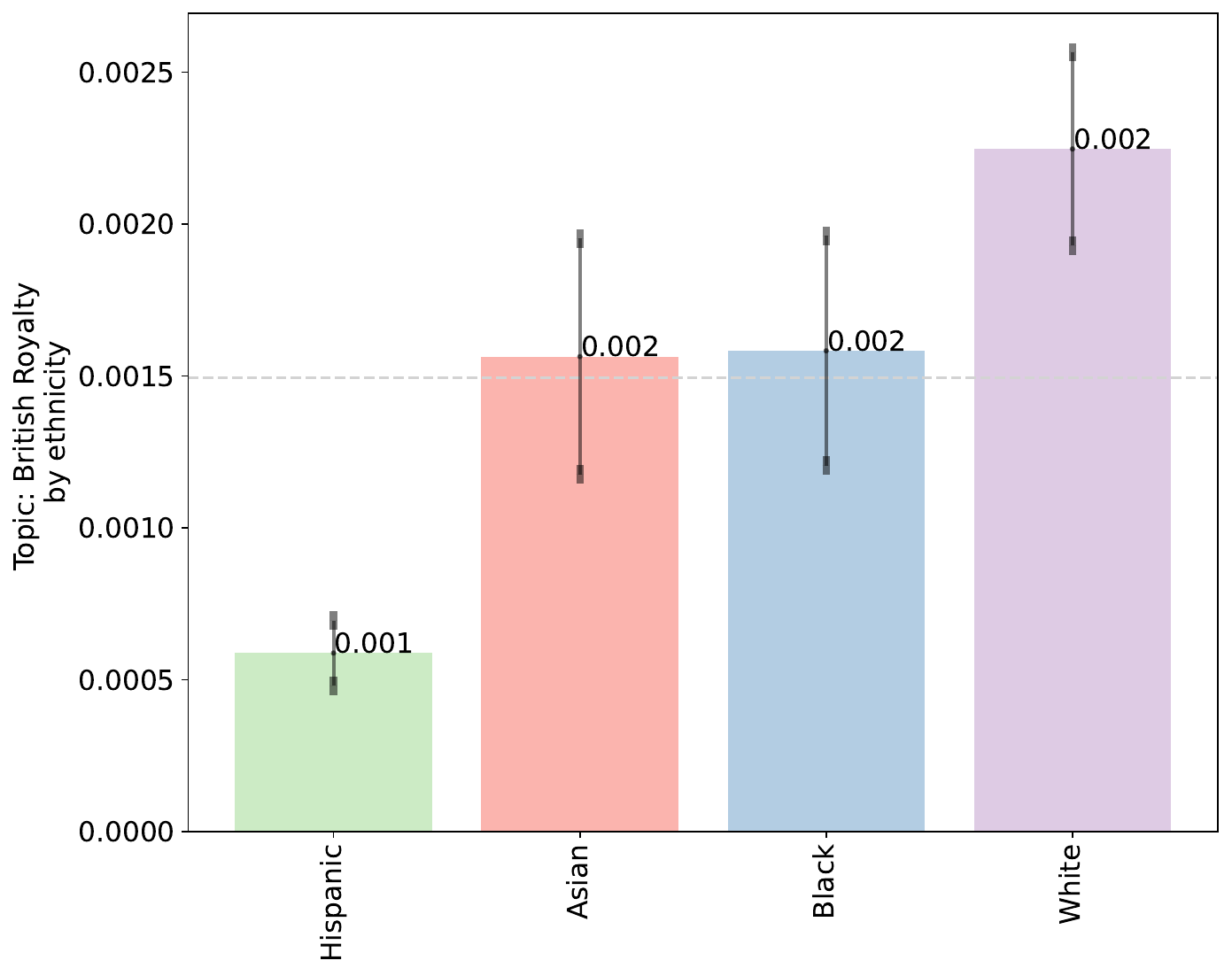}
    \caption{British royalty}
    \label{fig:4}
\end{subfigure}
\caption{Topic specific bar charts showing prevalence of various topics.}
\label{fig:topic_appendix1}
\vspace{-\baselineskip}
\end{figure*}

\begin{figure*}[ht]
\centering
\begin{subfigure}[b]{0.24\textwidth}
    \includegraphics[width=\textwidth]{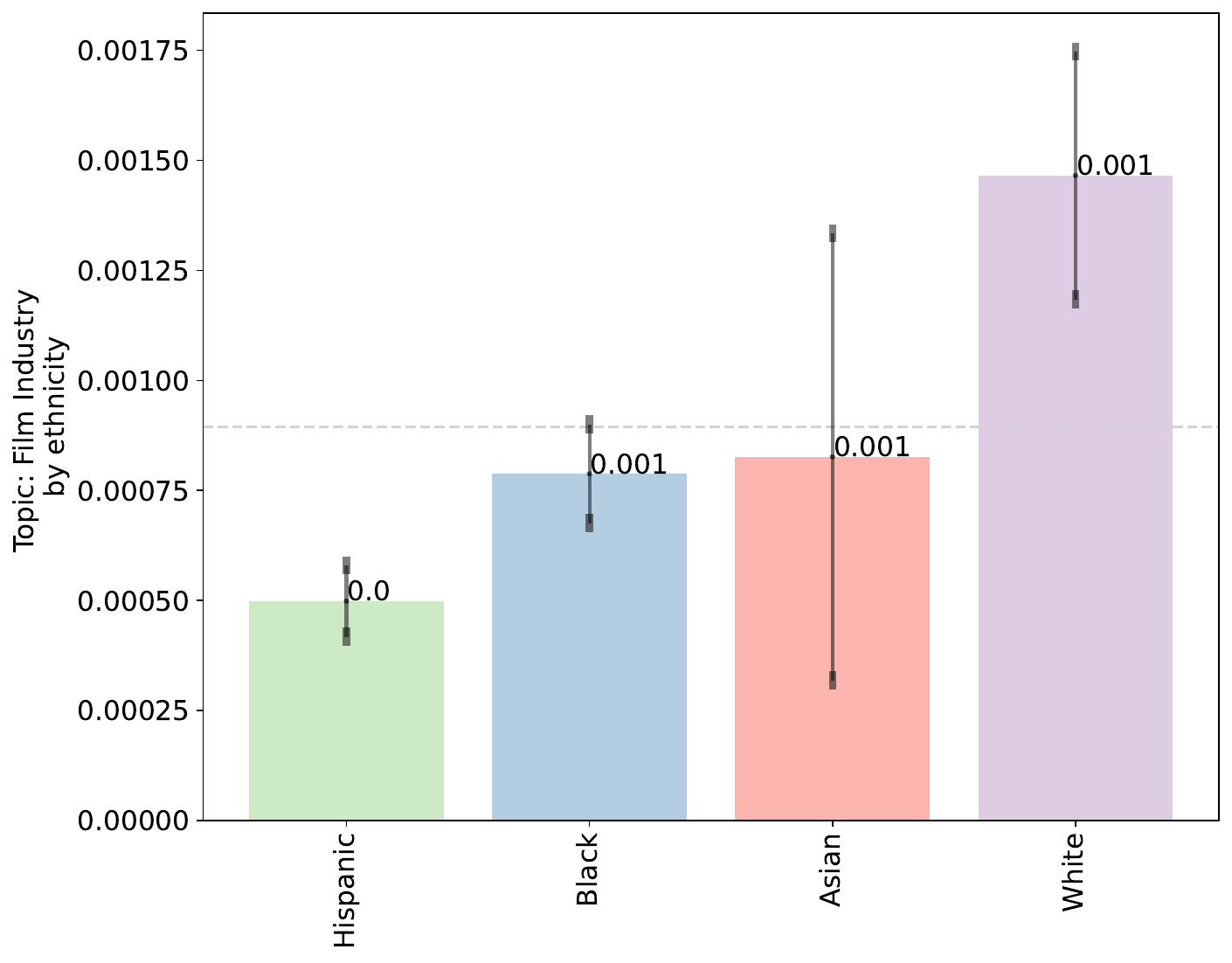}
    \caption{Film entertainment}
    \label{fig:1}
\end{subfigure}
\hfill
\begin{subfigure}[b]{0.24\textwidth}
    \includegraphics[width=\textwidth]{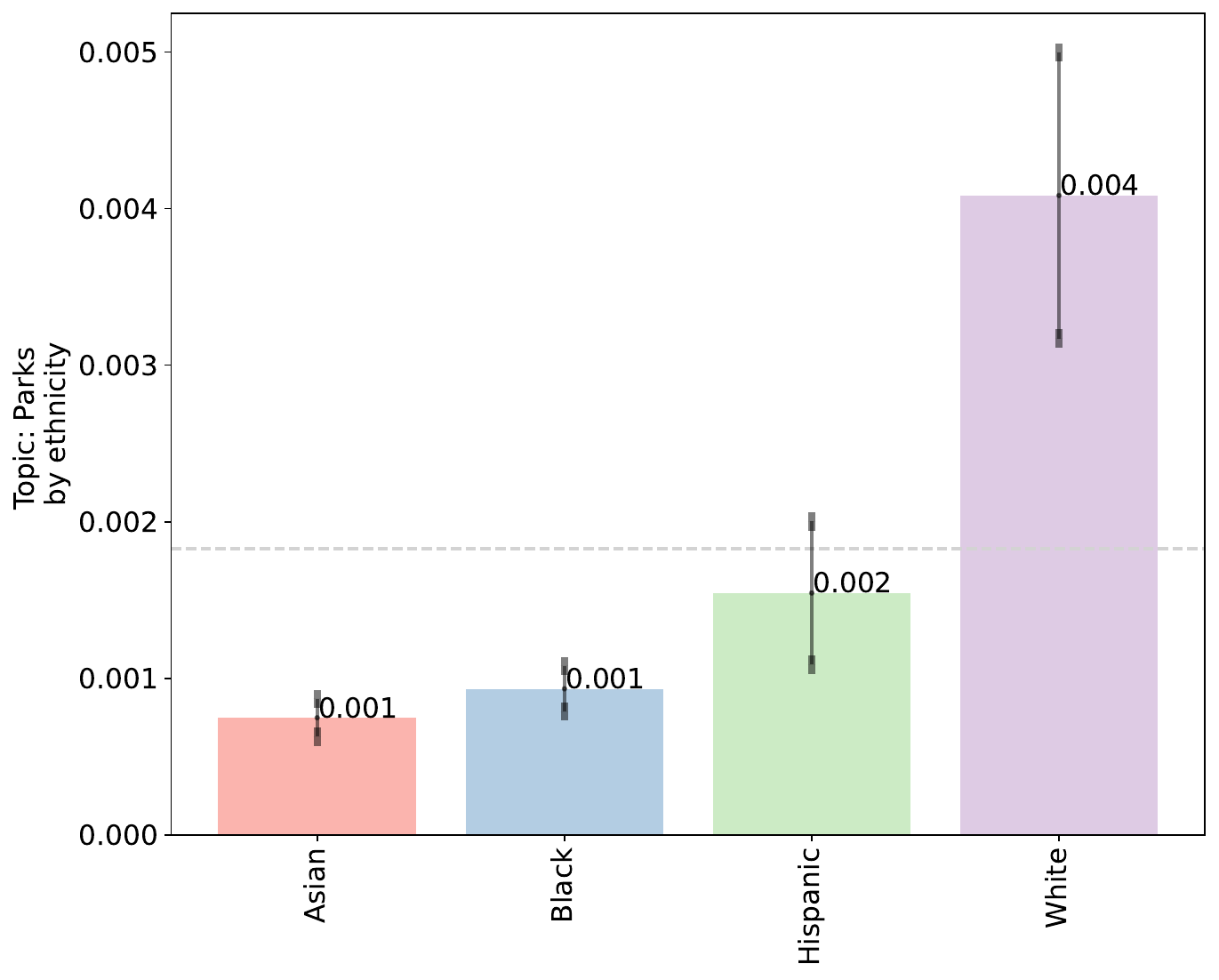}
    \caption{Parks}
    \label{fig:2}
\end{subfigure}
\hfill
\begin{subfigure}[b]{0.24\textwidth}
    \includegraphics[width=\textwidth]{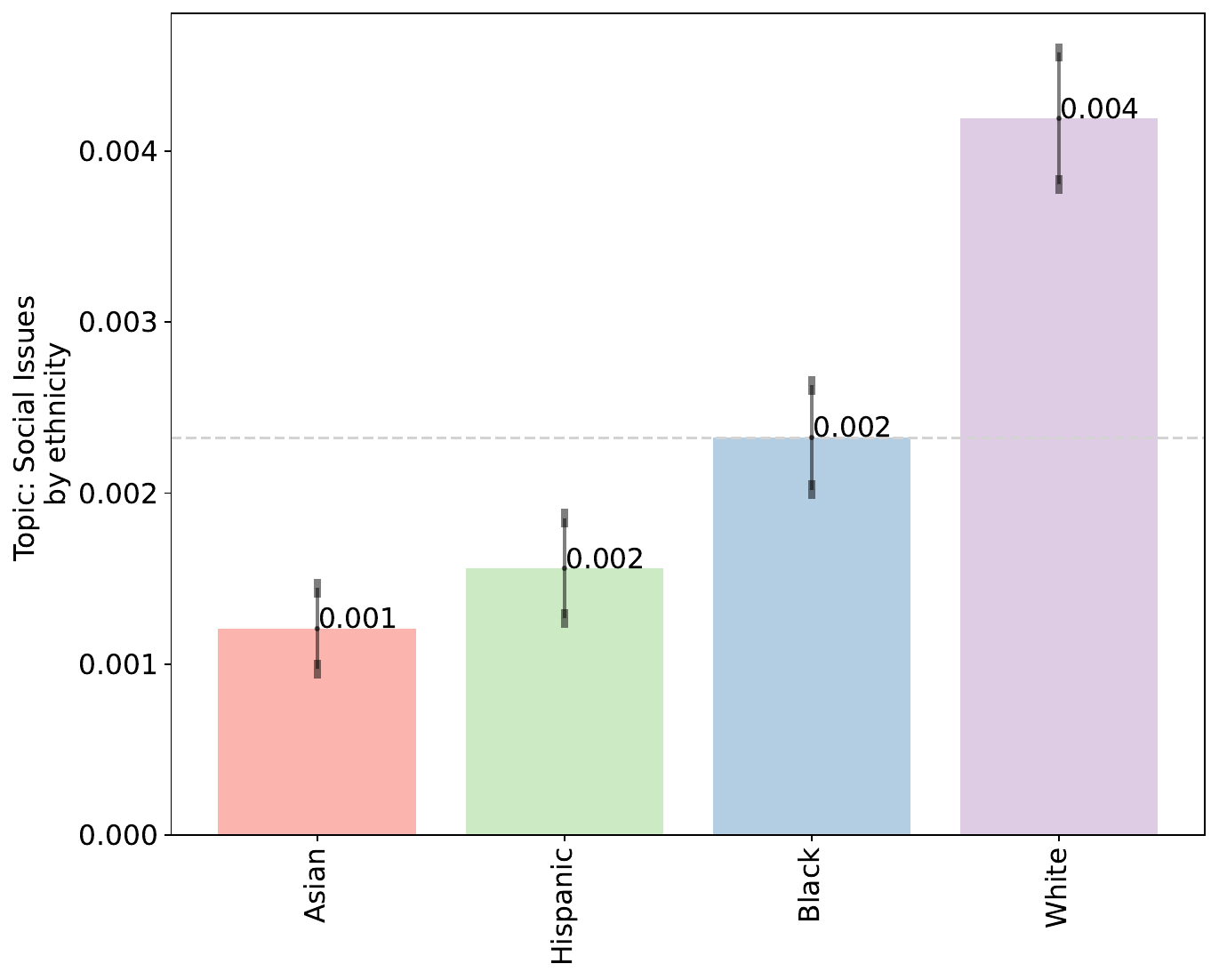}
    \caption{Social issues}
    \label{fig:3}
\end{subfigure}
\hfill
\begin{subfigure}[b]{0.25\textwidth}
    \includegraphics[width=\textwidth]{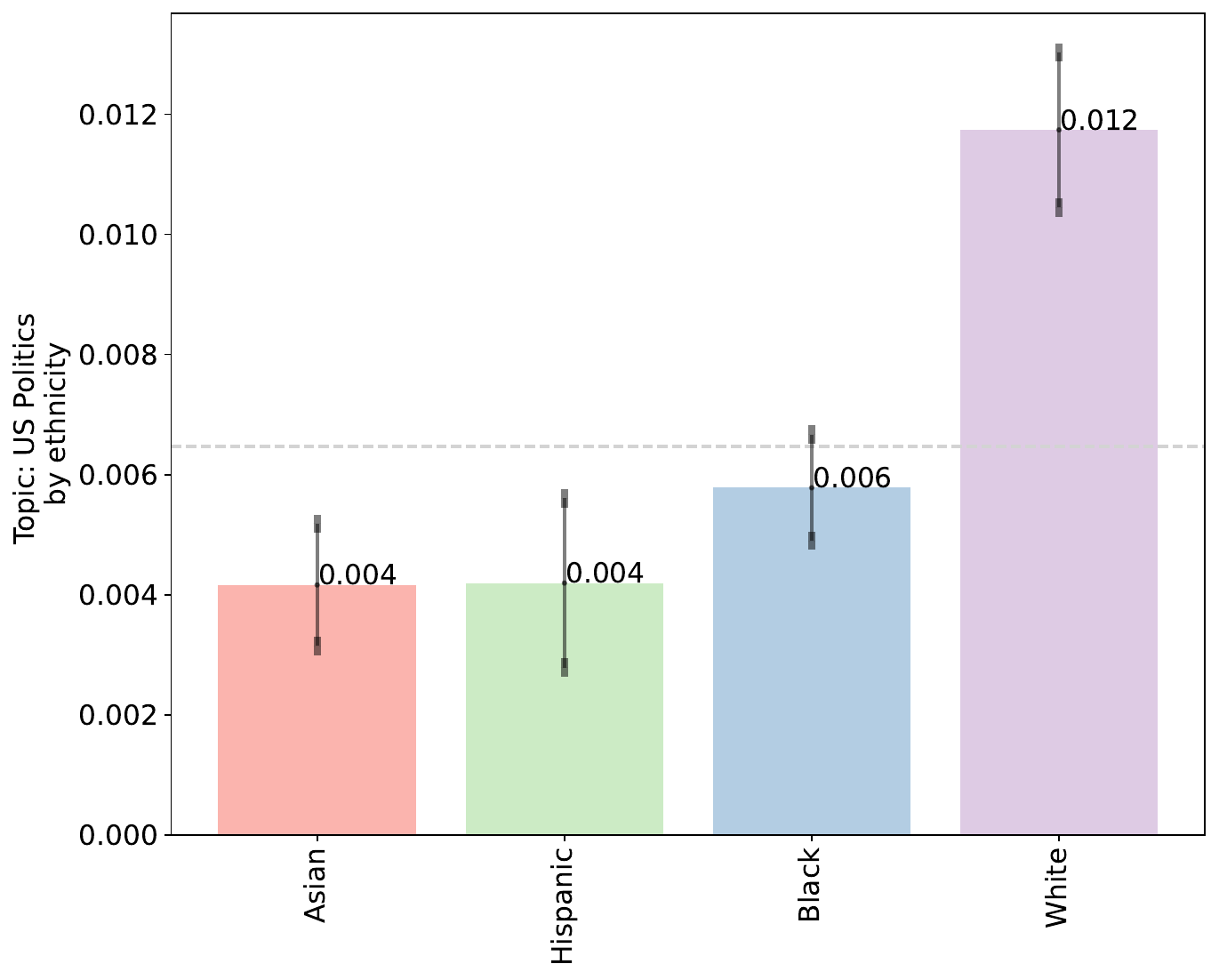}
    \caption{US Politics}
    \label{fig:4}
\end{subfigure}
\caption{Topic specific bar charts showing prevalence of various topics.}
\label{fig:topic_appendix2}
\vspace{-\baselineskip}
\end{figure*}

\begin{figure*}[ht]
\centering
\begin{subfigure}[b]{0.24\textwidth}
    \includegraphics[width=\textwidth]{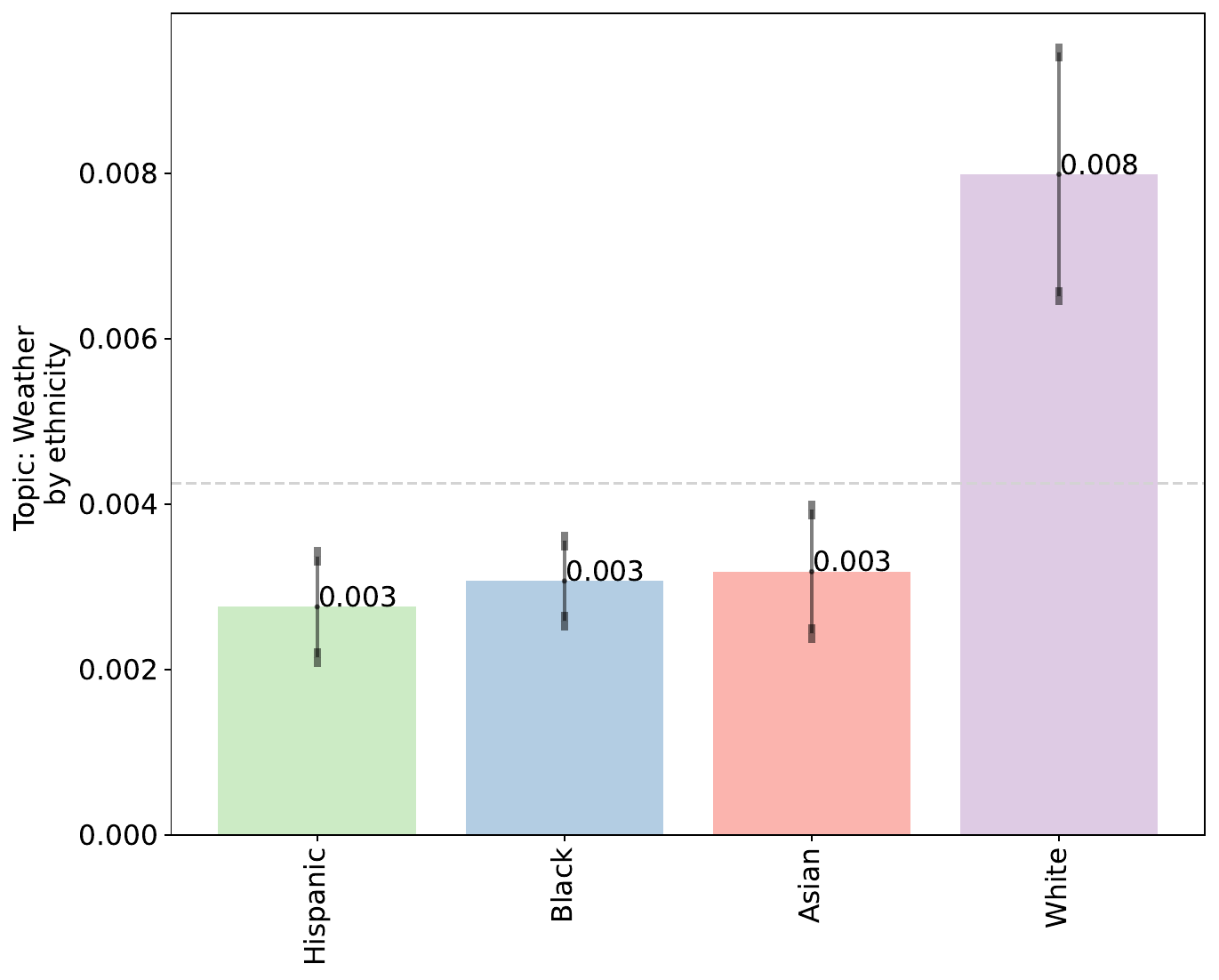}
    \caption{Weather}
    \label{fig:1}
\end{subfigure}
\hfill
\begin{subfigure}[b]{0.24\textwidth}
    \includegraphics[width=\textwidth]{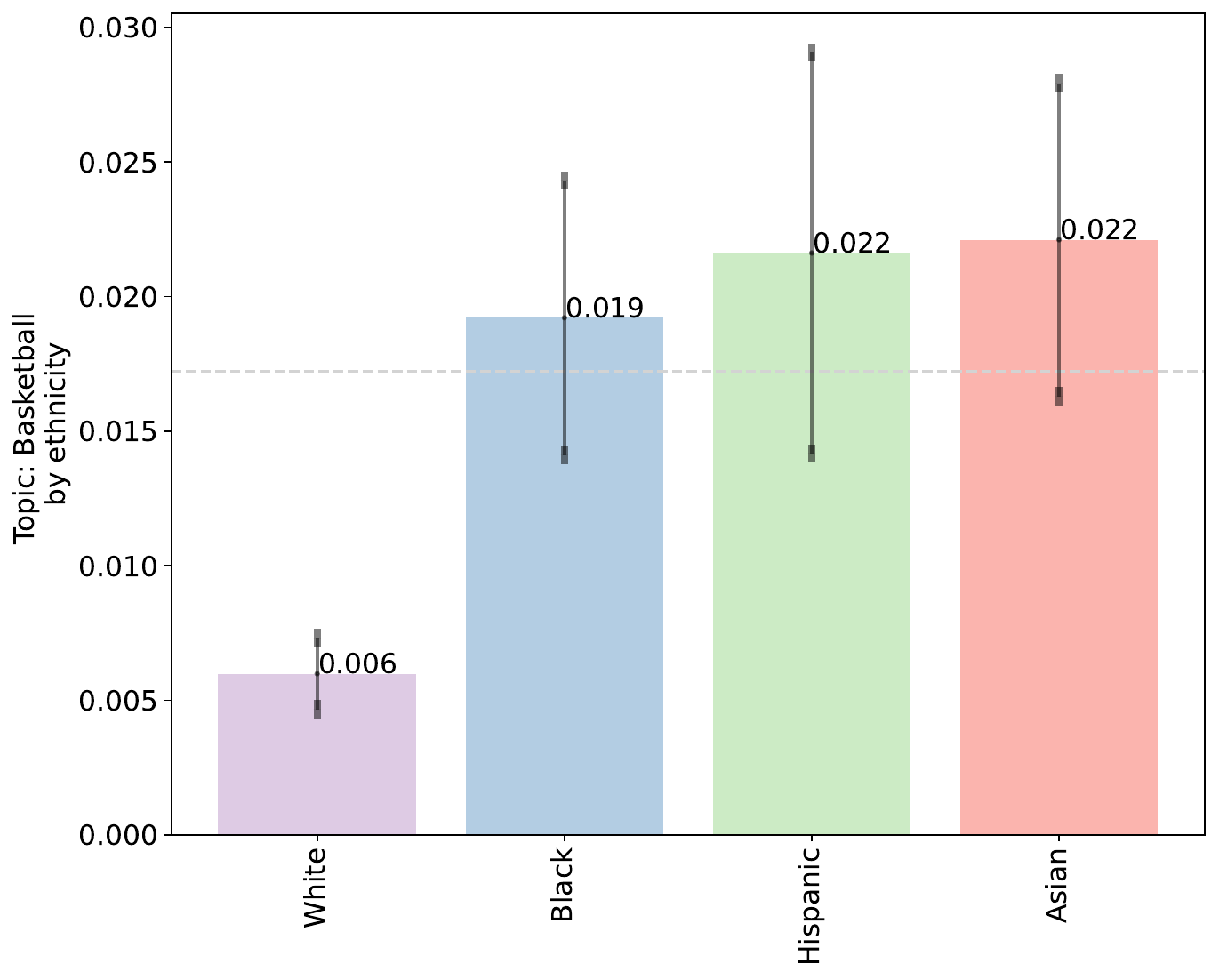}
    \caption{Basketball}
    \label{fig:2}
\end{subfigure}
\hfill
\begin{subfigure}[b]{0.24\textwidth}
    \includegraphics[width=\textwidth]{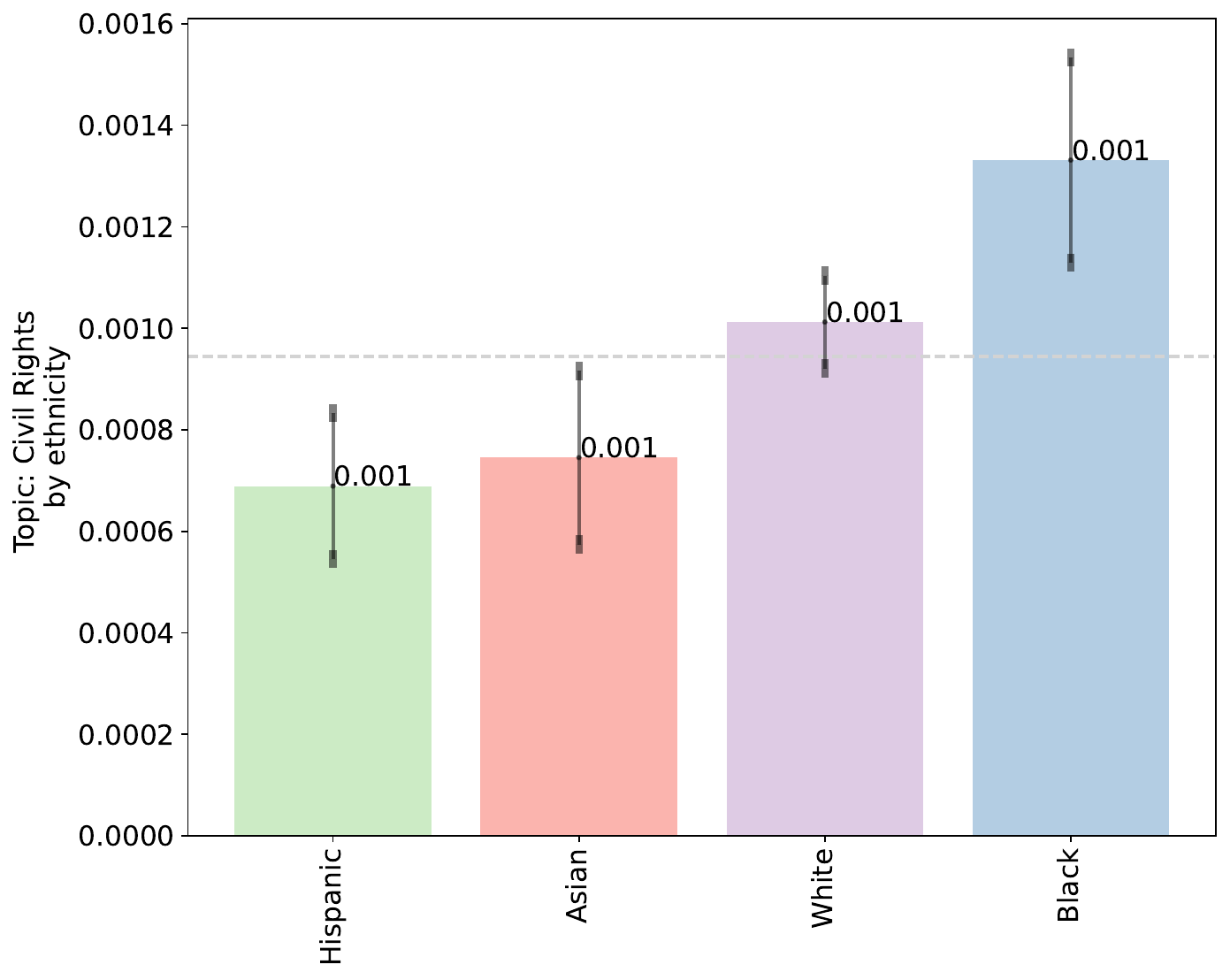}
    \caption{Civil rights}
    \label{fig:3}
\end{subfigure}
\hfill
\begin{subfigure}[b]{0.25\textwidth}
    \includegraphics[width=\textwidth]{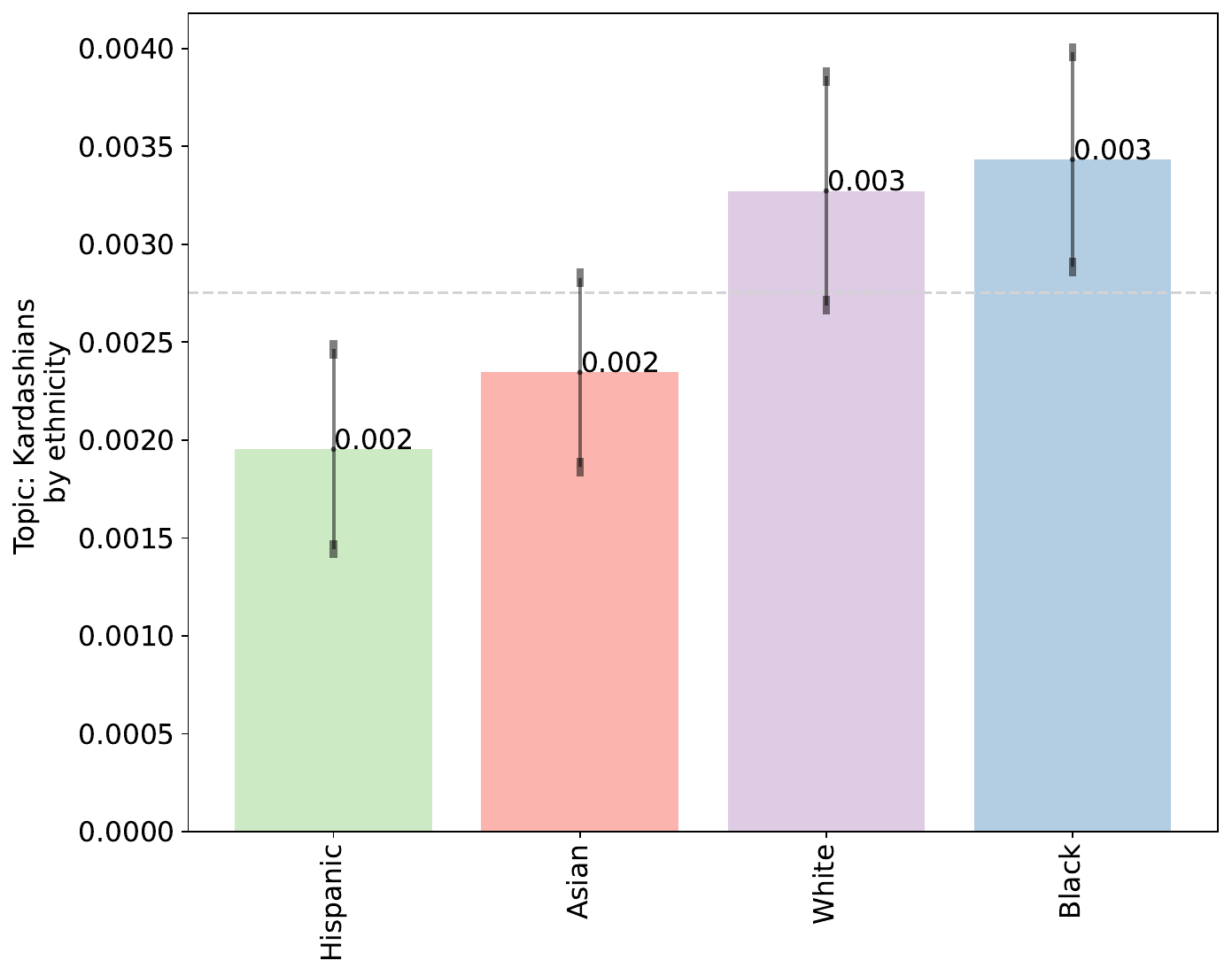}
    \caption{Kardashians}
    \label{fig:4}
\end{subfigure}
\caption{Topic specific bar charts showing prevalence of various topics.}
\label{fig:topic_appendix3}
\vspace{-\baselineskip}
\end{figure*}

\begin{figure*}[ht]
\centering
\begin{subfigure}[b]{0.24\textwidth}
    \includegraphics[width=\textwidth]{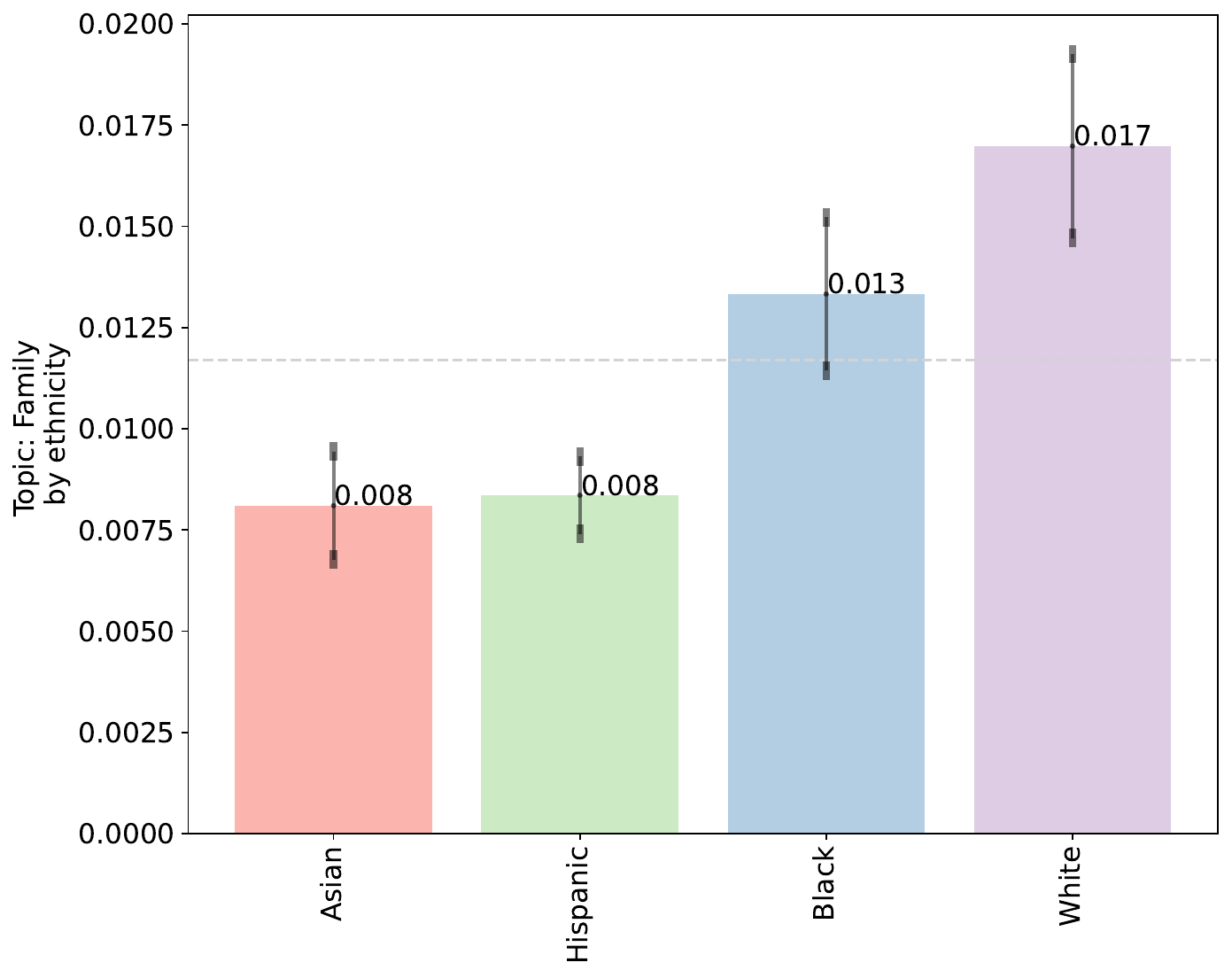}
    \caption{Family}
    \label{fig:1}
\end{subfigure}
\hfill
\begin{subfigure}[b]{0.24\textwidth}
    \includegraphics[width=\textwidth]{img/per_topic_bars/bar_Topic__Bollywood_per_Ethnicity.pdf}
    \caption{Bollywood}
    \label{fig:2}
\end{subfigure}
\hfill
\begin{subfigure}[b]{0.24\textwidth}
    \includegraphics[width=\textwidth]{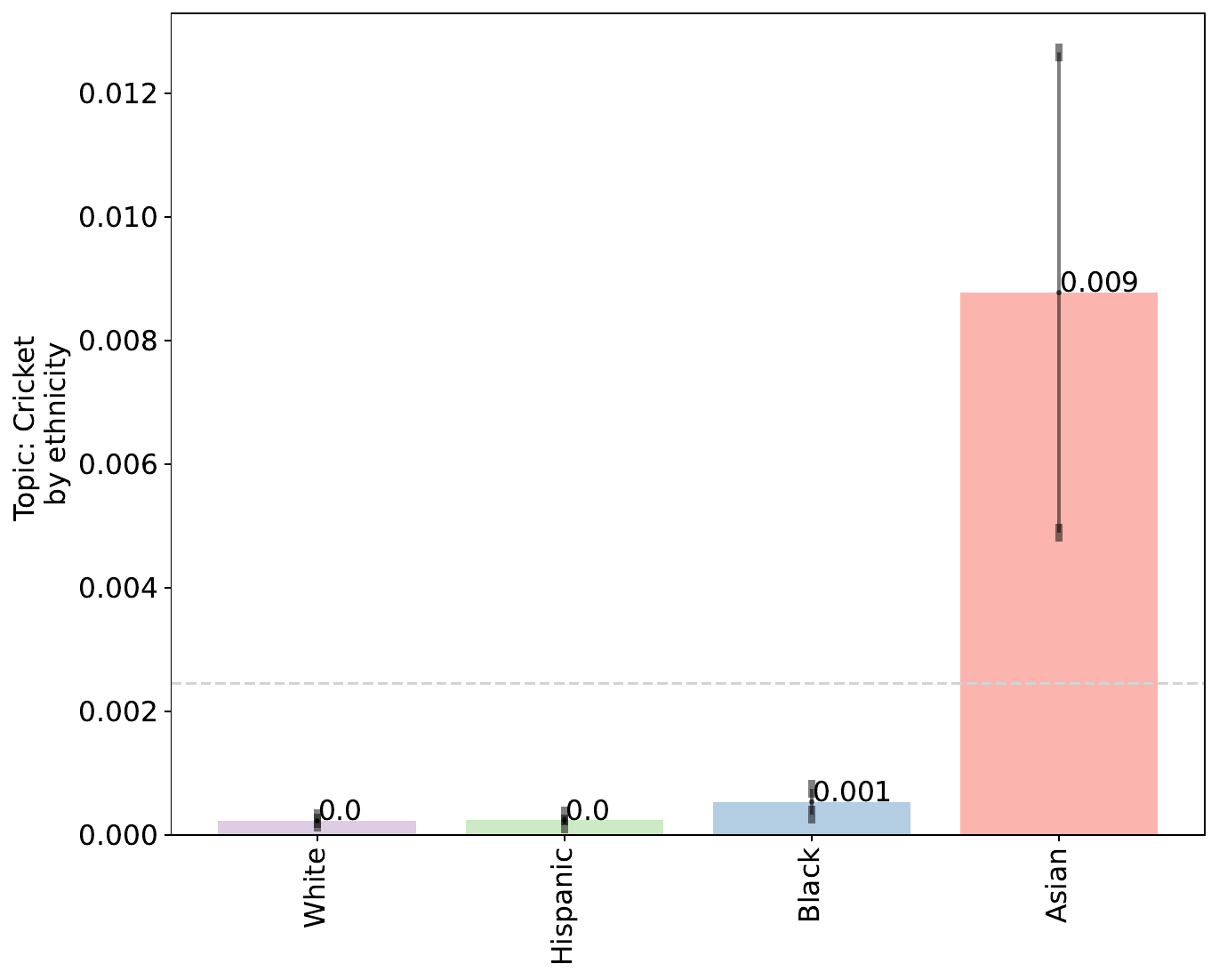}
    \caption{Cricket}
    \label{fig:3}
\end{subfigure}
\hfill
\begin{subfigure}[b]{0.25\textwidth}
    \includegraphics[width=\textwidth]{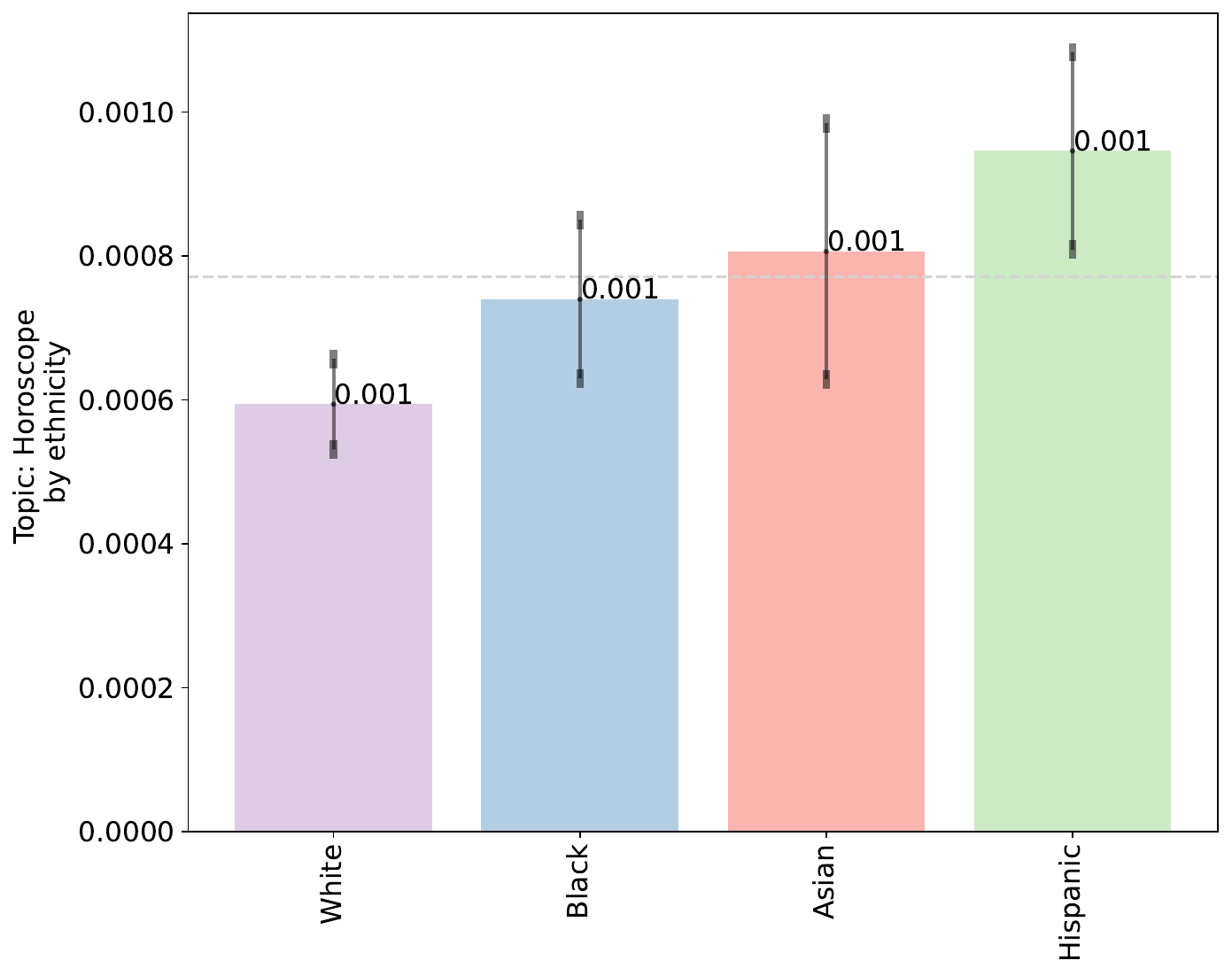}
    \caption{Horoscope}
    \label{fig:4}
\end{subfigure}
\caption{Topic specific bar charts showing prevalence of various topics.}
\label{fig:topic_appendix4}
\vspace{-\baselineskip}
\end{figure*}

\begin{figure*}[ht]
\centering
\begin{subfigure}[b]{0.24\textwidth}
    \includegraphics[width=\textwidth]{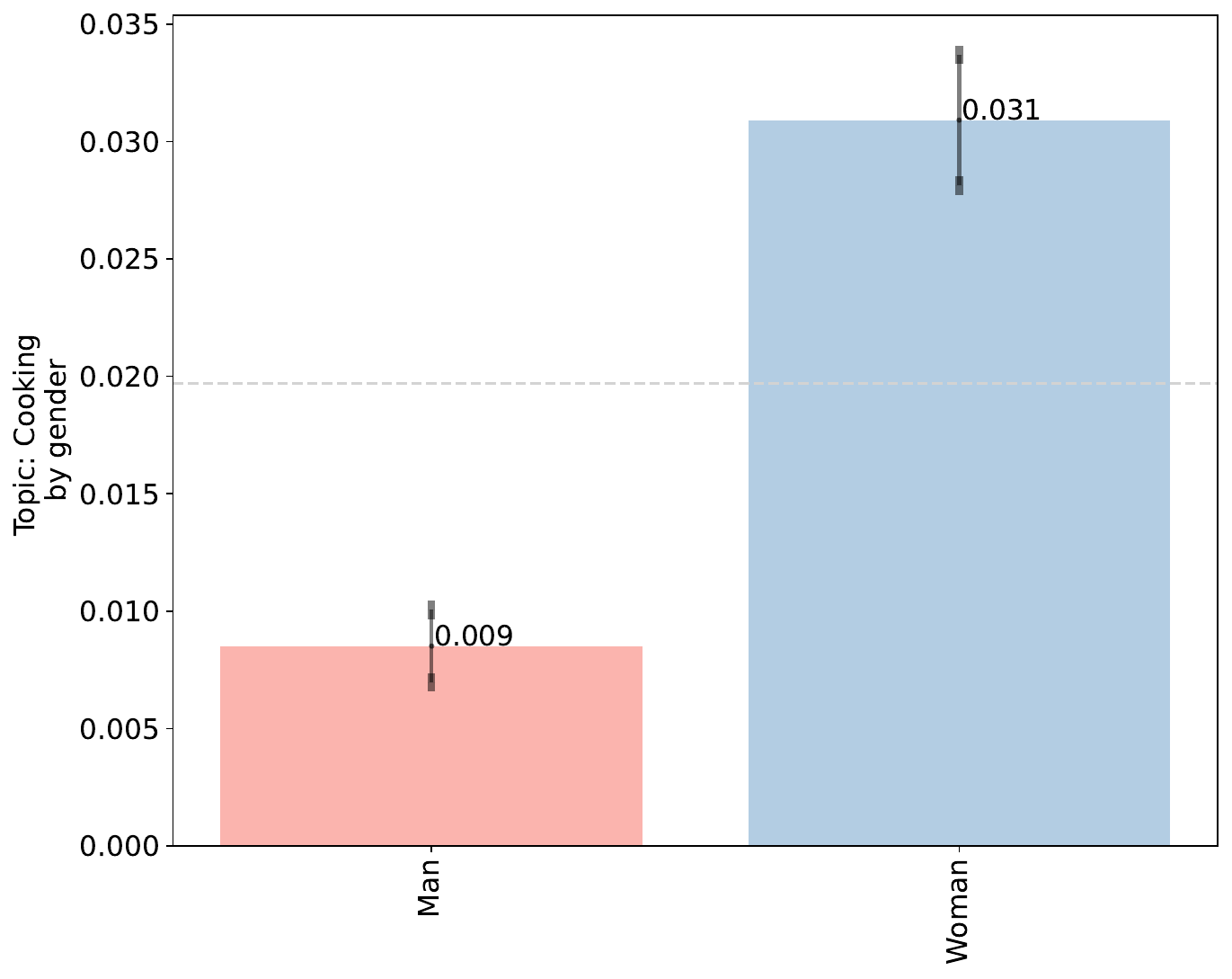}
    \caption{Cooking}
    \label{fig:1}
\end{subfigure}
\hfill
\begin{subfigure}[b]{0.24\textwidth}
    \includegraphics[width=\textwidth]{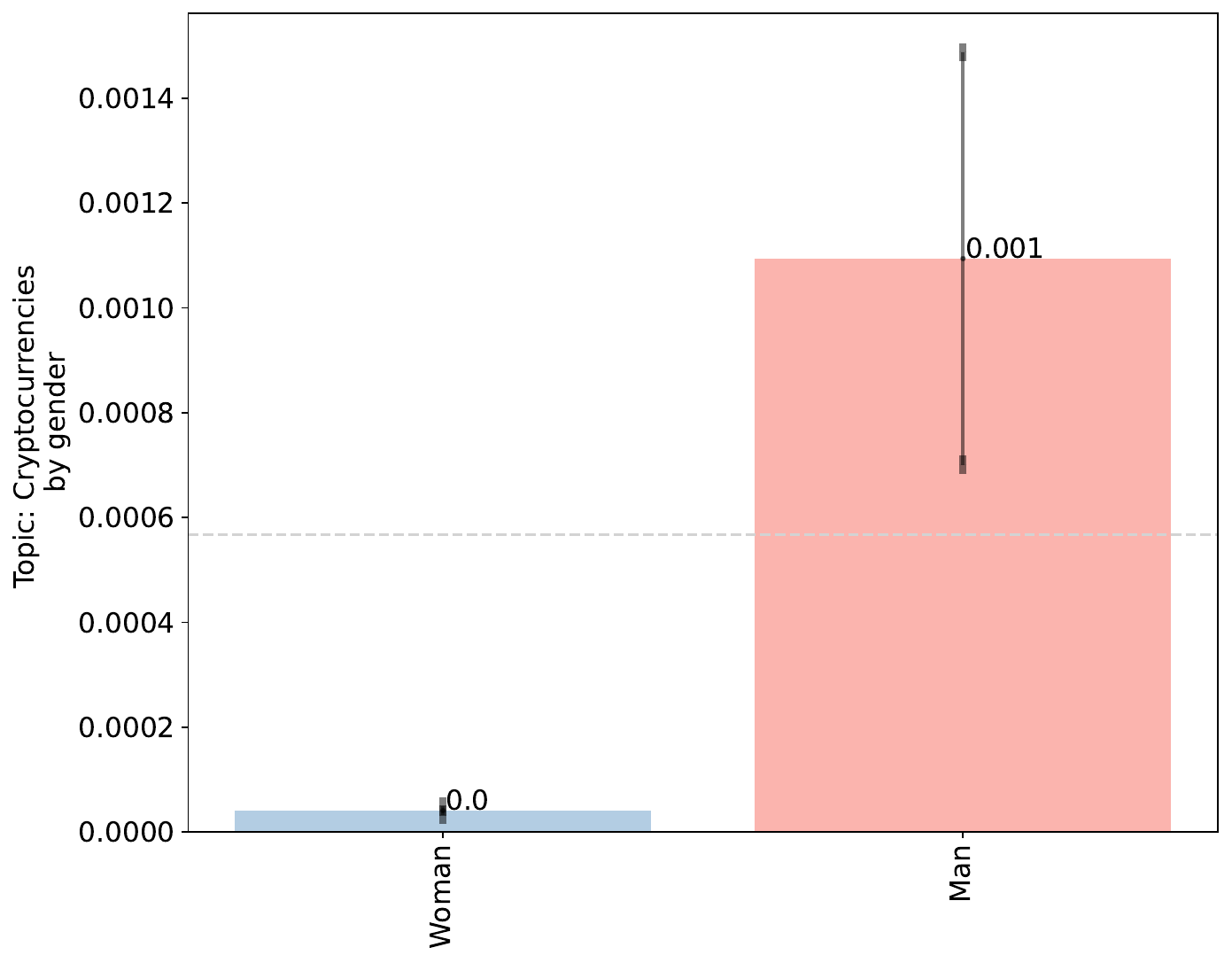}
    \caption{Crypto currencies}
    \label{fig:2}
\end{subfigure}
\hfill
\begin{subfigure}[b]{0.24\textwidth}
    \includegraphics[width=\textwidth]{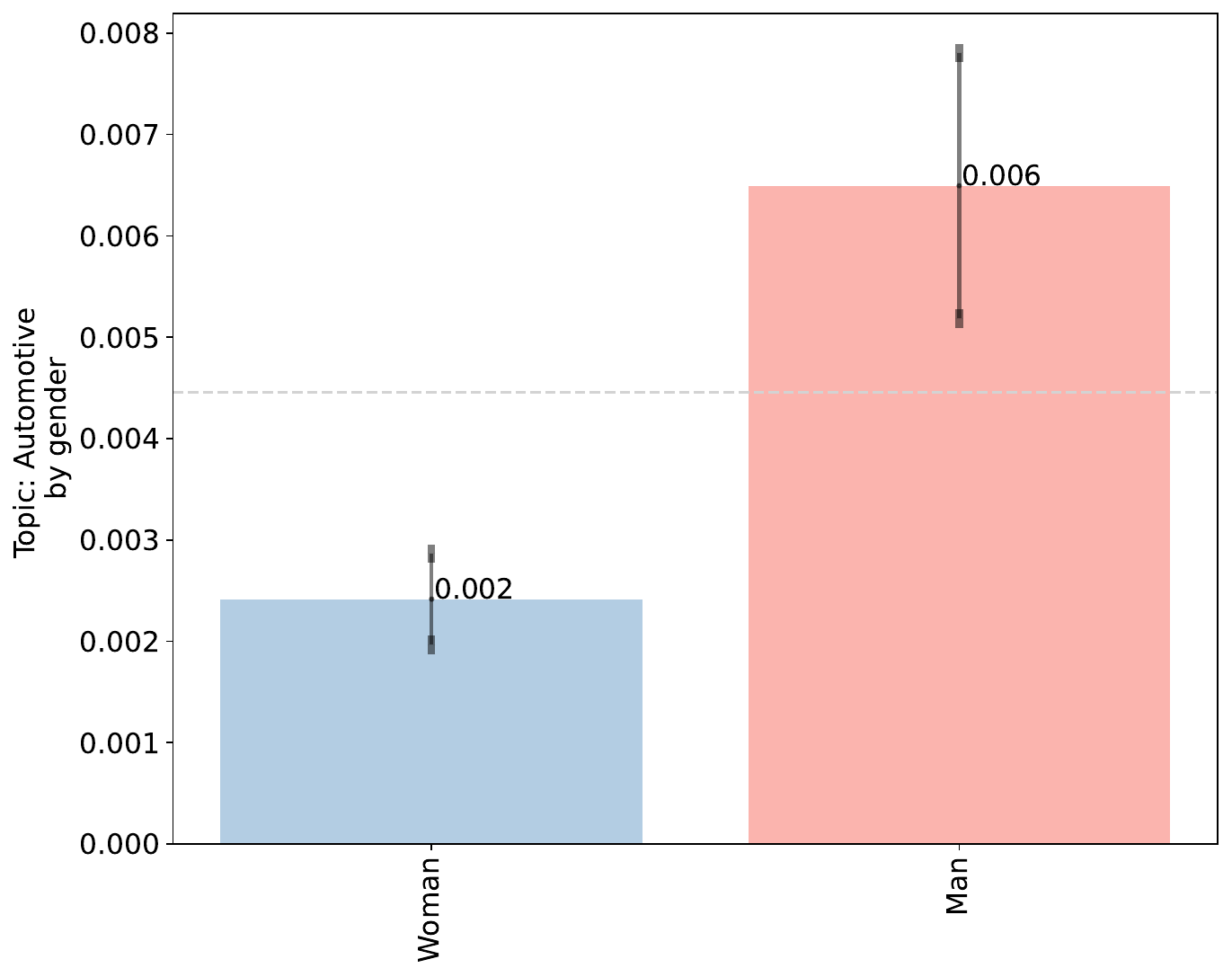}
    \caption{Automotives}
    \label{fig:3}
\end{subfigure}
\hfill
\begin{subfigure}[b]{0.25\textwidth}
    \includegraphics[width=\textwidth]{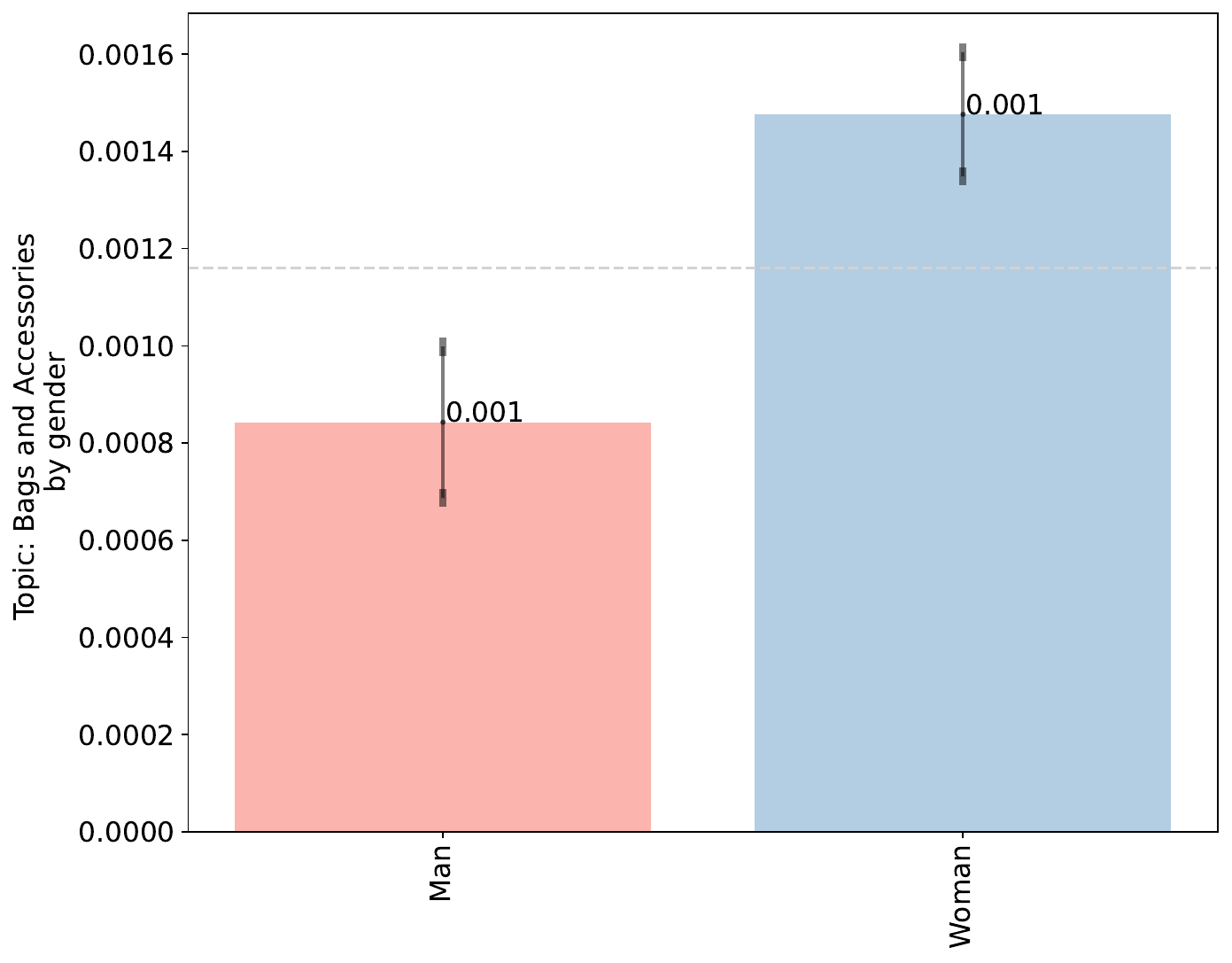}
    \caption{Bags and accessories}
    \label{fig:4}
\end{subfigure}
\caption{Topic specific bar charts showing prevalence of various topics.}
\label{fig:topic_appendix5}
\vspace{-\baselineskip}
\end{figure*}

\begin{figure*}[ht]
\centering
\begin{subfigure}[b]{0.24\textwidth}
    \includegraphics[width=\textwidth]{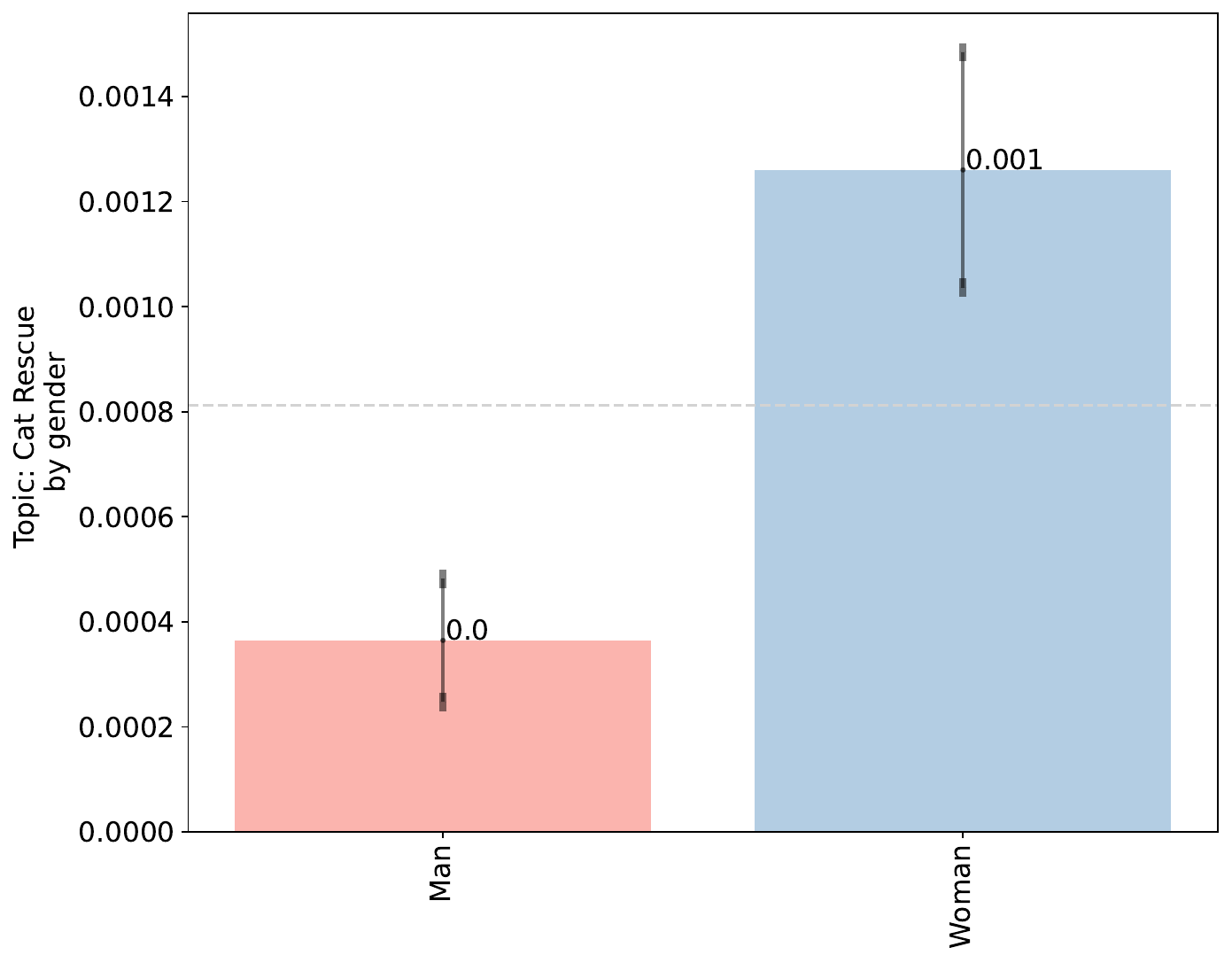}
    \caption{Cats}
    \label{fig:1}
\end{subfigure}
\hfill
\begin{subfigure}[b]{0.24\textwidth}
    \includegraphics[width=\textwidth]{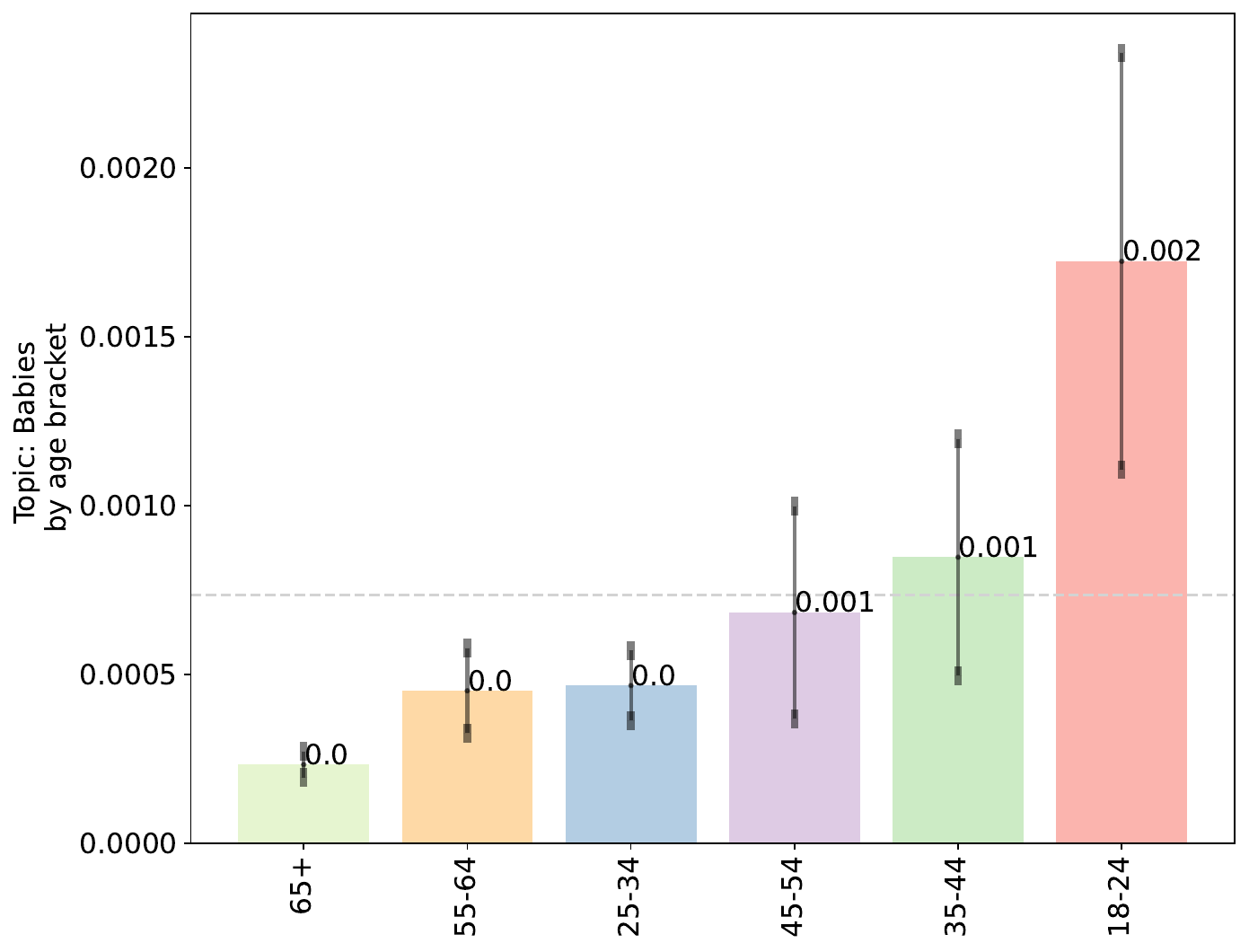}
    \caption{Cute babies}
    \label{fig:2}
\end{subfigure}
\hfill
\begin{subfigure}[b]{0.24\textwidth}
    \includegraphics[width=\textwidth]{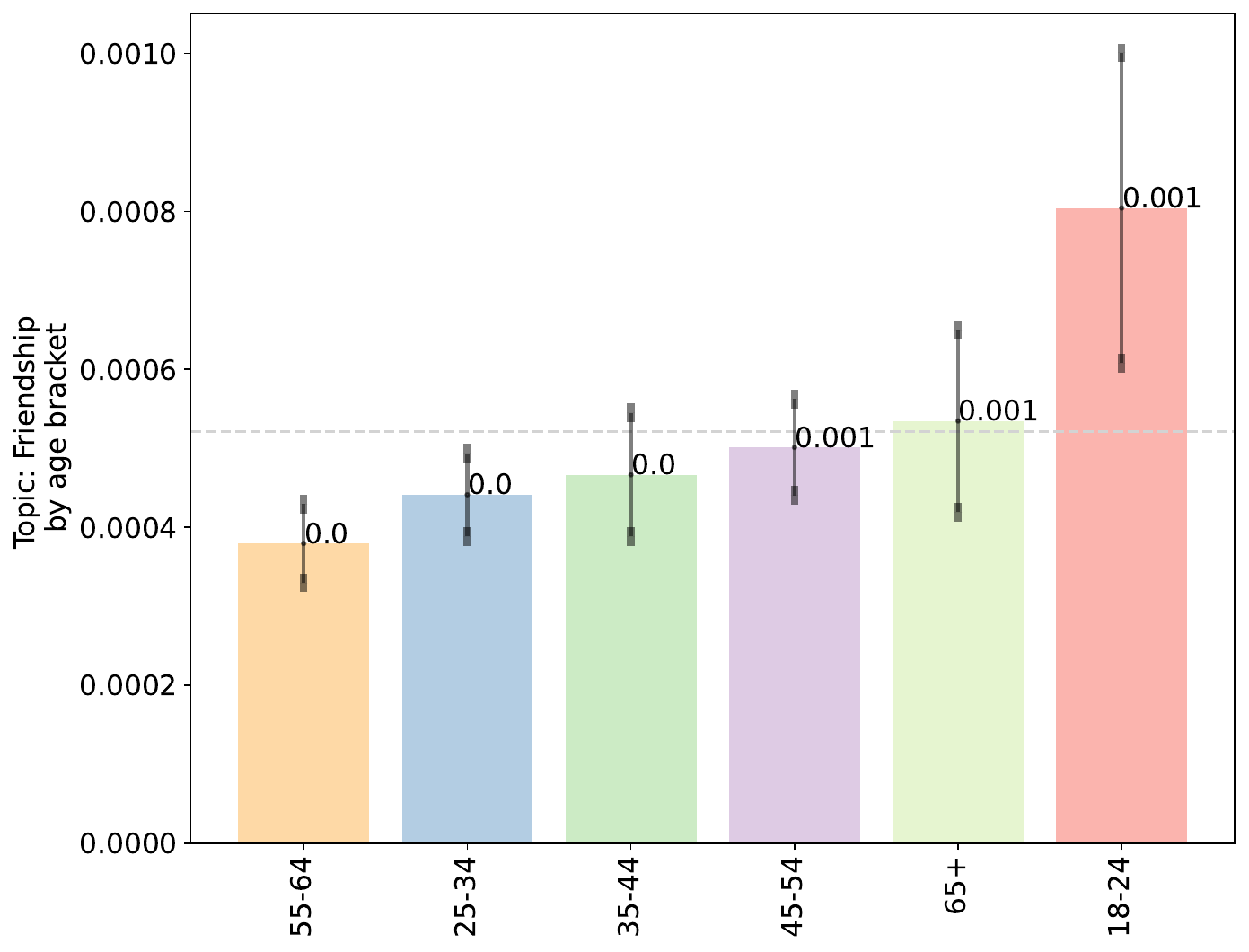}
    \caption{Friendship}
    \label{fig:3}
\end{subfigure}
\hfill
\begin{subfigure}[b]{0.25\textwidth}
    \includegraphics[width=\textwidth]{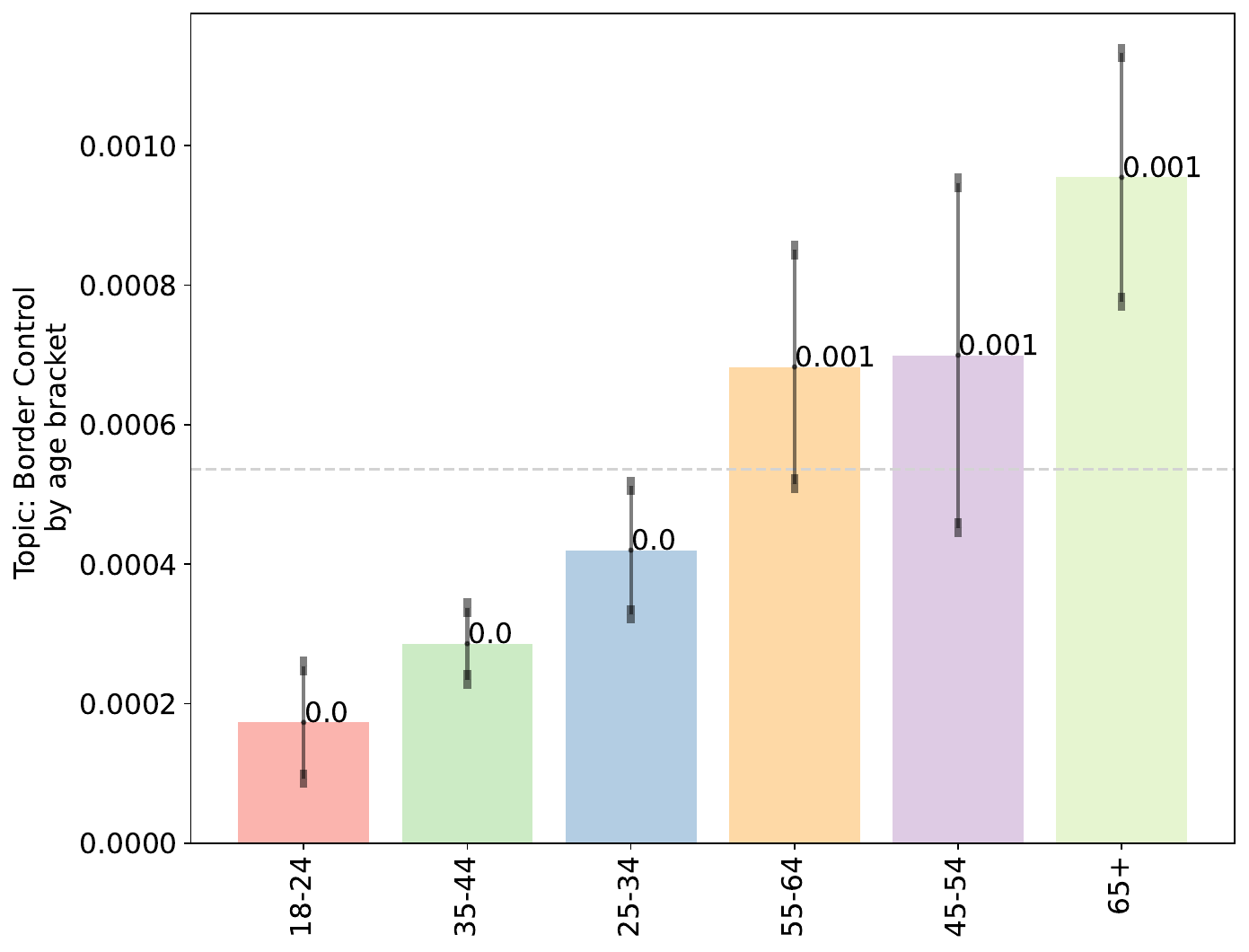}
    \caption{Migrants on the border}
    \label{fig:4}
\end{subfigure}
\caption{Topic specific bar charts showing prevalence of various topics.}
\label{fig:topic_appendix6}
\vspace{-\baselineskip}
\end{figure*}

\end{document}